\documentclass[twocolumn]{article}
\usepackage[utf8]{inputenc}
\usepackage[T1]{fontenc}
\usepackage[UKenglish]{babel}
\usepackage[english]{varioref}
\usepackage[nottoc]{tocbibind}
\usepackage[toc,page]{appendix}
\usepackage{caption}
\usepackage{amsmath}
\usepackage{amsthm}
\usepackage{amssymb}
\usepackage{siunitx}
\usepackage{adjustbox}
\usepackage{mathrsfs}
\usepackage{geometry}
\usepackage{float}
\usepackage[caption = true]{subfig}
\usepackage{caption}
\usepackage{subcaption}
\usepackage{graphicx}
\usepackage{array}

\usepackage{xcolor}

\usepackage{tikz}
\usetikzlibrary{arrows.meta, positioning, shapes.geometric, calc}

\usepackage{hyperref}
\hypersetup{
    colorlinks = True,
    linkcolor = blue,
    citecolor = magenta
}
\usepackage[font = small,labelfont = bf]{caption}
\usepackage{booktabs}

\usepackage{csquotes}
\usepackage{chngcntr}  

\usepackage{authblk}
\usepackage{subfiles}
\usepackage{graphbox}

\usepackage[table, svgnames, dvipsnames]{xcolor}
\definecolor{deepskyblue}{RGB}{0,191,255}

\usepackage{tikz}
\usetikzlibrary{positioning,fit}

\usetikzlibrary{arrows.meta, positioning, shapes.misc, calc}

\definecolor{MyGreenish}{RGB}{150,210,104}
\definecolor{MyLightGreenYellow}{RGB}{215, 241, 145}
\definecolor{MyYellowOrange}{RGB}{253, 202, 121}
\definecolor{MyReddishOrange}{RGB}{246, 122, 73}
\definecolor{Silver}{RGB}{220,220,220}

\definecolor{DEEPSKYBLUE}{RGB}{0,191,255}
\definecolor{LIME}{RGB}{0,255,0}
\definecolor{ORANGE}{RGB}{255,165,0}
\definecolor{MAGENTA}{RGB}{255,0,255}

\usepackage{pdflscape} 
\usepackage{booktabs}   
\usepackage{ragged2e}   
\usepackage{calc}       

\usepackage{multirow}

\usepackage{titlesec}

\titleformat*{\section}{\bfseries}
\titleformat*{\subsection}{\itshape}
\titleformat*{\subsubsection}{\itshape}

\titleformat{\paragraph}[runin]
  {\normalfont\itshape}   
  {\theparagraph.}       
  {1em}
  {}

\newcommand{\tpm}{%
  \mathbin{\vcenter{\offinterlineskip
    \hbox{$\scriptstyle\times$}
    \hbox{$\scriptstyle\div$}}}}

\newcommand{\Stot}{S_{\mathrm{tot}}}
\newcommand{\Vtot}{V_{\mathrm{tot}}}
\newcommand{\Vplasma}{V_{\mathrm{p}}}
\newcommand{\Splasma}{S_{\mathrm{p}}}
\newcommand{\Sdiv}{S_{\mathrm{div}}}
\newcommand{\Vdiv}{V_{\mathrm{div}}}
\newcommand{\Ldiv}{L_{\mathrm{div}}}
\newcommand{\Rdiv}{R_{\mathrm{div}}}
\newcommand{\ftor}{f_{\mathrm{tor}}}

\newcommand{\GammaQ}{\Gamma_{\mathrm{Q}}^{\mathrm{puff}}}
\newcommand{\GammaZ}{\Gamma_{\mathrm{Z}}^{\mathrm{puff}}}
\newcommand{\GammaQcore}{\Gamma_{\mathrm{Q}}^{\mathrm{core}}}
\newcommand{\GammaTcore}{\Gamma_{\mathrm{T}}^{\mathrm{core}}}
\newcommand{\GammaDcore}{\Gamma_{\mathrm{D}}^{\mathrm{core}}}
\newcommand{\GammaQex}{\Gamma_{\mathrm{Q}}^{\mathrm{ex}}}
\newcommand{\GammaTex}{\Gamma_{\mathrm{T}}^{\mathrm{ex}}}
\newcommand{\GammaDex}{\Gamma_{\mathrm{D}}^{\mathrm{ex}}}
\newcommand{\GammaHeCore}{\Gamma_{\mathrm{He}}^{\mathrm{core}}}
\newcommand{\GammaNeCore}{\Gamma_{\mathrm{Ne}}^{\mathrm{core}}}
\newcommand{\GammaTot}{\Gamma_{\mathrm{tot}}}
\newcommand{\GammaPuff}{\Gamma^\mathrm{puff}}
\newcommand{\GammaTsep}{\Gamma_\mathrm{T}^\mathrm{TP}}
\newcommand{\GammaQDIR}{\Gamma^\mathrm{DIR}_\mathrm{Q}}
\newcommand{\GammaTPuff}{\Gamma_\mathrm{T}^\mathrm{puff}}
\newcommand{\GammaDPuff}{\Gamma_\mathrm{D}^\mathrm{puff}}
\newcommand{\GammaCore}{\Gamma^\mathrm{core}}
\newcommand{\GammaTcorePI}{\Gamma_\mathrm{T,PI}^\mathrm{core}} 
\newcommand{\GammaTpuff}{\Gamma_\mathrm{T}^\mathrm{puff}}
\newcommand{\GammaNepuff}{\Gamma_\mathrm{Ne}^\mathrm{puff}}

\newcommand{\GammaPump}{\Gamma_{\mathrm{pump}}}

\newcommand{\GammaBypass}{\Gamma_{\mathrm{bp}}}

\newcommand{\GammaZges}{\Gamma_{\mathrm{Z,GES}}^{\mathrm{puff}}}
\newcommand{\GammaZsudo}{\Gamma_{\mathrm{Z,Sudo}}^{\mathrm{puff}}}
\newcommand{\GammaZstar}{\Gamma_{\mathrm{Z*}}^{\mathrm{puff}}}

\newcommand{\cZ}{c_{\mathrm{Z}}^{\mathrm{out}}}
\newcommand{\cHe}{c_{\mathrm{He}}^{\mathrm{out}}}
\newcommand{\fBurn}{f_{\mathrm{b}}}
\newcommand{\etaFuel}{\eta_\mathrm{fuel}}

\newcommand{\GammaQcoreNBI}{\Gamma_{\mathrm{Q,\,NBI}}^{\mathrm{core}}}
\newcommand{\GammaQcoreNBImax}{\Gamma_{\mathrm{Q,\,NBI}}^{\mathrm{core,max}}}
\newcommand{\GammaQcorePellet}{\Gamma_{\mathrm{Q,\,PI}}^{\mathrm{core}}}
\newcommand{\GammaQcorePelletNominal}{\Gamma_{\mathrm{Q,\,PI}}^{\mathrm{core,ref}}}
\newcommand{\GammaQcorePelletMax}{\Gamma_{\mathrm{Q,\,PI}}^{\mathrm{core,max}}}
\newcommand{\GammaQcoreSim}{\Gamma_{\mathrm{Q,\,true}}^{\mathrm{core}}}

\newcommand{\GammaQcorePhys}{\Gamma_{\mathrm{Q,\,phys}}^{\mathrm{core}}}
\newcommand{\etaPuffToCore}{\eta_{\mathrm{puff \rightarrow core}}}

\newcommand{\ratioPuffToCore}{\mathcal{R}^{\rm puff}_{\rm core}}

\newcommand{\nesep}{n_{\mathrm{sep}}}
\newcommand{\neavg}{n_{\mathrm{avg}}}
\newcommand{\neGW}{n_{\mathrm{GW}}}
\newcommand{\neSu}{n_{\mathrm{Su}}}
\newcommand{\neStar}{n_{\mathrm{*}}}
\newcommand{\tauIPB}{\tau_{\mathrm{IPB98(y,2)}}}
\newcommand{\tauISS}{\tau_{\mathrm{ISS04}}}
\newcommand{\tauStar}{\tau_{\mathrm{E}*}}
\newcommand{\tauP}{\tau_{\mathrm{p}}}
\newcommand{\tauE}{\tau_{\mathrm{E}}}

\newcommand{\lqEich}{\lambda^{\mathrm{E}}_q}
\newcommand{\fScara}{f^{\mathrm{S}}}
\newcommand{\BT}{B_{\mathrm{T}}}
\newcommand{\Pin}{P_{\mathrm{in}}}
\newcommand{\Pnbi}{P_{\mathrm{NBI}}}
\newcommand{\Enbi}{E_{\mathrm{NBI}}}

\newcommand{\TBE}{\mathrm{TBE}}
\newcommand{\pdiv}{p_\mathrm{div}}
\newcommand{\pChamber}{p_\mathrm{ch}}
\newcommand{\pPump}{p_\mathrm{pump}}
\newcommand{\Seff}{S_\mathrm{eff}}
\newcommand{\Spump}{S_\mathrm{pump}}
\newcommand{\tbrr}{\mathrm{TBR}_\mathrm{r}}
\newcommand{\Ist}{I_\mathrm{start-up}}
\newcommand{\fdir}{f_\mathrm{DIR}}
\newcommand{\fTdir}{f_\mathrm{T}^\mathrm{DIR}}
\newcommand{\fbp}{f_\mathrm{bp}}
\newcommand{\fTpuff}{f_\mathrm{T}^\mathrm{puff}}
\newcommand{\taudir}{\tau_\mathrm{DIR}}
\newcommand{\sne}{S_\mathrm{Ne}}
\newcommand{\necore}{N_\mathrm{Ne,core}}
\newcommand{\nediv}{N_\mathrm{Ne,div}}
\newcommand{\ftcore}{f_\mathrm{T}^\mathrm{core}}
\newcommand{\ftpuff}{f_\mathrm{T}^\mathrm{puff}}
\newcommand{\pfus}{P_\mathrm{fus}}
\newcommand{\ntdir}{N_\mathrm{T}^\mathrm{DIR}}
\newcommand{\nqdir}{N_\mathrm{Q}^\mathrm{DIR}}

\title{\textbf{On the Relationship Between Plasma and Tritium Fuel Cycle Through Matter Injection and Particle Exhaust}}

\author[1]{S. Meschini}
\author[2]{M. Moscheni}

\affil[1]{Politecnico di Torino, Corso Duca degli Abruzzi 24, Torino, Italy}
\affil[2]{Gauss Fusion GmbH, Parkring 29, 85748 Garching bei M\"unchen, Germany}

\date{}
\begin{document}

\twocolumn[
\begin{@twocolumnfalse}

\maketitle

\begin{abstract}

    
    This work identifies an inconsistency between plasma operating scenarios and the deliverable requirements of the tritium fuel cycle (TFC), calling for a re-examination of the traditional reactor-led design approach. The key is simple: with current TFC architectures, fuel puffing must contain tritium.
    Results from Moscheni \textit{et al} (2026 \textit{Nucl. Fusion} \textbf{66} 026008) investigated fuel puffing rates in detached divertor operation. Expanding the original database, the resulting figures are shown to exceed core fuelling by an order of magnitude in tokamaks and stellarators, from present-day experiments to next-step fusion devices. Though not unknown within the plasma community, TFC models instead assumed core fuelling to dominate particle balance.
    The implications are severe. In recent TFC architectures, direct internal recycling (DIR) is intended to minimise tritium inventory---but implicitly assumes near-50:50 D:T injection and exhaust compositions. This assumption may become self-defeating: a substantial fraction of the puffed fuel---now recognised to be far larger than anticipated---must be tritium. Tritium inventories, doubling time, required tritium breeding ratio, and pump sizing become critical once puffing is properly accounted for.
    Mitigation strategies are explored by extending the models of Meschini \textit{et al} (2023 \textit{Nucl. Fusion} \textbf{63} 126005). For a notional power plant, realistic TFC requirements can be satisfied by enforcing D-rich, T-lean fuel puffing ratios. However, the D:T imbalance in the exhaust, which eventually reaches the fuelling system through the DIR loop, propagates into the core plasma leading to a reduction of order $10\%$ in fusion power. Alternatively, for near-50:50 D:T puff ratio, reduced fuel puffing combined with stronger impurity seeding would maintain detachment while alleviating TFC constraints---albeit with increased core plasma contamination. The possible combined use of such strategies enables the identification of scenarios that minimise tritium inventory and throughput, while balancing the competing requirements of detached operation, acceptable impurity levels, and fusion performance.
    Ultimately, these results place renewed emphasis on the TFC as a central element of reactor design. This exemplifies the kind of plasma--TFC interface that the IAEA has highlighted as imposing important constraints on magnetic-confinement fusion.
    A viable fusion reactor requires integration of core plasma, edge plasma, and TFC, implying unavoidable trade-offs across all three. And although the present work provides possible implementations of these trade-offs, their broader exploration now requires targeted attention from the fusion community.

\end{abstract}

\vspace{1em}

\end{@twocolumnfalse}
]

\section{Introduction}\label{sec:intro}

    Historically, research emphasis in magnetic confinement fusion has been roughly inversely proportional to the distance from the core plasma. This prioritisation has served the field well during the physics exploration phase, where understanding and controlling the confined plasma was the primary objective. However, a nuclear fusion reactor ultimately represents a tightly integrated engineering system, composed of strongly interconnected subsystems whose constraints collectively determine the achievable operating point and performance.
    
    As the field transitions toward reactor design and system-level investigations advance \cite{COLEMAN201926, meneghini2024fuse, kovari2016process, DRAGOJLOVIC2010243, NAKAMURA2012864, coleman2025bluemira, FABLE2018131, Reux_2015, FRANZA20151767, HONG20081615}, it becomes increasingly evident that historical emphasis is not necessarily commensurate with system-level importance. Several elements located farther away from the core plasma (e.g. vacuum exhaust \cite{Zito_2025, varoutis_effect_2019, Tantos_2024}, breeding blanket \cite{IHLI2008912, FROIO2016116}, materials \cite{ROWCLIFFE2018290, GORLEY2021112513} and remote maintenance \cite{CROFTS2022113121, TESINI2008810}) have only recently come to be widely recognised as critical to reactor operation \cite{donne2019}---together with the importance of their integration \cite{KEMBLETON2022113120}. Consequently, these systems remain at a lower level of maturity, yet they may impose significant constraints on overall integration.

    The necessity of joint integration between the core and edge plasma regions is already well recognised, even if such integration is not yet fully achieved in practice \cite{Cowley_2026}. Similarly, the interaction between the edge plasma and plasma-facing components must be carefully controlled \cite{ZOHM2013428, PITTS201760, LOARTE2017256, PITTS2019100696}. These interactions are mediated through a set of engineering control inputs that regulate particle and power flows into and across the plasma. Among these, matter injection systems play a central role \cite{Ploeckl19052021}.

    Despite their importance in the quality of ``real controls'' \cite{Kukushkin_2016, PITTS2019100696}, the relationship between plasma conditions and matter injection controls has received relatively limited systematic investigation---much as particle balance has historically received less attention than power balance \cite{Na_2019}. Early work explored fuelling requirements and gas injection strategies in the context of ITER \cite{PACHER2003657}, including both gas puffing and core fuelling within scaling frameworks. Fuelling requirements have also been assessed through more in-depth studies \cite{Na_2019, Koechl_2020}. Broader comparative efforts do exist, notably \cite{PACHER2007400}, but cross-machine, database-wide investigations from which clear universal trends and reliable extrapolation can be readily established remain elusive.
    
    
    The involvement of matter injection systems has direct implications for their matter extraction counterparts---particle exhaust---and, ultimately, for the tritium fuel cycle (TFC). While tritium management was relevant in a limited number of past experimental devices \cite{Kappatou_2025, VILLARI2025115133, Strachan_1994}, it becomes central in power plant-relevant conditions. Future fusion reactors must achieve tritium self-sufficiency while minimising tritium inventories and complying with stringent regulatory constraints associated with radioactive materials. These considerations are not only critical for first-of-a-kind burning-plasma devices but also for establishing a sustainable fleet of future reactors that must ultimately be fuelled by bred tritium \cite{meschini2026tritium}.
    
    In this broader context, the International Atomic Energy Agency (IAEA) has highlighted the plasma--TFC interface as a \textit{neglected area despite its potentially large implications for the design and operation of magnetic confinement fusion systems} \cite{IAEA_TECDOC_2049_2024}. Concepts such as direct internal recycling (DIR) thus attract interest as promising strategies to minimise tritium inventories within the fuel cycle \cite{day_smart_2019}.
    
    
    Nevertheless, a coherent integration of the tritium fuel cycle together with core and edge plasma operation remains largely unexplored, though recent works are beginning to bridge these fields \cite{whyte_tritium_2023, masline_modeling_2025, hattab_analysis_2025, Meschini_2025_SOFE, Morandi_2025_SOFE}. Multi-machine investigations transparently linking fuelling/impurity injections and the TFC have only recently begun to emerge for reactor-relevant detached-plasma scenarios \cite{Moscheni_2026}.
    
    Because gas puffing still remains comparatively under-investigated, and its relationship with core fuelling even less so, a critical uncertainty arises. As reported by the IAEA, \textit{overambitious plasma requirements may render the fuel cycle plant unattractively large} \cite{IAEA_TECDOC_2049_2024}---similarly noted by Henderson \textit{et al.} \cite{Henderson_2025}. Given the complexity of the power-exhaust problem \cite{LOARTE2017256}, and the extent to which its mitigation relies on fuel and impurity puffing strengths, elevated injection rates are naturally expected in reactor-relevant operating regimes. In this context, \textit{if puffing demand exceeds that associated with core fuelling---contrary to historical assumptions---new and severe constraints would emerge for the fuel cycle}. Examples include an higher tritium throughput for D:T gas puffing, highly unbalanced D:T ratios through the DIR loop for D-dominated puffing, high impurity fraction in the exhaust for an impurity-dominated approach. As the DIR was shown to be the most effective way to improve the fuel cycle performance in DEMO-like architectures \cite{day_smart_2019, abdou_physics_2020, meschini_modeling_2023, clark_breeder_2025, jin2026cfedr}, although yet to be demonstrated at a reactor scale, \textit{any deviation from the assumption of negligible gas puffing challenges fuel cycle designs}.
    
    Recent works, namely \cite{Moscheni_2026} and---crucially within the TFC modelling realm---Hattab \textit{et al.}~\cite{hattab_analysis_2025}, have begun to explicitly highlight the potentially dominant role of gas puffing relative to core fuelling. The possibility itself is known within the plasma community at least since 1992 \cite{PETRIE1992848}. Yet, it has not been established across machines in a form that yields a general system-level picture. Here, by analysing a wide multi-machine database and extracting the underlying trends, we address one concrete instance of the broader plasma-TFC integration challenge identified by the IAEA \cite{IAEA_TECDOC_2049_2024}.
    
    In this work, we do not derive detailed design formulae for a specific reactor. Instead, we characterise the magnitude of the issue, and explore mitigation strategies together with the trade-offs that must be sought within an integrated framework.

    The challenge is to ensure a gas puffing rate high enough to sustain satisfactory edge plasma conditions, while keeping the tritium content of the puffed gas manageable. At the same time, the resulting exhaust composition must remain compatible with efficient DIR operation. This is essential to limit tritium demand and inventories, and must avoid excessive impurity accumulation in the core plasma.
    
    Specifically, section \ref{sec:intro_matter_injection_plasma} provides an overview of matter injection in fusion devices and its effects, with particular focus on core fuelling as well as fuel and impurity gas puffing. These systems constitute bridge between plasma physics and the tritium fuel cycle, and therefore must be framed in a way accessible to readers approaching the problem from either community. Section \ref{sec:intro_matter_injection_TFC} examines their implications for the tritium fuel cycle, before section \ref{sec:intro_conundrum} clarifies the central issue. Section \ref{sec:methods} describes the methods and tools used to systematise and generalise the available data. The results are presented in section \ref{sec:results}, followed by a discussion of their broader implications in section \ref{sec:discussion}. Finally, section \ref{sec:conclusion} summarises the conclusions of this work and outlines future directions that we believe are crucial for the community to address.

\section{Matter Injection in Plasmas}\label{sec:intro_matter_injection_plasma}
    
    \subsection{Overview}\label{sec:intro_matter_injection_plasma__overview}
    
        Matter injection systems govern all particle streams entering the vacuum vessel, thereby constituting the direct interface between engineering systems and the physics of core and edge plasma \cite{Ploeckl19052021}. Among these are: cryogenic D--T pellets for core fuelling and density control \cite{geulin_pellet_2022, Panadero_2023, Valovic_2024}; smaller cryogenic D pellets for ELM pacing in tokamaks \cite{Lang_2015, Wilcox_2022}; neutral beam injectors \cite{Hemsworth_2017, Singh_2017, Shikhovtsev_2024}; shattered pellets and dedicated gas manifolds for disruption mitigation in tokamaks \cite{Lee_2024}; and gas puffing systems for divertor impurity seeding and edge density control \cite{Kallenbach_2015, Baiocchi_2023}. 
        
        Through the imposed particle fluxes, matter injection systems set the total hydrogen (D and T) and impurity throughput circulating in the machine. In turn, this defines boundary conditions for plasma operation as well as tritium inventory requirements (start-up and steady-state) and the processing load on the inner fuel cycle.
        
        In steady-state reactor scenarios, this role is dominated by core fuelling and gas puffing.

    \subsection{Core fuelling}\label{sec:intro_matter_injection_plasma__core_fuelling}

        Neutral beam injection (NBI) contributes a finite particle source to the confined plasma and must therefore be accounted for in the global particle balance. Additional details on the treatment of NBI within the present study are provided in appendix \ref{apx:estimation_GammaQcore}. Pellet injection (PI) is generally expected to represent the more reactor-relevant core fuelling mechanism in future reactor-scale devices \cite{geulin_pellet_2022}---and is therefore the main focus here.
        
        The required D--T pellet injection rate is set by the amount of fuel that must be deposited in the hot core plasma region (i.e.\ within the active core region). It therefore depends on two key efficiencies \cite{Ploeckl19052021}: the fraction of pellets that reach the vacuum vessel intact ($c_\mathrm{ext}$), and the fraction of pellet mass effectively deposited inside the core plasma ($\eta_\mathrm{fuel}$). The fraction $c_\mathrm{ext}$ can be kept close to unity by properly tuning the pellet speed to maintain the pellet intact while traveling through the guide tubes \cite{combs_pellet-injector_2018}. This requirement is in competition with high pellet speeds to improve the fuelling efficiency. Although experiments report satisfactory fuelling efficiency on the order of $\eta_\mathrm{fuel} = 80$--$100\%$ \cite{baylor_comparison_2003}, extrapolations to tokamak pilot plants with $c_\mathrm{ext}\eta_\mathrm{fuel} \sim 20$--$30\%$ (because of the ejection of some fuel due to ELMs \cite{abdou_physics_2020}) would lead to a substantial fraction of injected fuel that does not directly contribute to core density control and instead recirculates ``unburnt'' through the inner fuel cycle. In stellarators, by contrast, this specific degradation pathway is expected to be less prominent due to shallower gradients \cite{Nespoli2022}, potentially favouring higher fuelling efficiency \cite{Panadero2023, Panadero_2023}.
        
        Increasing pellet velocity enhances penetration depth, as described by advanced ablation and deposition models \cite{lang_considerations_2015}. However, high-field-side launch---generally favoured for improved deposition efficiency---requires curved guide tubes \cite{geulin_pellet_2022}, which impose practical limits on survivable pellet speed and introduce additional mass loss \cite{combs_pellet-injector_2018, lang_considerations_2015}.
        

        From a throughput perspective, pellet penetration depth directly influences the required particle injection rate \cite{McClenaghan_2023}: insufficient penetration must be compensated by higher fuelling rates to sustain the desired core density, with potential variations of up to a factor $\sim 4$ depending on deposition characteristics \cite{lang_considerations_2015}. Regardless of the precise value or D:T mixture, pellet injection remains a dominant, burdensome driver of tritium throughput and therefore of fuel cycle load in reactor scenarios \cite{lang_considerations_2015}.
        
        Irrespective of its engineering complexity, core fuelling should be expected to follow a relatively well-defined system-level scaling, as preliminarily speculated in footnote \#13 of \cite{Moscheni_2026}. This motivates the systematic analysis undertaken in the present study. At the same time, that same observation hinted at the relative magnitude of core fuelling compared to gas puffing, motivating the detailed assessment of puffing presented in the following.

    \subsection{Gas puffing}\label{sec:intro_matter_injection_plasma__puffing}

        In contrast to the technological complexity of core fuelling, gas puffing is functionally simple: injection lines connect room-temperature reservoirs of gaseous fuel and extrinsic impurities to the vacuum vessel.
        
        Yet, incommensurably to its simplicity, its role is paramount. Coordinated injection of fuel species Q and impurity species Z is required to establish and sustain divertor detachment \cite{krasheninnikov_physics_2017, Kallenbach_2016}. Detachment is presently the prime candidate solution to fulfil the divertor’s multi-functional mission of power and particle exhaust while limiting plasma-facing component erosion \cite{YOU2022113010}. Despite all the possible combinations of fuel and impurity puffing rates \cite{Lore_2022}, and the corresponding injection locations \cite{McCracken_1997, Pitcher_2000, PETRIE2007416, ZHOU2022113222, KAVEEVA2023101424, Park_2024_ITER, Osawa_2024, Lee_2026}, some are found as more beneficial to plasma operation than others. Indeed, through edge density control and radiative cooling, puffing directly shapes edge plasma conditions and core performance. Physics implications are well-recognised and, to the leading order, independent on the D:T puffing composition \cite{Kirov_2025, leoni2024scrape, Goldston_2012}.
        
        The need for a quantitative assessment of gas puffing as a primary engineering input has indeed been emphasised \cite{Kukushkin_2016, PITTS2019100696, Lore_2022}. Initial efforts in this direction have begun \cite{Moscheni_2026, Moscheni_2026_PSI}. Yet, the topic remains comparatively under-explored, especially in terms of system-level implications and scaling to reactor conditions---which motivated this very work. In the following, the salient features relevant to the present analysis are outlined.
        
    \subsubsection{Fuel puffing}\label{sec:intro_matter_injection_plasma__puffing__fuel}

        \paragraph{Plasma viewpoint}\label{sec:intro_matter_injection_plasma__puffing__fuel__plasma}

            The impact of fuel puffing on core and pedestal plasma \cite{SENGOKU199065, Kreter_2003, Joffrin_2017} is well illustrated by Na \textit{et al.} \cite{Na_2019} (section 3.4 therein), and by Kaveeva \textit{et al.} \cite{KAVEEVA2023101424}---who report a ``beneficial effect of high fuel throughput on separatrix and pedestal plasma composition''.
            
            From an edge-plasma perspective \cite{PITTS2019100696}, the role of fuel puffing in reactor-relevant scenarios with concomitant impurity seeding is exemplified by Lore \textit{et al.}~\cite{Lore_2022} for ITER.
            
            In the square series (figures~4a--4b \cite{Lore_2022}), deuterium puffing increases more slowly than impurity seeding. Accordingly, upstream separatrix conditions remain approximately constant while detachment is strengthened downstream. In the diamond series, fuel and impurity puff increase in parallel at constant ratio. As a result, detachment is enhanced while upstream impurity concentration is reduced (figure~4a) and plasma density increases (figure~4b). Similar behaviour has been observed elsewhere \cite{Kawashima_1999}, and experimentally in \cite{Schaffer_1995} where increased fuel puffing led to a reduction of core argon density by a factor of 20.
            
            In addition, the enhanced plasma density following fuel puffing strengthens the radiative effectiveness of impurities through the quadratic dependence of radiation on density \cite{Kukushkin_2016, KAVEEVA2023101424, Putterich_2019}.
            
            These effects depend not only on the puffing rate, but also on the injection location and, in coupled fuel-impurity control, on the relative positioning of the two systems \cite{McCracken_1997, Pitcher_2000, PETRIE2007416, KAVEEVA2023101424, Park_2024_ITER, Osawa_2024, Lee_2026}.
            
            Although the quantitative mapping between imposed puffing rates and separatrix conditions remains an active area of research~\cite{Moscheni_2026_PSI, Silvagni_2026, SILVAGNI2025101867, Henderson_2025, Kallenbach_2018, Lomanowski_2026_PSI, Eich_2026}, the general conclusion is robust: \textit{externally actuated fuel puffing is a key control knob to access and sustain detached conditions while maintaining acceptable upstream plasma performance.} This is especially important for stable, long-pulse operation \cite{Tao_2024}.
            
            A natural question follows: does fuel puffing contribute meaningfully to fuelling the core plasma? In reactor-relevant H-mode conditions, the answer is largely negative---``For core fuelling [...] conventional gas puffing most probably has to be disregarded'' \cite{lang_considerations_2015}, a conclusion supported by multiple studies \cite{Polevoi_2005, Hughes_2007, kukushkin_physics_2011, Kukushkin_2016, Na_2019, geulin_pellet_2022}. Fuel puffing predominantly fuels the plasma periphery \cite{REKSOATMODJO2021100971, Park_2024_ITER} and pedestal region \cite{Mordijck_2020}, while penetration to the hot core is strongly limited by ionisation- and charge-exchange-driven opacity \cite{MAYER1982204, Schuster_2022, Miller_2025, Wilkie_2026}. Consistently, AUG experiments indicate that the neutral flux reaching confined regions is small and that core observables can be relatively insensitive to puffing variations \cite{Luda_2020}---although a finite contribution cannot be entirely excluded. Figure 2 of \cite{Kirov_2025} (top panels b and c) clearly illustrates the above.
            
            Importantly, while the contribution of fuel puffing to core fuelling becomes negligible as reactor-relevancy is approached, the absolute magnitude of the injected flux does not.

        \paragraph{Particle exhaust}\label{sec:intro_matter_injection_plasma__puffing__fuel__exhaust}

            From the particle exhaust perspective, elevated gas fuelling can be beneficial to assist pumping. In the conditions characteristic of divertor and sub-divertor regions, fuelling sets a pressure leading to viscous/transitional flow regimes, which improves the divertor and duct conductance \cite{Tantos_2024, Varoutis_2024, TANTOS2025115021, Zito_2025}.
            
            In steady state, the divertor neutral pressure (units [Pa]) is expressed as \cite{Kallenbach_2018, Henderson_2025, Moscheni_2026, Silvagni_2026}:
            \begin{equation}\label{eq:pdiv_divertor_pressure}
                \pdiv = \frac{\GammaPump}{\Seff} \, k_{\rm B} T ,
            \end{equation}
            where the $T \, [\rm{K}]$ refers to the local neutral temperature, $k_{\rm B}$ is the Boltzmann constant, and $\GammaPump = \sum_i \Gamma_i \, [\mathrm{s^{-1}}]$ is the total pumped particle flux\footnote{In this work, when the throughput $Q = \pdiv \Seff = \GammaPump k_{\rm B} T$ is computed from particle flow rates, a reference temperature $ T = 273.15$ K is used if not differently stated.}. In steady conditions, this balances the net sum of all particle source/sink rates (fuel and impurity puffing, core fuelling, burn consumption, helium production, etc.). The total effective pumping speed seen by the divertor is:
            \begin{equation}\label{eq:effective_pumping_speed}
                \Seff = \frac{\Spump \, C}{\Spump + C} \,,
            \end{equation}
            where $\Spump \, [\rm{m^3 \, s^{-1}}]$ is the effective pumping speed (i.e. accounting for the baffles configuration, sticking coefficients of the different species, etc.) of the primary pumps and $C \, [\rm{m^3 \, s^{-1}}]$ is the vacuum duct conductance from the pump inlet to the location at which $\Seff$ (and all the other quantities) are evaluated \cite{Litovoli_2026}. Most commonly the reference location is taken at the divertor plenum and the regime is conductance-limited, whereby $C < \Spump$ and $\Seff$ primarily depends on $C$ \cite{PEARCE2013809}. For fusion technologies (cryopumps and vapour diffusion pumps) the pumping speed depends on the molar mass of the species pumped, and further species-dependent mechanisms could play a role depending on the technology, like the gas capacity and the sticking coefficient of He and hydrogenic species in cryopumps \cite{scannapiego_experimental_2017, haas2003performance}. 
            
            Once \textit{passive} pressure-enhancement strategies are maximised---e.g.\ divertor closure optimisation via dome installation and/or baffle shaping~\cite{Loarte_2001, Moulton2015Pumping, Kukushkin_2016, Reimerdes_2021, Sun_2023}---active fuel puffing remains the primary lever to increase $\pdiv$ through an increase of $\sum_i \Gamma_i$ \cite{Kallenbach_2018, Moscheni_2026}. \textit{Although enhanced fuel puffing also raises the total particle throughput to be exhausted, this is not necessarily self-defeating}. A sufficiently high neutral density (i.e.\ pressure) is required so that the pressure-driven flux through a conductance-limited path can sustain the required exhaust rate \cite{Pitcher_1997, DAY20141505, IAEA_TECDOC_2049_2024, PEARCE2013809, zito_investigation_2023}. In tokamaks, experiments have observed conductance to increase by up to a factor 2 \cite{Pitcher_2001}, and effective pumping speed by 50\% \cite{Bosch_1997, Bosch_1996, ROHDE2009474}, with increasing divertor pressure due to viscous effects. Figure 9.9 of \cite{Van_Oost_2023} ($0.1 \, \mathrm{Pa} \lesssim \pdiv \lesssim 10 \, \rm Pa$) and figure 4 of \cite{Litovoli_2026} ($1 \lesssim \delta \lesssim 100$) and of \cite{KOVARI20133293} demonstrate the increase in conductance with pressure \cite{Zito_2025}.

            Crucially, enhanced pressure---by means of enhanced fuel throughput---favours helium exhaust\footnote{And is also correlated to lower heat fluxes \cite{PITTS2019100696, Kaveeva_2020}.} \cite{PITTS2019100696, Kukushkin_2002, pacher_impurity_2015, IAEA_TECDOC_2049_2024, park_full_2024}, partly by entrainment in the main fuel flow \cite{Reiter_1991, Zito_2025} and by the increase in effective pumping speed \cite{Bosch_1997}. The benefit of enhanced fuel throughput, however, is expected to persist only up to the point where ``plasma plugging'' / ``self-baffling'' remains effective \cite{Reiter_1991, PITTS2019100696, Cowley_2026}, and the conductance is the limiting term in equation \eqref{eq:effective_pumping_speed}. At sufficiently deep detachment, the plasma in the divertor can become too tenuous to retain neutrals therein, leading to increased upstream leakage \cite{PITTS2019100696}.

        \subsubsection{Impurity seeding}\label{sec:intro_matter_injection_plasma__puffing__impurities}

            In contrast to fuel puffing, which primarily controls particle density and neutral pressure, impurity seeding acts mainly through radiation and momentum balance---with disproportionately large plasma effects despite its small absolute flux.
    
            Figure 1 of P\"utterich \textit{et al.} \cite{Putterich_2019} illustrates the strong dependence of radiative cooling efficiency on plasma temperature, highlighting how different impurities can be selected to radiate preferentially in specific plasma regions. Seeding extrinsic impurities is indeed the most effective means of mitigating divertor power loads and is widely regarded as a mandatory ingredient of reactor operation \cite{Kallenbach_2016}.
            
            In absolute terms, however, the injected impurity flux is small compared to the total particle throughput. The geometric-mean impurity-to-fuel puffing ratio is $\langle \GammaZ / \GammaQ\rangle \sim 2\%$ across the database \cite{zenodo_repo_v1} in \cite{Moscheni_2026}. As a result, impurity seeding contributes negligibly to $\GammaPump$ in equation (\ref{eq:pdiv_divertor_pressure}), and the divertor pressure $\pdiv$ remains approximately unchanged even for large variations in $\GammaZ$. For instance, all triangular markers cluster around $\sim$\,11~Pa in figure~3a of \cite{Lore_2022}, despite $\GammaZ$ increasing by more than a factor 10. The consequences, however, are substantial.
            
            The principal concern with impurity seeding is plasma contamination. Once ionised, impurities are subject to friction against the main ion species and to thermal forces, the latter driving them upstream with a strength scaling approximately as their squared charge \cite{Keilhacker_1991, Schaffer_1995, Senichenkov_2019, Makarov_2021}. Together with first-flight penetration of impurity neutrals, this makes impurity behaviour strongly dependent not only on species choice, but also on injection location---which affects where radiation is preferentially emitted and how efficiently impurities stagnate, penetrate, or are transported upstream \cite{Schaffer_1995, Wade_1998, WEST199944, Senichenkov_2019, ZHOU2022113222, Emdee_2023, KAVEEVA2023101424, Park_2024_ITER, Tao_2024, Meng_2022}. Consequently, even when impurities are injected in the divertor---where their radiative contribution is desired---they remain subject after ionisation to a net transport towards the core. The effect is illustrated, once more, in \cite{Lore_2022}: for the triangular series (constant $\GammaQ$ and increasing $\GammaZ$, figure~4a of \cite{Lore_2022}), the upstream impurity concentration rises rapidly, approximately as $\propto (\GammaZ)^{0.74}$, while the plasma density decreases (figure~4b of \cite{Lore_2022}). Although the precise exponent is expected to be machine- and scenario-dependent (ongoing work in \cite{Moscheni_2026_PSI}), the qualitative trend is robust. The outcome is increased core dilution and enhanced radiative losses.

            Understanding how impurity injection scales across devices has been systematised in \cite{Moscheni_2026}, whereas its quantitative comparison with fuel injection is addressed and further assessed in the present study. While minimising impurity concentration is desirable for core performance, operating at very low levels inherently implies high sensitivity: even modest absolute variations can result in disproportionately large changes in plasma contamination and radiation losses.
    
    \subsection{Synthesis}\label{sec:intro_matter_injection_plasma__synthesis}

        As distilled by Kukushkin and Pacher \cite{Kukushkin_2016}, ``gas puffing at impurity concentrations as low as possible is the preferable means for the control of target loading and detachment''. In other words: \textit{if plasma allowed, no impurities would be seeded---at all.} It is therefore unsurprising that such conditions have even motivated explicit ``puff-and-pump'' strategies involving extreme levels of fuel throughput \cite{Schaffer_1995, WEST199944, Wade_1998, PETRIE2007416}. In this sense, puffing a substantial amount of fuel is a rational response to divertor constraints.

        As a consequence, \textit{the ever-increasing fuel circulation in a fusion reactor appears not to be the one producing fusion, but the one sustaining divertor performance}---an enabling, not power-generating, subsystem. From a narrow energy-production viewpoint, this would constitute a substantial overhead under present detachment paradigms: large circulating flows would be required without proportionally increasing fusion output.

        And, crucially: \textit{in a D:T reactor the fuel puffed stream must contain a fraction of tritium}, as explained in the next section. The resulting TFC burden is therefore not avoided by simply regarding puffing as a deuterium-only injection. To compensate for this, impurity injection may be required ``as much as necessary''---rather than strictly ``as low as possible''---to alleviate TFC constraints. This represents a central question addressed in the present study.

\section{Matter Injection in Tritium Fuel Cycle}\label{sec:intro_matter_injection_TFC}

The analysis and design of the matter injection systems from the TFC perspective is the complement to one of what described for plasma in section \ref{sec:intro_matter_injection_plasma}. Whether the goal is to compute tritium self-sufficiency metrics (required TBR ($\tbrr$), start-up inventory ($\Ist$), doubling time ($t_\mathrm{d}$)) \cite{abdou_physics_2020, meschini_modeling_2023, coleman_demo_2019, clark_breeder_2025}, or component and system design \cite{PLOECKL2017186, Ploeckl19052021, day_pre-concept_2022, lang_flexible_2019}, the plasma is either treated as a boundary conditions for the fuelling system and the pumping system, or simple proxies are used to lump the plasma behaviour and quantify the dominant throughput (e.g. by using a fuelling efficiency $\eta_\mathrm{f}$ and a tritium burnup fraction $f_\mathrm{b}$ \cite{abdou_physics_2020}, a tritium burn efficiency $\TBE$ \cite{whyte_tritium_2023}, or the particle confinement time in a particle transport term in the form $\neavg \Vplasma/\tauP$ \cite{morandi2026multispecies}). Furthermore, tritium self-sufficiency studies are general in nature, providing wide design space explorations, while component and system design usually revolve around a point-design scenario for plasma. 

In the former analyses \cite{abdou_physics_2020, meschini_modeling_2023, coleman_demo_2019, clark_breeder_2025, chen_tritium_2016} gas puffing is neglected because it is commonly assumed that (a) tritium is not intentionally injected in gas form (i.e. gas puffing uses D$_2$ and seeding impurities only); (b) the global tritium balance is dominated by breeding, burnup, retention, and reprocessing; (c) the exhaust composition is close to 50:50 D:T so that a DIR can be implemented. 

However, considering gas puffing becomes essential when the puff rate is comparable or greater than the core fuelling rate. Even when the puffed gas contains no tritium, continuous D puffing and impurity seeding increase the total throughput that must be pumped and processed, and they can substantially alter the composition of the exhaust stream delivered to the DIR and the IFC. Conversely, if D:T gas puffing is employed, the contribution must be included explicitly in both inventory and tritium self-sufficiency analysis.

A notable example is ITER, in which the required gas puff rate strongly vary by order of magnitudes depending on the plasma scenario. In ITER stationary $Q=10$ flat-top scenario, Eriksson \textit{et al.} \cite{Eriksson_2024} adopt a fuelling scheme combining 45:55 D:T pellets with pure D gas puffing to maintain an approximately 50:50 core mixture. The gas puff rate is relatively low in this scenario ($2 \times 10^{22} \, \rm at \,s^{-1}$) and comparable to the core fuelling contribution ($0.9 \times 10^{22} \, \rm at \,s^{-1}$ for tritium and $0.75 \times 10^{22} \, \rm at \,s^{-1}$ for deuterium). Koechl \textit{et al.} \cite{Koechl_2020} instead employ 50:50 D:T gas puffing, highlighting that the gas puff rate required for simultaneous density and divertor heat load management is strongly time dependent and must be tuned across operational phases. The required gas puff rate in \cite{Koechl_2020} is $1.2 \times 10^{22} \, \rm at \,s^{-1}$ at most, hence still manageable by the pumping system, although it doubles the tritium throughput (considering a T fuelling rate of $\sim 1 \times 10^{22} \, \rm at \,s^{-1}$ as in \cite{Eriksson_2024}). Lastly, Lore \textit{et al.} \cite{Lore_2022} use D puffing together with Ne seeding in a range $(0.1$--$6.0)\times 10^{23} \, \rm at \, s^{-1}$, with half of the scenarios above $1 \times 10^{23} \, \rm at \,s^{-1}$. Crucially, these scenarios are the same that satisfy all divertor requirements \cite{YOU2022113010}. While ITER will rely on an open fuel cycle, and the impact of a high gas puff rate on tritium self-sufficiency is limited, the highest-throughput case (section 3.1 of \cite{Lore_2022}) could challenge the pumping capability.

Taken together, these ITER examples demonstrate that gas puffing can range from a secondary input comparable to pellet fuelling to a dominant throughput driver set by divertor control. Although FPPs will likely operate around a single-point plasma scenario, the preliminary design of TFC should be flexible enough to accommodate uncertainties and variations of plasma conditions.

    \subsection{DIR: working principle and limitations}\label{sec:intro_matter_injection_TFC__DIR}
    
    The DIR is a proposed fuel cycle architecture element in which a fraction of the unburnt fuel in the exhaust is separated at the primary vacuum pump stage (as for multi-stage cryopumps (MSC) \cite{haertl_design_2022}) or downstream of it (as for metal foil pumps (MFP) \cite{peters_metal_2018}). Such fraction is then routed back to the matter injection system, short-cutting the tritium processing plant (exhaust purification and isotope separation system) \cite{day_smart_2019}. Although studies exist on the separation capabilities of multi-stage cryopumps \cite{haertl_design_2022, lin2025design} and metal foil pumps \cite{li_direct_2024, li_high_nodate} at laboratory scale, no DIR loop has ever been test in a fusion reactor. For this reason, the DIR fraction assumed in fuel cycle analysis should be regarded as a desired performance metric, and the fuel cycle design should be assessed for robustness even at low DIR fractions.
    
    Slightly different variants of DIR have been proposed, from those directly routing the separated gas to the matter injection system (through the gas distribution system) \cite{day_pre-concept_2022}, to those considering an additional bypass loop and an isotope rebalancing system \cite{hattab_analysis_2025}. Employing a DIR is not mandatory for operations, but it can be advantageous because it reduces the throughput and tritium inventory across the IFC \cite{abdou_physics_2020, meschini_modeling_2023}. From the residence time perspective, it introduces an effective residence time in the IFC as the weighted average: 
    \begin{equation}\label{eq:tau_ifc_eff}
        \tau_\mathrm{IFC,eff} = \tau_\mathrm{DIR}f_\mathrm{DIR} + (1-f_\mathrm{DIR})(\tau_\mathrm{ISS} + \tau_\mathrm{TEP})
    \end{equation}
    \noindent where $\taudir$ and $\fdir$ are the residence time for the DIR loop and the fraction of exhaust processed by the DIR, $\tau_\mathrm{ISS}$ and $\tau_\mathrm{TEP}$ are the isotope separation system and tritium exhaust processing residence times.
    
    Purifying the exhaust and separating the hydrogen isotope is a lengthy process, with characteristic times on the order of $\tau_\mathrm{ISS} + \tau_\mathrm{TEP} \sim 2$--5h \cite{day2013direct, coleman_demo_2019, abdou_physics_2020}. In contrast, an ideal DIR can operate on much shorter timescales \cite{coleman_demo_2019}, down to the order of seconds for metal foil pumps (MFP) \cite{peters_metal_2018}. In the limiting case of $f_\mathrm{DIR} \rightarrow 1$ most of the unburnt fuel is quickly recirculated to the fuelling system, drastically reducing the inventories in the IFC, thus $\Ist$, and, ultimately, $\tbrr$. This explains why a DIR would be so effective in enabling tritium self-sufficiency, especially at low TBE, where the majority of tritium exits the vacuum chamber unburnt and must be reprocessed. 
    
    However, to work so efficiently a DIR relies on the exhaust D:T ratio remaining close to the fuelling D:T ratio. Any imbalance cannot be corrected without introducing processing components along the DIR loop, thus increasing $\tau_\mathrm{DIR}$ and eroding the advantage. Puffing can strongly imbalance the exhaust isotopic composition when the puffed mixture has a D:T ratio different from the core fuelling ratio, and a magnitude comparable to it. Therefore, there exists an optimal tritium fraction in the puffed gas, where the penalty of higher tritium throughput (and the associated increase in circulating inventory) is offset by the ability to sustain a higher DIR fraction. This trade-off is discussed in section \ref{subsec:optimal_T_fraction_gaspuffing}.
    
    An additional issue is impurity accumulation in the core plasma, which depends on the techonology used to implement a DIR. Two leading options are metal foil pumps \cite{li_high_nodate, peters_metal_2018} and multi-stage cryopumps \cite{haertl_design_2022}. MFPs provide very sharp separation between hydrogenic species and other species due to the low impurity permeability of the foil, but they do not separate hydrogen isotopes, thus leading to potential protium accumulation into the core plasma \cite{hattab_analysis_2025}. MSCs equipped with cryosorbent panels can, in principle, separate H$_2$ from D$_2$ and T$_2$ because the saturation curves differ by about 1.5 K, corresponding to roughly four orders of magnitude in saturation pressure at a fixed temperature. However, the T$_2$ saturation curve lies very close to that of Ne, likely resulting in a non-unitary separation sharpness for Ne. Since protium is continuously produced by secondary DD reactions and outgassing from the wall chambers, while Ne is commonly used---or assumed to be---as seeding gas \cite{Wigram_2019, Liu_2020, Lore_2022, Miller_2024, Bader}, both DIR technology face challenges on impurity control.

    \subsection{Consequences of high gas puffing rates on core plasma with a TFC implementing a DIR loop}\label{sec:intro_matter_injection_TFC__consequences}
    This section introduces key concepts associated to high gas puff rates and the presence of seeding impurity (assumed as Ne) in the puffed gas. These features are then applied to fuel cycle simulations in section \ref{sec:results}. If a fraction of the exhaust $\fdir$ is routed to the SDS and the core fuelling system, the evolution of the Ne core fraction can be described by the following system of equations:

    \begin{equation}
        \begin{split}
            \frac{d\necore}{dt} &= \fdir(1-\sne)\frac{\nediv}{\taudir} - \frac{\necore}{\tauP} \\
            \frac{d\nediv}{dt} &= \GammaNepuff + \frac{\necore}{\tauP} - \frac{\nediv}{\taudir}
        \end{split}
    \end{equation}
    
    Where the Ne concentration in the core, $\necore$, increases with time due to an imperfect separation by MSCs of Ne from the DIR stream. The term $\fdir(1-\sne)\frac{\nediv}{\taudir}$ thus represents the fraction of Ne from the exhaust stream that is recirculated through the DIR and eventually injected into the core by the D:T (plus remaining impurities) pellets. For very high DIR fractions ($\fdir \geq 0.9$) and high gas puffing rates with seeding impurities ($\GammaZ = 5 \times 10^{20} \, \rm at \,s^{-1}$), the impurity concentration in the core plasma reaches high values within times comparable to a single pulse. 
    
\begin{figure}
        \centering
        \includegraphics[width=0.95\linewidth]{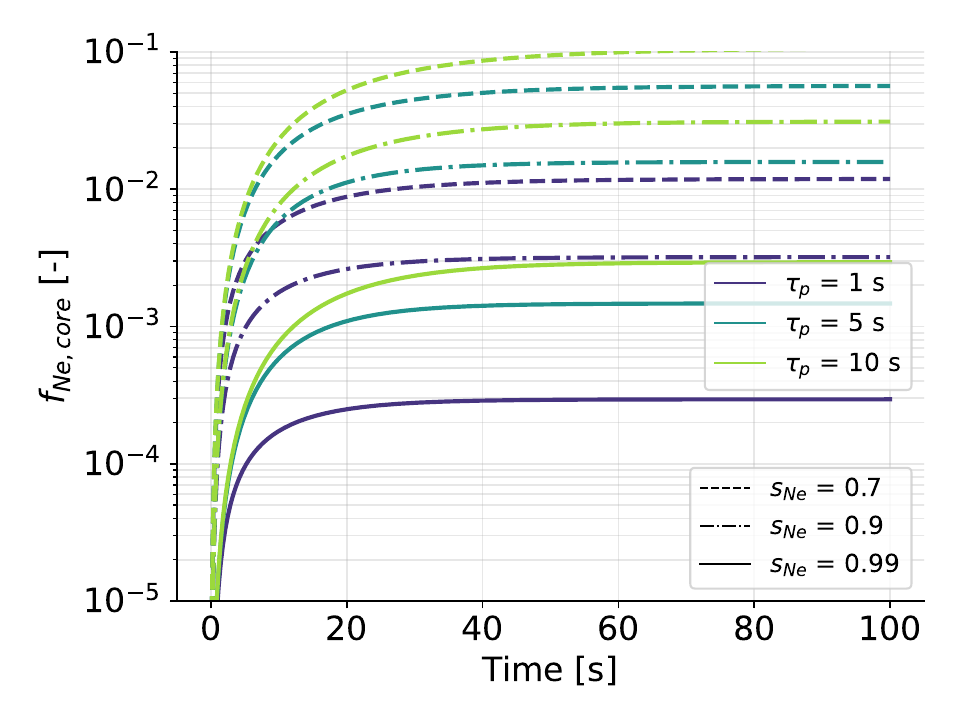}
        \caption{Neon fraction in the core plasma due to imperfect separation, as a function of the particle confinement time $\tauP$ and the separation sharpness $S_\mathrm{Ne}$, for $f_\mathrm{DIR} = 0.9$, $\tau_\mathrm{DIR} = 10 \,\rm s$, $\GammaNeCore = 5 \times 10^{20} \, \mathrm{at \, s^{-1}}$, $n_e = 10^{20} \, \rm m^{-3}$.}
        \label{fig:Ne_fraction_DIR}
    \end{figure}

    Figure \ref{fig:Ne_fraction_DIR} shows the evolution of the Ne core fraction for $\GammaNepuff = 5 \times 10^{20} \, \rm at \,s^{-1}$. Regardless the particle confinement time and the separation sharpness, a steady state concentration is reached within minutes. Indicative, acceptable values of the average Ne core fraction ($f_\mathrm{Ne,core} \lesssim 10^{-3}$) requires almost perfect separation sharpness ($S_\mathrm{Ne} > 0.99$). Such a high separation sharpness could in principle be achieved by MFPs since only hydrogen permeates through the foil (although the presence of impurities, which is in any case inevitable due to the presence of He, can lower the performance and lifetime of the metal foil \cite{li_impact_2024}), while it looks more challenging in MSCs due to the precise temperature control required. If $S_\mathrm{Ne} > 0.99$ cannot be achieved, then a DIR loop could increase the impurity fraction in the core to unacceptable levels. A similar buildup of protium is expected, although the source term is different (degassing from the first wall and production from D-D fusion reactions), as described in \cite{hattab_analysis_2025}.
    
     Suppose next that, instead, a scenario compatible with the divertor requirements exists with vanishing impurity seeding ($\Gamma_Z \to 0$), compensated by increased fuel puffing. Since adding tritium to this compensating puff would increase the tritium throughput, we first consider the limiting case of pure deuterium puffing. In the absence of processing components downstream of the DIR loop (whose very purpose is to avoid exhaust processing), the D-rich exhaust is routed through the DIR and eventually to the fuelling system. As in the impurity case discussed above, this progressively increases the core deuterium fraction and correspondingly decreases the core tritium fraction, $\ftcore$. Within a short time, the core fuelling mixture shifts toward pure deuterium, leading to a reduction in fusion power. This is, of course, an extreme case. In principle, the core fuelling tritium fraction could be rebalanced by introducing a dedicated component downstream of the DIR loop, drawing on tritium intentionally added to the start-up inventory for this purpose; alternatively, gas puffing could be performed with a 50:50 D:T mixture. However, the former solution reintroduces precisely the issues that the DIR is meant to avoid. The latter would be acceptable either if the gas puffing rate were lower than the core fuelling rate---whereas Section \ref{sec:methods} shows it to be much higher---or if the DIR fraction were extremely high, so that the additional tritium throughput associated with gas puffing became negligible because the exhaust is rapidly recycled. Yet the DIR loop itself and DIR fractions approaching unity remain to be demonstrated in a fusion reactor.
    
    This situation motivates extending the residence-time fuel cycle model of \cite{meschini_modeling_2023} to include gas puffing (Section \ref{sec:methods__TFC_model}), and underlies the analyses presented in Sections \ref{subsec:tss_puffing_DT} and \ref{subsec:optimal_T_fraction_gaspuffing}.

    \subsection{Consequences of high gas puffing rates on exhaust pumping}

    The issues reported before are associated to the mixture in the gas puff. Furthermore, regardless of the species mix, increasing the puffing rate increases the total exhaust throughput that must be pumped, and therefore drives the required pumping capacity to be installed. Species effects can also be important (e.g. if the pumping speed differs markedly between H isotopes and seeded impurities), but for the purpose of the following discussion they are neglected.
    
    In many pre-conceptual designs the pumping system is sized assuming that the dominant contribution to exhaust throughput is the core fuelling stream (see, for example, table 10 in \cite{noauthor_pre-concept_2022}). When realistic gas puffing levels are included, however, the difference can be substantial. As the authors of \cite{noauthor_pre-concept_2022} point out, the design proposed is not a point design but gives order of magnitudes estimations, acknowledging uncertainties in the core fuelling and gas puffing streams. Nonetheless, the upper bound of the gas puff rate is $\GammaPuff \simeq 50 \, \mathrm{Pa \, m^3 \, s^{-1}}$ in \cite{noauthor_pre-concept_2022}, while simulated edge plasma scenarios for the same machine consider $\GammaPuff = 10^{23} \, \mathrm{at \, s^{-1}}$, or $\simeq 370 \, \mathrm{Pa \, m^3 \, s^{-1}}$ \cite{subba_solps-iter_2021}, which, in turn, is below the average for EU-DEMO (table \ref{tab:fueling_stats}). This is approximately seven times larger than the value used for fuel cycle design in \cite{noauthor_pre-concept_2022} and is of the same order as the core fuelling throughput. Similarly, the vacuum system for CFETR is sized for a throughput of $290 \, \mathrm{Pa \, m^3 \, s^{-1}}$ \cite{HU2022113058} by scaling ITER baseline value of $200 \, \mathrm{Pa \, m^3 \, s^{-1}}$, although we estimate more than twice that value in table \ref{tab:fueling_stats}.
    
    In practical terms, the required pumping capacity, and associated footprint, can increase by roughly a factor of two compared with a sizing performed under the assumption of a core fuelling-dominant scenario.
    
\section{The Conundrum}\label{sec:intro_conundrum}

\begin{figure*}[t]
\centering
\begin{tikzpicture}[
    node distance=1.5cm and 2.4cm,
    every node/.style={font=\small},
    box/.style={
        draw,
        rounded corners=2pt,
        align=center,
        minimum width=4.5cm,
        minimum height=1.0cm
    },
    critical/.style={
        box,
        draw=black,
        text=black,
        fill=red!75
    },
    group/.style={
        draw,
        dashed,
        rounded corners=3pt,
        inner sep=8pt
    },
    line/.style={-Latex, thick},
    bothline/.style={<->, thick, >=Latex},
    branch/.style={thick},
]

\node[box] (goal)
    {DIR must minimise\\T inventory};

\node[box, below=of goal] (method)
    {DIR minimises\\TFC processing time};

\node[box, below=of method] (req)
    {DIR favours\\50{:}50 D:T throughput};

\node[box, below=of req] (plasma)
    {Puffing favours\\high D:T throughput};

\node[box, below=of plasma] (plasmagoal)
    {Puffing must maximise\\plasma purity, detachment\\and helium exhaust};

\node[
    group,
    fit=(goal)(method)(req),
    label={[font=\small]above:{\shortstack{Direct internal recycling (DIR)\\in tritium fuel cycle (TFC)}}}
] {};

\node[
    group,
    fit=(plasma)(plasmagoal),
    label={[font=\small]above:Edge plasma and divertor}
] (edgeplasmadivertor) {};

\node[critical] (balance) at ($(req)!0.5!(plasma)+(7.0cm,0)$)
    {Favourite puffing is\\high and 50{:}50 D:T};

\node[critical, above=of balance] (tritium)
    {High puffing-driven\\total T throughput};
    
\node[box, right=of edgeplasmadivertor, below=of balance] (core)
    {Fuel puffing dominates\\particle balance:\\$\GammaQ \sim 10 \times \GammaQcore$};

\draw[line] (goal.south) -- (method.north);
\draw[line] (method.south) -- (req.north);

\draw[line] (plasmagoal.north) -- (plasma.south);

\coordinate (merge) at ($(balance.west)+(-1.1cm,0)$);

\draw[branch, red] (req.east) -- ++(0.8,0) |- (merge);
\draw[branch, red] (plasma.east) -- ++(0.8,0) |- (merge);
\draw[line, red]   (merge) -- (balance.west);
\draw[line, red]   (core.north) -- (balance.south);

\draw[line, dashed] (edgeplasmadivertor) -- (core.west);

\draw[line, red] (balance.north) -- (tritium.south);
\draw[line, red] (tritium.north) |- (goal.east);

\end{tikzpicture}
\caption{The conundrum arising from compartmentalised optimisation of plasma operation and the tritium fuel cycle (TFC). In red the specific criticality highlighted in the present work.}
\label{fig:plasma_tfc_conundrum}
\end{figure*}
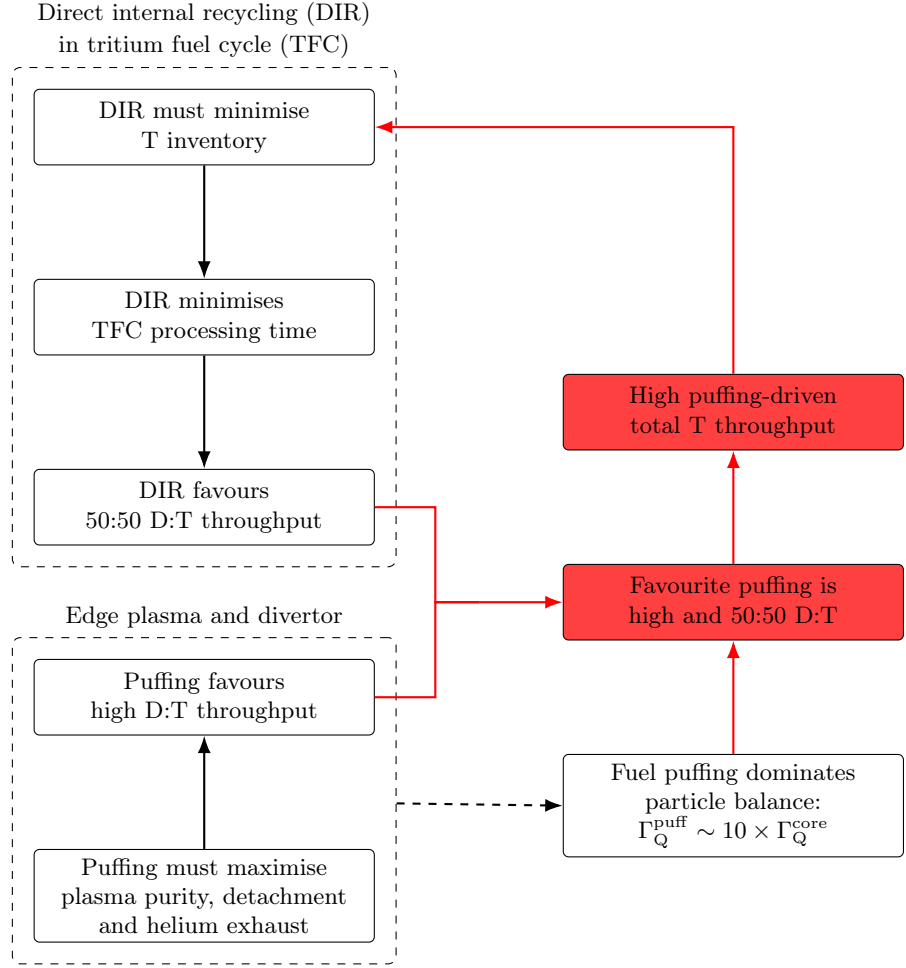

   Figure \ref{fig:plasma_tfc_conundrum} highlights a possible misalignment between state-of-the-art fuel cycle design and plasma operating scenarios. Strategies aimed at reducing tritium processing time, most notably the various DIR configurations, have been developed to lower the total tritium inventory across the plant \cite{day_smart_2019,cui_new_2022,hattab_analysis_2025}, and minimise the required TBR to achieve tritium self-sufficiency at fixed doubling time, or to reduce the doubling time at fixed TBR. However, these strategies become less effective when the puffing rate exceeds the core fuelling rate. If puffing is performed with little tritium, the resulting imbalance in the D:T ratio of the exhaust stream becomes incompatible with a high DIR fraction, unless additional tritium is added to $\Ist$ for the purpose of rebalancing the core fuelling mixture. Conversely, if puffing is performed with a 50:50 D:T ratio to restore this balance, the higher tritium content of the puffed gas undermines the original purpose of DIR, unless extremely high DIR fraction are achieved. This can complicate the start-up and make tritium self-sufficiency more difficult to achieve (see Section \ref{subsec:tss_puffing_DT} and \ref{subsec:optimal_T_fraction_gaspuffing}).
    

\section{Methods}\label{sec:methods}

    In this section we describe the toolkit used to systematise and assess the semi-quantitative discussions in the sections above. Symbols used throughout the manuscript are listed in appendix \ref{apx:list_of_symbols}.

    \subsection{Database}\label{sec:methods__database}

        A multi-machine analysis is conducted within a 0D framework. While this approach has obvious limitations, it offers the advantage of enabling systematic assessments over large parameter domains and across multiple devices, as discussed in section 1.1 of Angioni \textit{et al.} \cite{Angioni_2026}. Therein, and similarly to the methodology also adopted in \cite{Moscheni_2026}, both tokamaks and stellarators are considered---therefore included simultaneously in the present work as well. At the level of the present analysis, no clear evidence is found to justify treating the two configurations separately, at least from the perspectives of plasma, and fuelling and particle exhaust rates.

        \subsubsection{Overview}\label{sec:methods__database__overview}

            Because fuel puffing was found in \cite{Moscheni_2026} to strongly correlate with divertor and plasma volumes, analogous correlations are here sought for core fuelling. This enables a direct and consistent comparison between the two.
            
            Detached cases are exclusively included, in order to focus on reactor-relevant operating conditions. This choice is particularly important in the present context, since in attached regimes the required gas puffing rate may approach very low values and would therefore not be representative.
            
            The same set of cases constituting the original puffing database \cite{zenodo_repo_v1}, as employed in \cite{Moscheni_2026}, has therefore been supplemented with core plasma quantities. The resulting, newly-compiled database is available at \cite{zenodo_repo_v2} and summarised in table \ref{tab:db_minmax}. Overall 354 entries span 25 machines, corresponding to the subset of devices for which core plasma data could be consistently assembled. It includes both experimental and numerical entries, covering conventional tokamaks (CTKs), spherical tokamaks (STKs), high-field tokamaks (HTKs), and stellarators (STLs), with fuelling provided by neutral beam injection (NBI) and/or pellet injection (PI).

            The main variables used throughout the study are detailed below.

            \begin{table}[t]
\centering
\renewcommand*\arraystretch{1.05}
\small
\begin{tabular}{lccc}
\toprule
Quantity & Units & Minimum & Maximum \\
\midrule
$a$ & $\mathrm{m}$ & $1.6 \times 10^{-1}$ & $2.9 \times 10^{0}$ \\
$R_0$ & $\mathrm{m}$ & $3.6 \times 10^{-1}$ & $2.0 \times 10^{1}$ \\
$\Vplasma$ & $\mathrm{m^3}$ & $6.5 \times 10^{-1}$ & $2.5 \times 10^{3}$ \\
$\Vdiv$ & $\mathrm{m^3}$ & $5.8 \times 10^{-3}$ & $4.8 \times 10^{2}$ \\
$\Vtot$ & $\mathrm{m^3}$ & $7.0 \times 10^{-1}$ & $2.9 \times 10^{3}$ \\
$\Splasma$ & $\mathrm{m^2}$ & $4.6 \times 10^{0}$ & $1.6 \times 10^{3}$ \\
$\Sdiv$ & $\mathrm{m^2}$ & $5.2 \times 10^{-1}$ & $8.0 \times 10^{2}$ \\
$\Stot$ & $\mathrm{m^2}$ & $5.3 \times 10^{0}$ & $1.8 \times 10^{3}$ \\\midrule
$\GammaQ$ & $\mathrm{at \, s^{-1}}$ & $1.0 \times 10^{20}$ & $2.5 \times 10^{24}$ \\
$\GammaZ$ & $\mathrm{at \, s^{-1}}$ & $1.2 \times 10^{18}$ & $7.0 \times 10^{22}$ \\
$\GammaQcore$ & $\mathrm{at \, s^{-1}}$ & $4.0 \times 10^{19}$ & $1.2 \times 10^{23}$ \\
$\GammaHeCore$ & $\mathrm{at \, s^{-1}}$ & $1.0 \times 10^{20}$ & $1.1 \times 10^{21}$ \\\midrule
$\ratioPuffToCore$ & $-$ & $1.8 \times 10^{-1}$ & $2.8 \times 10^{2}$ \\
$\cZ$ & $-$ & $1.0 \times 10^{-4}$ & $9.7 \times 10^{-1}$ \\
$\cHe$ & $-$ & $1.4 \times 10^{-4}$ & $1.5 \times 10^{-2}$ \\
\bottomrule
\end{tabular}
\caption{Database-wide minimum and maximum values for geometry, fueling, and derived ratios/concentrations.}
\label{tab:db_minmax}
\end{table}

        \subsubsection{Geometrical parameters}\label{sec:methods__database__geometry}

            A set of geometrical parameters is considered. The plasma volume $V \, [\mathrm{m}^3]$ and plasma surface $S \, [\mathrm{m}^2]$ are computed as:
            \begin{equation}\label{eq:plasma_volume_and_surface}
                \begin{split}
                    \Vplasma &= 2\pi^2 R_0 a^2 \kappa \\
                    \Splasma &= 4\pi^2 R_0 a \,\sqrt{\frac{1 + \kappa^2}{2}} \;,
                \end{split}
            \end{equation}
            where $a \,[\mathrm{m}]$ and $R_0 \,[\mathrm{m}]$ are the plasma minor and major radii, respectively, and $\kappa \,[-]$ its elongation\footnote{For stellarators, $\kappa=1$ and the complexity of the cross-sectional shape is implicitly embedded within the \textit{average} major and minor radii.}.
            
            With divertor volume and surface defined in equation (1) of \cite{Moscheni_2026}, the total volume and total surface (proxying those of the chamber as a whole) can be expressed as:
            \begin{equation}\label{eq:total_volume_and_surface}
                \begin{split}
                    \Vtot &= \Vplasma + \Vdiv \\
                    \Stot &= \Splasma + \Sdiv - 2\pi \Rdiv \Ldiv  (1 + \ftor) \;,
                \end{split}
            \end{equation}
            where the subtraction---a correction of $\sim 10\%$ on average---stems from having the plasma and divertor volumes sharing a common surface of effective length $\Ldiv$, defined in figure 1 of \cite{Moscheni_2026}. This surface is always toroidally-continuous for the plasma while it covers a fraction $\ftor$ of the circumference for the divertor---hence the $(1 + \ftor)$ factor. In particular, $\ftor = 50\%$ for the island divertor of W7-AS \cite{McCormick_2002_PRL}, W7-X \cite{Effenberg_2019} and Infinity-Two \cite{Bader}, and 100\% otherwise.

            In the following, $\Vtot$ and $\Stot$ are adopted as the reference geometrical variables unless otherwise specified. Empirically, they tend to provide marginally better fits than plasma-/divertor-only quantities, although the difference remains well within the overall uncertainty level. Their use is nevertheless preferable in the present context, since the analysis simultaneously involves puffing---primarily linked to divertor geometry \cite{Moscheni_2026}---and core fuelling---which instead depends on the confined plasma.

            The translation between surface-based and volume-based representations is operated via the mapping:
            \begin{equation}\label{eq:volumte_to_surface_conversion}
                \Vtot = \Stot^{3/2} \,.
            \end{equation}

        \subsubsection{Engineering particle rates in core plasma}\label{sec:methods__database__core__fuelling}
                        
            The only new devices added to the updated database \cite{zenodo_repo_v2} are ARC \cite{sorbom_arc_2015, Hillesheim_2026}, GIGA \cite{gaussfusion_cdr_exec_2026}, Infinity Two \cite{Guttenfelder_2025}, MANTA \cite{Rutherford_2024} and Stellaris \cite{LION2025114868}. Existing entries have been supplemented with quantities related to the core plasma, here defined in a simplified manner as the plasma region enclosed by the magnetic separatrix (last closed flux surface), and in particular with estimates of the core fuelling rate $\GammaQcore$.
            
            Unless otherwise specified, $\GammaQcore$ represents the \textit{engineering} core fuelling rate, i.e.\ the particle throughput that the TFC must deliver to the plasma chamber. This quantity is to be distinguished from the \textit{physical} core fuelling rate, which accounts for the amount of fuel actually depositing in the confined region (section \ref{sec:intro_matter_injection_plasma__core_fuelling}). The latter may include contributions from mechanisms such as penetration of puffed gas and/or of wall outgassing. Further discussion of this distinction is provided in section \ref{sec:discussion__core_fuelling__physics__penetration} and in appendix \ref{apx:estimation_GammaQcore__puffing_to_core_unimportant}.
            
            Even when considering the engineering definition of $\GammaQcore$, an unavoidable level of uncertainty remains. Core fuelling is not always explicitly reported in edge plasma models---e.g. when fixed-density boundary conditions are adopted instead of flux-based ones \cite{Zhang_2022}. In other cases its contribution is negligible compared to other quantities, and therefore not accounted for \cite{GUILLEMAUT2013S638}. For future devices different values of $\GammaQcoreSim$ may be obtained across design iterations \cite{Henderson_2025} and/or assumptions on the tritium burn-up fraction---for instance, $3.5\times10^{22} \, \mathrm{s^{-1}}$ reads in the edge models of \cite{Osawa_2023, Osawa_2024}, $6$--$8 \times 10^{22} \, \mathrm{s^{-1}}$ in \cite{Henderson_2025}, while $1\times10^{22} \, \mathrm{s^{-1}}$ figures in the core plasma assessment of \cite{Tholerus_2024} (table 6 therein). Overall, this does not imply any inconsistency in the underlying models; rather, it reflects the limited availability of directly comparable data and an intrinsic uncertainty.
            
            These aspects are therefore accounted for and an approximate but systematic procedure is devised. Details are reported in appendix \ref{apx:estimation_GammaQcore}. Here it suffices to note that, when the value $\GammaQcoreSim$ is not explicitly available, the total core particle source from neutral beam injection (NBI) and pellet injection (PI), namely $\GammaQcoreNBI + \GammaQcorePelletNominal$, is evaluated.
            
            The resulting engineering $\GammaQcore$ therefore provides an approximate quantification of the particle source delivered by core fuelling systems, spanning the operational range up to the maximum deliverable pellet source $\GammaQcorePelletMax$. This estimate is sufficient for the purposes of the present study, which compares the order of magnitude of core fuelling with the fuel gas puffing---itself affected by a factor-of-two uncertainty \cite{Moscheni_2026}.
            
            The resulting $\GammaQcore$ are collated in table \ref{tab:db_minmax} and span nearly three orders of magnitude across 25 machines, approximately uniformly distributed over the considered range. The treatment of machine-to-machine variability is described in section \ref{sec:methods__database__statistics}.

            For ARC \cite{sorbom_arc_2015, Hillesheim_2026}, GIGA \cite{gaussfusion_cdr_exec_2026}, MANTA \cite{Rutherford_2024} and Stellaris \cite{LION2025114868}---future power plants not explicitly reporting core fuelling rates---$\GammaQcore$ is estimated as:
            \begin{equation}\label{eq:gammaQcore_from_helium_production}
                \GammaQcore = \GammaDcore + \GammaTcore = \frac{2 P_{\mathrm{fus}}}{\fBurn E_{\mathrm{fus}}}
                = \frac{2 \GammaHeCore}{\fBurn} \,,
            \end{equation}
            where:
            \begin{equation}\label{eq:helium_production_rate}
                \GammaHeCore = \frac{P_{\mathrm{fus}}}{E_{\mathrm{fus}}}
            \end{equation}
            is the helium production rate for D–T fusion, $E_{\mathrm{fus}} = 17.6 \,\mathrm{MeV} = 2.8 \times 10^{-12} \,\mathrm{J}$ is the D–T fusion energy and $P_{\mathrm{fus}} \,[\mathrm{W}]$ the fusion power corresponding to the chosen reactor operating point. The factor 2 reflects a nominal 50:50 D:T composition of the incoming fuel ($\GammaDcore = \GammaTcore$), and $\fBurn$ is the burn-up fraction\footnote{For nominal equimolar D:T core fuelling, the present definition $\fBurn = 2 \GammaHeCore / \GammaQcore$ is numerically identical to conventions that normalise by the tritium component of the same core-fuelling stream, $\fBurn = \GammaHeCore / \GammaTcore$ \cite{Xie_2020, McClenaghan_2023, Tholerus_2024}, since $\GammaQcore = \GammaDcore + \GammaTcore = 2\GammaTcore$. The apparent factor-of-two difference is therefore conventional rather than physical.}. An indicative value $\fBurn=3\%$ is assumed, consistent with \cite{Xie_2020, McClenaghan_2023} and \cite{Guttenfelder_2025} in CFETR and Infinity Two, respectively, and consistent with a 2-GW EU-DEMO \cite{lang_considerations_2015}. Still, the sensitivity of $\fBurn$ caused, at least, by pellet penetration \cite{lang_considerations_2015} and transport conditions (appendix of \cite{Tholerus_2024}) ought to be acknowledged.
            
            Notably, equation (\ref{eq:gammaQcore_from_helium_production}) is a \emph{core-fuelling} quantity: it measures the fraction of the engineering core-fuelling stream that is ultimately burned. This should not, however, be conflated with the tritium burn efficiency (TBE) of Whyte \textit{et al.} \cite{whyte_tritium_2023}, which is defined at the vacuum-vessel boundary as the fraction of tritium entering the principal vessel that undergoes fusion. TBE therefore embeds the fuelling efficiency, edge fuelling, recycling, divertor exhaust and helium-ash removal.

        \subsubsection{Puffing injections and stream concentrations}\label{sec:methods__database__puffing_and_concentration}

            Actual puffing rates for fuel, $\GammaQ \,[\mathrm{at \, s^{-1}}]$, and for impurities, $\GammaZ \,[\mathrm{at \, s^{-1}}]$, are retrieved from the database \cite{zenodo_repo_v2} and here expressed in atomic equivalents.
            
            The fuel puffing rate $\GammaQ$ may correspond to any D:T mixture; in present-day devices, where tritium is absent, $\GammaQ = \Gamma_{\mathrm{D}}^{\mathrm{puff}}$. Different D:T mixtures may have secondary consequences (e.g. section 3 of \cite{Kappatou_2025} and isotopic mass-dependences discussed in \cite{Kirov_2025, leoni2024scrape, Goldston_2012}), which are not accounted for here. Consistent with \cite{Moscheni_2026}, a factor-of-two accuracy is assigned to all puffing-related quantities.

            Actual impurity puffing rates $\GammaZ$ are intrinsically difficult to handle because of the complexities associated with impurity seeding (sections 3.2 and 4.8 of \cite{Moscheni_2026}). In particular, $\GammaZ$ depends sensitively on the degree of detachment, which is not quantified across the database. For this reason, the following scaling estimates are also considered \cite{Moscheni_2026}:
            \begin{equation}\label{eq:seeding_rate_GES_Sudo}
                \begin{split}
                    \GammaZges &\sim 3.6 \times 10^{21} \, Z^{-1} \, a^{1.51} \, (\lqEich \fScara)^{-0.27} \,,\\
                    \GammaZsudo &\sim 2.4 \times 10^{21} \, Z^{-1} \, a^{1.51} \, \left(\Pin \BT / R_0 \right)^{0.16} \,,
                \end{split}
            \end{equation}
            which are meant to approximate the non-linear, general formulation in equation (8) of \cite{Moscheni_2026}. Overall they are denoted as:
            \begin{equation}\label{eq:effective_seeding_rate}
                \GammaZstar \in \{\GammaZges \,, \GammaZsudo\} \,.
            \end{equation}
            Here, $\lqEich$ represents Eich’s regression \#14 for the power fall-off length \cite{Eich_2013}, and $\fScara$ the Scarabosio correction factor \cite{SCARABOSIO2013S426}. The only modification compared to \cite{Moscheni_2026} is here the explicit inclusion of the case-specific atomic number $Z$, which converts the expressions to $[\mathrm{at \, s^{-1}}]$. The first formula is specific to tokamaks and is implicitly defined via the Greenwald density limit $\neGW$ \cite{Greenwald_1988}, whereas the second applies to stellarators through the Sudo limit $\neSu$ \cite{Sudo_1990}, such that:
            \begin{equation}\label{eq:effective_density_limit}
                \neStar \in \{\neGW, \, \neSu\} \,.
            \end{equation}
            Further details are given in section 2.2.2 of \cite{Moscheni_2026} but, notably, while the GES scaling for tokamaks has been successfully---albeit preliminarily---validated, the Sudo-based stellarator version lacked validation in the original work due to limited data.

            ARC \cite{sorbom_arc_2015, Hillesheim_2026}, GIGA \cite{gaussfusion_cdr_exec_2026}, Infinity Two \cite{Guttenfelder_2025}, MANTA \cite{Rutherford_2024} and Stellaris \cite{LION2025114868} represent devices relevant for the present analysis but lacking explicit puffing figures in the literature. Therefore, $\GammaQ$ and $\GammaZ$ are taken from the approximate values reported in table 3 of \cite{Moscheni_2026}: equations (10)--(12) estimate the uppermost expected $\GammaQ$ sufficient to only access detachment, but not the absolute upper bound; equations (11)--(13) (i.e. Equations (\ref{eq:seeding_rate_GES_Sudo}) herein) approximate\footnote{Also in the SOLPS5.0 \cite{Schneider_2000} simulations of \cite{MAO20151233} for CFETR \cite{Song_2014} where carbon was used as a substitute for impurity seeding.} the seeding rate $\GammaZ$.
            
            Finally, the dimensionless puff-to-core fuelling ratio is introduced:
            \begin{equation}
                \ratioPuffToCore = \frac{\GammaQ}{\GammaQcore} \,,
            \end{equation}
            together with:
            \begin{equation}\label{eq:impurity_stream_concentrations}
                [\cZ, \, \cHe] = \frac{[\GammaZ, \, \GammaHeCore]}{\GammaQ + (1-0.5\fBurn) \GammaQcore + \GammaZ} \,.
            \end{equation}
            
            Here, $[\cdot, \cdot]$ provides a compact notation for the two quantities defined in parallel. These represent the concentrations of extrinsically seeded impurities and helium in the overall outgoing particle stream, leaving the divertor and heading towards the pumps. Generally speaking, these concentrations vary in space, and shall not be assumed to be representative of other domains in the reactor (e.g. core/edge plasma). The $0.5\fBurn$ factor represents 1 triton and 1 deuteron burning and resulting in 1 helium atom.

            Possible contributions from wall outgassing and sputtered intrinsic impurities are neglected. Despite its relevance from a TFC perspective \cite{hattab_analysis_2025}, protium is also neglected in the present analysis. Its absolute concentration is typically low and, to first order, expected to scale similarly to helium for being a D--D fusion product itself. Nevertheless, its inclusion may represent a relevant refinement in future studies.

        \subsubsection{Statistical methods}\label{sec:methods__database__statistics}

            To quantify $\GammaQcore$ scaling trends while limiting bias from uneven sampling across devices, a machine-level approach is adopted. Each of the 25 machines contributes a single representative datapoint, irrespective of the number of database entries available for $\GammaQcore$, thereby preventing over-weighting of machines with richer coverage. Using individual operating points instead of machine-averaged values modifies the fitted exponents by only $\lesssim 5\%$. Figure 11 of \cite{Angioni_2026} clarifies on the visual differences between the two approaches.
            
            However, an un-biased approach clarifies that the regression addresses the specific question: \textit{how does a characteristic machine-level core fuelling rate scale with characteristic machine-level parameters?} This is conceptually distinct from asking how $\GammaQcore$ scales across all individual operating points in the full dataset. As discussed similarly in section 4.1 of \cite{Moscheni_2026}, the former probes cross-device \textit{macroscopic} trends between machines, whereas the latter would emphasise within-machine \textit{microscopic} variability and operating-point-dependent correlations, which are outside the scope of the present study.

            Concretely, for each machine, the values of $\GammaQcore$ defined in section \ref{sec:methods__database__core__fuelling} are geometrically averaged\footnote{Arithmetically averaging does not lead to an appreciable difference.}. The same procedure is applied to all other machine-specific variables introduced in sections \ref{sec:methods__database__geometry} and \ref{sec:methods__database__puffing_and_concentration}, and is generically denoted by $\langle \cdot \rangle$. 
            
            The associated within-machine variability is represented by a multiplicative\footnote{An alternative would be to report asymmetric variation ranges, similarly to the rescaling factors used in section 6 of \cite{Verdoolaege_2021}, but this would be unnecessarily cumbersome for the present purpose.} $\tpm 1\sigma$ factor derived in log-space and used when reporting values in tables and for statistical purposes. For visualisation, the full variability across database entries is instead illustrated using min/max bars, in order to display the complete spread of operating points associated with each machine.
            
            The resulting set of machine-level datapoints is collated in table \ref{tab:fueling_stats} and then fitted, in log-log space, using a weighted least-squares regression to a classical power-law model. The goodness-of-fit is quantified through $R^2$ computed in log-space and the constant pre-factors implicitly contain all the dimensional adjustments.
            
    \subsection{TFC system-level model}\label{sec:methods__TFC_model}
\begin{figure}
        \centering
        \includegraphics[width=0.99\linewidth]{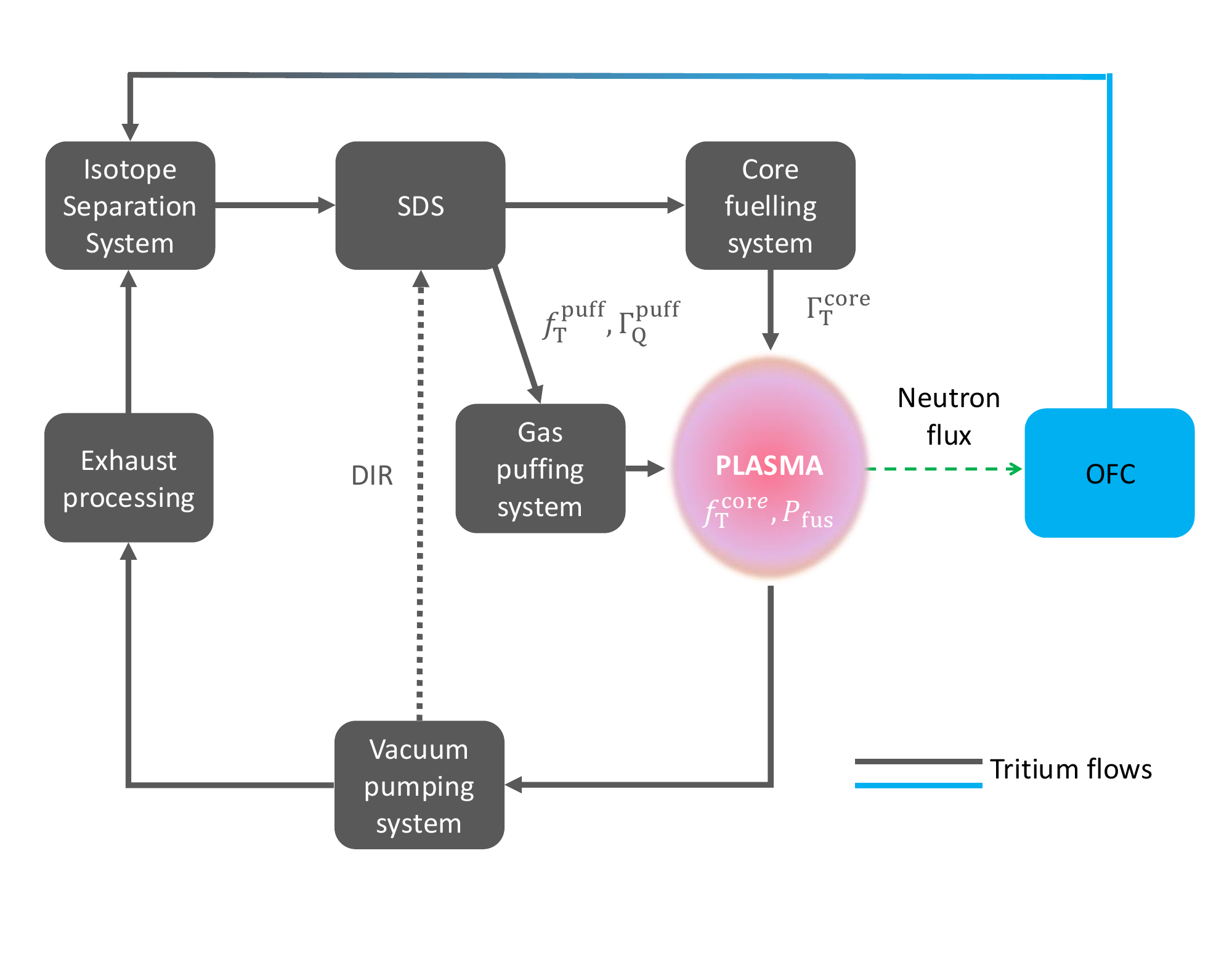}
        \caption{Simplified layout of the fuel cycle model. The outer fuel cycle components have been collapsed in a single block. The storage and delivery system (SDS) provides tritium for both the gas puffing and the core fuelling system.}
        \label{fig:fuel_cycle_layout}
    \end{figure}

    The tritium fuel cycle model in \cite{meschini_modeling_2023} has been extended to include the tritium throughput from the gas puffing system by adding a tritium fuelling term (whenever the gas puff rate contains tritium) that delivers tritium to the edge plasma (assuming no penentration into the core plasma), according to the layout in figure \ref{fig:fuel_cycle_layout}. 
    The reactor and fuel cycle parameters (residence times, extraction efficiency, etc.) used in the model are the same as in \cite{meschini_modeling_2023}, unless otherwise specified. The results presented in section \ref{fig:tss_puffing_DT} are based on this layout.
    
    To isolate the effect of gas puffing on tritium self-sufficiency at varying $\fdir$, we consider also cases in which the core tritium fraction is driven by the combination of puffed stream, DIR stream and tritium stream from the reprocessing plant, without actively controlling core fuelling by providing additional tritium from the storage system. This is non-optimal from the operation standpoint, but allows to clearly highlight how high gas puffing rates impact the TFC, accounting also for the uncertainty on the achievable DIR fraction. Optimization studies accounting for additional by-pass loops (as done, for instance, in \cite{hattab_analysis_2025}) or a variable tritium fraction in the gas puff stream require design-dependent considerations and are not addressed in this work.
    For our purposes, we make the following simplifying assumptions that preserve the essential features of the model, reducing its complexity while retaining its generality:
    
    \begin{itemize}
        \item Low $\TBE$ (TBE $=2\%$)\footnote{Section \ref{sec:methods__database__core__fuelling} makes use of $\fBurn$ because it is the most natural choice given the nature of the estimation of $\GammaQcore$ (i.e., core fuelling is usually prescribed in simulations as the amount of fuel delivered into the plasma). The fuelling efficiency, $\etaFuel$, is implicitly assumed 100\%. Since the following analysis is instead focused on a notional pilot plant, we make use of TBE that embeds the fuelling efficiency (among other factors), which is still assumed to be unitary (i.e. $\Gamma_\mathrm{T,PI} = \GammaTcore$). We consider cases with TBE = 2\% for consistency and ease of comparison with previous works using the same model \cite{meschini_modeling_2023, meschini_impact_2025}.}. This way the dominant tritium stream in the fuel cycle is the one related to the exhaust, and the contribution of the bred tritium is not significant to provide a sizable tritium rich stream to feed into the fuelling system to offset the DIR. This assumption is in line with the projected TBE for near term devices (e.g. ITER \cite{abdou_physics_2020}) and limitations due to excessive fuel dilution and He concentration in the core plasma \cite{whyte_tritium_2023};
        \item $\ratioPuffToCore > 1$. This means that the gas puff rate is at least equal to the core fuelling rate, up to gas puffing dominant cases, as identified by table \ref{tab:fueling_stats} for detached regimes;
        \item Non-negligible $\fdir$. Since fuel cycle design for FPP commonly rely on a DIR to process a significant fraction of the exhaust \cite{abdou_physics_2020,coleman_demo_2019, meschini_modeling_2023, hattab_analysis_2025, clark_breeder_2025, jin2026cfedr}, the cases of interest are those with non-negligible $\fdir$. In such cases, gas puffing becomes particularly consequential, as it unbalances the D:T exhaust ratio and thereby reduces DIR effectiveness. Specifically, we consider $\fdir \geq 0.3$;
        \item The core fuelling mixture is obtained by mixing a 100\% T stream from the tritium plant with the fraction of the DIR stream required to match the total core fuelling rate and achieve a core tritium fraction the closest to $\ftcore=0.5$. The remaining fraction of the DIR stream is routed to the IFC;
        \item The core tritium fraction change consistently with the D:T ratio of the core fuelling. Any delay associated to particle transport in the core is neglected for simplicity. Therefore, $\ftcore = \GammaTcore/\GammaQcore$;
        \item The gas puff injections operate at a single point scenario. Neither $\GammaQ$ nor the tritium fraction in the puffed gas change during operations.
        
        Similarly, the total core fuelling rate, $\GammaQcore$, is kept constant to balance particle losses. A particular case in which $\GammaQ$ and $\fTpuff$ are changed during operations is instead discussed in Section \ref{sec:discussion__different_plasma_scenarios}.

        
    \end{itemize}
        
        Under this assumptions, the evolution of the core fuelling rate and the tritium fuelling fraction can be described by the following equations which apply for 100:0 D:T puffing: 

        \begin{equation}\label{eq:gamma_vs_ftdir}
            \GammaTcore(t) = \frac{\fTdir(t) \GammaQcore}{1-k_1(1-\fTdir(t))} \,,
        \end{equation}   
        
        \begin{equation}\label{eq:ftdir}
            \frac{d\fTdir(t)}{dt} = \fdir \frac{\left(\GammaTex(t) - \fTdir(t)\GammaQex(t)\right)}{N_\mathrm{DIR}} \,.
        \end{equation}

    \noindent Here, $\fTdir$ is the tritium fraction in the D:T mixture through the DIR loop, hence $\fTdir \equiv N_\mathrm{T,DIR}/N_\mathrm{Q,DIR}$; $\GammaTcore$ is the tritium core fuelling rate, which is a a function of time due to its dependency on $\fTdir (t)$ (i.e. $ \GammaTcore(t) = \GammaTcore(\fTdir(t))$); $k_1 \equiv (1-\TBE)(1-\fdir)$; $\GammaTex (t)$ and $\GammaQex (t)$ are the (time dependent) exhaust rate for tritium and hydrogenic species. $N_\mathrm{DIR}$ is the total, steady state inventory in the DIR loop, that is, $N_\mathrm{Q,DIR} = \fdir \GammaQex \taudir$. In the remainder of the text we omit the time dependency of the variables introduced by equations \eqref{eq:gamma_vs_ftdir} and \eqref{eq:ftdir} for simplicity. The derivation of these equations can be found in appendix \ref{appendix:DIR_model}. Lastly, the fusion power as a function of $\ftcore$:
    \begin{equation}\label{eq:pfus_vs_ftcore}
        \frac{\pfus(\ftcore)}{\pfus(\ftcore=0.5)}
        = 4\ftcore(1 - \ftcore)
    \end{equation}
    
    equations \eqref{eq:gamma_vs_ftdir} and \eqref{eq:ftdir}, together with \eqref{eq:pfus_vs_ftcore} provide the time evolution of the tritium core fraction and the fusion power, providing a useful metric to evaluate the effect of 100\%D gas puffing. 

    As explained in Section \ref{sec:intro_matter_injection_TFC}, the impact of the gas puffing on the DIR and the fusion power can be mitigated by adding some tritium to the gas puff mixture. Equation \eqref{eq:gamma_vs_ftdir} can be generalised by introducing the tritium fraction in the gas puff rate, $\fTpuff \equiv \GammaTpuff/(\GammaTpuff + \GammaDPuff)$. The tritium core fuelling rate then becomes:

    \begin{equation}\label{eq:gamma_vs_ftdir_ftpuff}
        \GammaTcore(t) = \frac{\fdir \GammaQcore + k_2 \GammaQ}{1-k_1(1-\fTdir(t))}
    \end{equation}
    
    \noindent where $k_2 \equiv (1-\fdir)(1-\fTdir)\fTpuff$.  Equation (\ref{eq:gamma_vs_ftdir_ftpuff}) encapsulates how the DIR establishes a material coupling between the fuel puffing stream and the core fuelling rate. In other words, even if the direct penetration of puffed gas into the core is negligible \textit{per se}, puffing still influences $\GammaTcore$ indirectly through the recirculated exhaust.
    
    The equations introduced in this section allows to assess the dynamic effect of different gas puffing strategies, as shown in Section \ref{subsec:optimal_T_fraction_gaspuffing}.
    The extension to the cases where a fraction of seeding gas is present in the gas puffing stream follows the same logic.  The model is run starting from the steady-state solution of \eqref{eq:gamma_vs_ftdir_ftpuff} for $\GammaTcore$, and the fusion power is adjusted according to the resulting $\ftcore(t\rightarrow\infty)$.

\section{Results}\label{sec:results}

    The results are organised around the particle-balance hierarchy that determines the exhaust stream entering the TFC. Scaling laws extracted from heterogeneous multi-machine databases are necessarily uncertain. Here, they are therefore not used to prescribe design values, but to identify robust long-range trends and order-of-magnitude separations. This is sufficient to assess the severity of the challenge considered, even if the exact coefficients of the regressions remain uncertain, and inefficiencies unaddressed (appendix \ref{apx:estimation_GammaQcore__puffing_to_core_unimportant}).
    
    The analysis first compares fuel puffing with core fuelling across the available database and examines how their relative importance evolves with device size in section \ref{sec:results__how_high_fuel_puffing}. These trends are then propagated to the TFC interface. Section \ref{subsec:tss_puffing_DT} shows the increase in $\Ist$ that would be needed to sustain operation at increasing values of $\GammaQ$ with 50:50 D:T. Several scenarios in the intermediate range $0.3 < \fdir < 0.7$ require a several-fold increase of $\Ist$, which could be unacceptable from a tritium self-sufficiency and tritium availability point of view. Hence, Section \ref{subsec:optimal_T_fraction_gaspuffing} explores the fuel cycle operations at the steady-state defined by a given puffing scenario with 90:10 D:T and 75:25 D:T,with and without including additional tritium in $\Ist$ to match the core fuelling D:T imbalance - this way, the impact of gas puffing on operation can be isolated. The doubling time in all these simulations is kept fixed to allow meaningful comparisons. 

    Lastly, the same particle balances used in section \ref{sec:results__how_high_fuel_puffing} are applied to seeded impurity and helium ash concentrations in section \ref{sec:results__how_low_impurities}.

    \subsection{Fuel puffing---the dominant particle input stream: comparison with core fuelling}\label{sec:results__how_high_fuel_puffing}

        Table \ref{tab:fueling_stats} reports average values across the 25 machines, ordered with ascending $\Vtot$. Fuel puffing figures in grey cells are estimated as per section \ref{sec:methods__database__puffing_and_concentration}.
        
        Irrespective of device class---conventional (CTK, blue), spherical (STK, green), high-field (HFTK, orange) tokamaks or stellarators (STL, magenta)---the fuel puffing rate $\GammaQ$ is invariably in significant excess of the core fuelling rate $\GammaQcore$. The puff-to-core ratio $\ratioPuffToCore$ attains a multi-machine geometric-mean value of $\sim 10$ (black), with a minimum value of $\sim 2$ in Globus-M2 \cite{Sorokina_2018}. Importantly, the same conclusion also holds (within a wider $\tpm 4$) only for cases specifically at detachment onset \cite{Moscheni_2026}. Puffing therefore manifests, on average, as the dominant source of externally injected fuel in detached regimes.

        \begin{landscape}
            \begin{table}[t]
\centering
\small
\begin{tabular}{lcccccccc}
\toprule
Machine & ID & $\langle\GammaQcore\rangle \, [\mathrm{at \, s^{-1}}]$ & $\langle\GammaQ\rangle \, [\mathrm{at \, s^{-1}}]$ & $\langle\GammaZ\rangle \, [\mathrm{at \, s^{-1}}]$ & $\langle\GammaHeCore\rangle \, [\mathrm{at \, s^{-1}}]$ & $\langle\ratioPuffToCore\rangle \,[-]$ & $\langle \cZ \rangle \,[-]$ & $\langle \cHe \rangle \,[-]$ \\
\midrule
Globus-M2 & \cellcolor{LIME}GM & $(1.8 \times 10^{20}) \tpm 1.0$ & $(3.5 \times 10^{20}) \tpm 1.0$ & $(5.0 \times 10^{19}) \tpm 1.0$ & -- & $(1.9 \times 10^{0}) \tpm 1.0$ & $(8.6 \times 10^{-2}) \tpm 1.0$ & -- \\
W7-AS & \cellcolor{MAGENTA}WA & $(8.7 \times 10^{19}) \tpm 1.0$ & $(7.5 \times 10^{20}) \tpm 2.7$ & -- & -- & $(8.5 \times 10^{0}) \tpm 2.7$ & -- & -- \\
TCV & \cellcolor{DEEPSKYBLUE}TC & $(1.1 \times 10^{20}) \tpm 1.5$ & $(7.4 \times 10^{20}) \tpm 2.1$ & $(4.1 \times 10^{20}) \tpm 2.3$ & -- & $(6.2 \times 10^{0}) \tpm 1.9$ & $(3.5 \times 10^{-1}) \tpm 2.1$ & -- \\
JFT-2M & \cellcolor{DEEPSKYBLUE}J2 & $(1.0 \times 10^{20}) \tpm 1.7$ & $(1.4 \times 10^{21}) \tpm 1.4$ & -- & -- & $(1.4 \times 10^{1}) \tpm 1.2$ & -- & -- \\
MAST-U & \cellcolor{LIME}MU & $(2.0 \times 10^{20}) \tpm 1.3$ & $(4.1 \times 10^{21}) \tpm 2.3$ & $(3.9 \times 10^{21}) \tpm 6.5$ & -- & $(1.8 \times 10^{1}) \tpm 2.0$ & $(4.2 \times 10^{-1}) \tpm 2.6$ & -- \\
NSTX-U & \cellcolor{LIME}NU & $(1.9 \times 10^{20}) \tpm 1.0$ & $(7.0 \times 10^{21}) \tpm 1.2$ & -- & -- & $(3.7 \times 10^{1}) \tpm 1.2$ & -- & -- \\
EAST & \cellcolor{DEEPSKYBLUE}EA & $(2.6 \times 10^{20}) \tpm 1.1$ & $(1.4 \times 10^{21}) \tpm 1.4$ & $(7.5 \times 10^{18}) \tpm 1.5$ & -- & $(4.9 \times 10^{0}) \tpm 1.3$ & $(3.1 \times 10^{-2}) \tpm 1.5$ & -- \\
AUG & \cellcolor{DEEPSKYBLUE}AU & $(6.6 \times 10^{20}) \tpm 2.1$ & $(1.3 \times 10^{22}) \tpm 2.5$ & $(5.3 \times 10^{20}) \tpm 5.5$ & -- & $(2.2 \times 10^{1}) \tpm 2.3$ & $(3.7 \times 10^{-2}) \tpm 6.0$ & -- \\
NSTX & \cellcolor{LIME}NS & $(4.3 \times 10^{20}) \tpm 1.2$ & $(7.3 \times 10^{21}) \tpm 2.2$ & -- & -- & $(1.7 \times 10^{1}) \tpm 2.0$ & -- & -- \\
SPARC & \cellcolor{ORANGE}SP & $(1.5 \times 10^{21}) \tpm 1.0$ & $(2.6 \times 10^{22}) \tpm 1.2$ & $(1.9 \times 10^{20}) \tpm 1.5$ & -- & $(1.7 \times 10^{1}) \tpm 1.2$ & $(6.9 \times 10^{-3}) \tpm 1.4$ & -- \\
HL-2M & \cellcolor{DEEPSKYBLUE}HM & $(4.7 \times 10^{20}) \tpm 1.5$ & $(6.0 \times 10^{21}) \tpm 3.8$ & $(1.2 \times 10^{18}) \tpm 1.0$ & -- & $(1.3 \times 10^{1}) \tpm 5.4$ & $(3.7 \times 10^{-4}) \tpm 1.9$ & -- \\
DIII-D & \cellcolor{DEEPSKYBLUE}DD & $(5.8 \times 10^{20}) \tpm 1.4$ & $(9.3 \times 10^{21}) \tpm 2.1$ & $(1.2 \times 10^{20}) \tpm 4.9$ & -- & $(1.4 \times 10^{1}) \tpm 2.5$ & $(5.2 \times 10^{-2}) \tpm 11$ & -- \\
LHD & \cellcolor{MAGENTA}LH & $(6.5 \times 10^{21}) \tpm 1.0$ & $(3.9 \times 10^{22}) \tpm 3.7$ & $(9.2 \times 10^{20}) \tpm 2.4$ & -- & $(5.9 \times 10^{0}) \tpm 3.9$ & $(9.6 \times 10^{-2}) \tpm 3.1$ & -- \\
DTT & \cellcolor{DEEPSKYBLUE}DT & $(5.8 \times 10^{20}) \tpm 1.5$ & $(1.9 \times 10^{22}) \tpm 3.0$ & $(3.0 \times 10^{20}) \tpm 3.8$ & -- & $(3.3 \times 10^{1}) \tpm 2.6$ & $(1.0 \times 10^{-2}) \tpm 2.0$ & -- \\
JET & \cellcolor{DEEPSKYBLUE}JE & $(1.2 \times 10^{22}) \tpm 1.0$ & $(2.4 \times 10^{22}) \tpm 1.9$ & $(1.2 \times 10^{21}) \tpm 5.6$ & -- & $(2.0 \times 10^{0}) \tpm 1.9$ & $(3.2 \times 10^{-2}) \tpm 5.9$ & -- \\
JT-60SA & \cellcolor{DEEPSKYBLUE}J6 & $(2.5 \times 10^{21}) \tpm 3.1$ & $(2.4 \times 10^{22}) \tpm 1.9$ & $(5.1 \times 10^{20}) \tpm 2.4$ & -- & $(9.6 \times 10^{0}) \tpm 4.3$ & $(1.9 \times 10^{-2}) \tpm 1.8$ & -- \\
ARC & \cellcolor{ORANGE}AR & \cellcolor{Silver}$(1.2 \times 10^{22}) \tpm 1.0$ & \cellcolor{Silver}$(1.7 \times 10^{23}) \tpm 2.0$ & \cellcolor{Silver}$(4.8 \times 10^{20}) \tpm 2.0$ & $(1.9 \times 10^{20}) \tpm 1.0$ & $(1.4 \times 10^{1}) \tpm 1.0$ & $(2.6 \times 10^{-3}) \tpm 1.0$ & $(1.0 \times 10^{-3}) \tpm 1.0$ \\
MANTA & \cellcolor{ORANGE}MA & \cellcolor{Silver}$(1.1 \times 10^{22}) \tpm 1.0$ & \cellcolor{Silver}$(1.5 \times 10^{23}) \tpm 2.0$ & \cellcolor{Silver}$(4.9 \times 10^{20}) \tpm 2.0$ & $(1.6 \times 10^{20}) \tpm 1.0$ & $(1.4 \times 10^{1}) \tpm 1.0$ & $(3.0 \times 10^{-3}) \tpm 1.0$ & $(9.9 \times 10^{-4}) \tpm 1.0$ \\
Infinity Two & \cellcolor{MAGENTA}I2 & $(2.2 \times 10^{22}) \tpm 1.0$ & \cellcolor{Silver}$(1.3 \times 10^{23}) \tpm 2.0$ & \cellcolor{Silver}$(5.2 \times 10^{20}) \tpm 2.0$ & $(2.8 \times 10^{20}) \tpm 1.0$ & $(5.9 \times 10^{0}) \tpm 1.0$ & $(3.4 \times 10^{-3}) \tpm 1.0$ & $(1.9 \times 10^{-3}) \tpm 1.0$ \\
STEP & \cellcolor{LIME}ST & $(3.5 \times 10^{22}) \tpm 1.0$ & $(4.7 \times 10^{23}) \tpm 2.0$ & $(6.5 \times 10^{21}) \tpm 1.3$ & $(3.5 \times 10^{20}) \tpm 1.0$ & $(1.3 \times 10^{1}) \tpm 2.0$ & $(1.2 \times 10^{-2}) \tpm 2.3$ & $(6.8 \times 10^{-4}) \tpm 1.9$ \\
Stellaris & \cellcolor{MAGENTA}SL & \cellcolor{Silver}$(6.4 \times 10^{22}) \tpm 1.0$ & \cellcolor{Silver}$(2.2 \times 10^{23}) \tpm 2.0$ & \cellcolor{Silver}$(3.5 \times 10^{20}) \tpm 2.0$ & $(9.6 \times 10^{20}) \tpm 1.0$ & $(3.4 \times 10^{0}) \tpm 1.0$ & $(1.2 \times 10^{-3}) \tpm 1.0$ & $(3.4 \times 10^{-3}) \tpm 1.0$ \\
CFETR & \cellcolor{DEEPSKYBLUE}CF & $(1.5 \times 10^{22}) \tpm 1.0$ & $(1.7 \times 10^{23}) \tpm 2.2$ & \cellcolor{Silver}$(1.5 \times 10^{21}) \tpm 2.0$ & $(2.1 \times 10^{20}) \tpm 1.0$ & $(1.2 \times 10^{1}) \tpm 2.2$ & $(7.8 \times 10^{-3}) \tpm 2.0$ & $(1.1 \times 10^{-3}) \tpm 2.0$ \\
ITER & \cellcolor{DEEPSKYBLUE}IT & $(1.7 \times 10^{22}) \tpm 2.8$ & $(1.1 \times 10^{23}) \tpm 1.9$ & $(1.1 \times 10^{20}) \tpm 3.0$ & $(1.0 \times 10^{20}) \tpm 1.0$ & $(6.1 \times 10^{0}) \tpm 2.8$ & $(8.5 \times 10^{-4}) \tpm 2.8$ & $(7.5 \times 10^{-4}) \tpm 1.9$ \\
GIGA & \cellcolor{MAGENTA}GI & \cellcolor{Silver}$(7.1 \times 10^{22}) \tpm 1.0$ & \cellcolor{Silver}$(4.0 \times 10^{23}) \tpm 2.0$ & \cellcolor{Silver}$(1.0 \times 10^{21}) \tpm 2.0$ & $(1.1 \times 10^{21}) \tpm 1.0$ & $(5.6 \times 10^{0}) \tpm 1.0$ & $(2.1 \times 10^{-3}) \tpm 1.0$ & $(2.3 \times 10^{-3}) \tpm 1.0$ \\
EU-DEMO & \cellcolor{DEEPSKYBLUE}ED & $(1.3 \times 10^{22}) \tpm 2.8$ & $(5.7 \times 10^{23}) \tpm 3.0$ & $(5.0 \times 10^{20}) \tpm 4.5$ & $(6.2 \times 10^{20}) \tpm 1.3$ & $(7.1 \times 10^{1}) \tpm 2.4$ & $(1.7 \times 10^{-3}) \tpm 6.6$ & $(2.1 \times 10^{-3}) \tpm 4.3$ \\
\midrule
Multi-machine & -- & $(2.0 \times 10^{21}) \tpm 9.2$ & $(2.1 \times 10^{22}) \tpm 9.0$ & $(3.2 \times 10^{20}) \tpm 6.9$ & $(3.2 \times 10^{20}) \tpm 2.3$ & $(1.0 \times 10^{1}) \tpm 2.3$ & $(1.1 \times 10^{-2}) \tpm 6.9$ & $(1.4 \times 10^{-3}) \tpm 1.7$ \\
\bottomrule
\end{tabular}
\caption{Machine-wise geometric means $\langle\cdot\rangle$ with multiplicative $1\sigma$ spread. From left to right: core fueling rate, fuel and impurity puffing rates, helium production rate, ratio of fuel puffing and core fueling rates, and impurity concentrations in the outgoing stream. The colour-coding replicates the legend in figure \ref{fig:scaling_GammaQcore_and_GammaQpuff_Vtot}, differentiating between different device types. Greyed entries are estimates.}
\label{tab:fueling_stats}
\end{table}

        \end{landscape}
        
        The ordering with volume apparent in the raw data is substantiated by regressions. Core fuelling rates scale with total plasma surface as ($R^2 = 0.86$):
        
        \begin{equation}\label{eq:scaling_GammaQcore_Stot}
            \GammaQcore = 4.1 \times 10^{18} \times \Stot^{1.30} \,,
        \end{equation}

        with units being those in table \ref{tab:db_minmax}. An equivalent representation in terms of total plasma volume yields ($R^2 = 0.82$):
        
        \begin{equation}\label{eq:scaling_GammaQcore_Vtot}
            \GammaQcore = 6.6 \times 10^{19} \times \Vtot^{0.89} \,.
        \end{equation}
        
\begin{figure*}[htbp]
            \centering
            \subfloat{\includegraphics[width=0.45\textwidth]{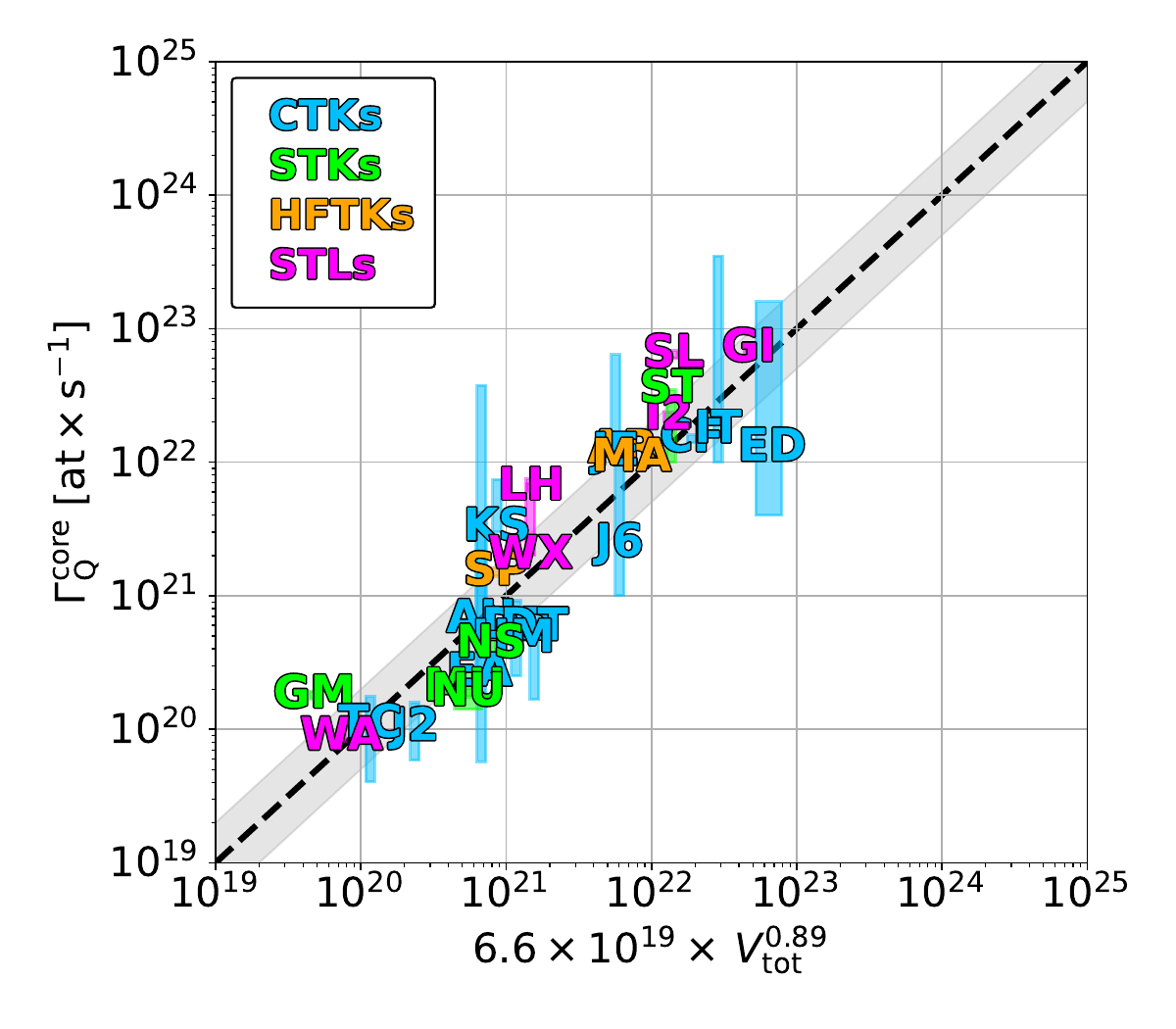}\label{fig:scaling_GammaQcore_Vtot}}
            \hspace{0.15cm}
            \subfloat{\includegraphics[width=0.45\textwidth]{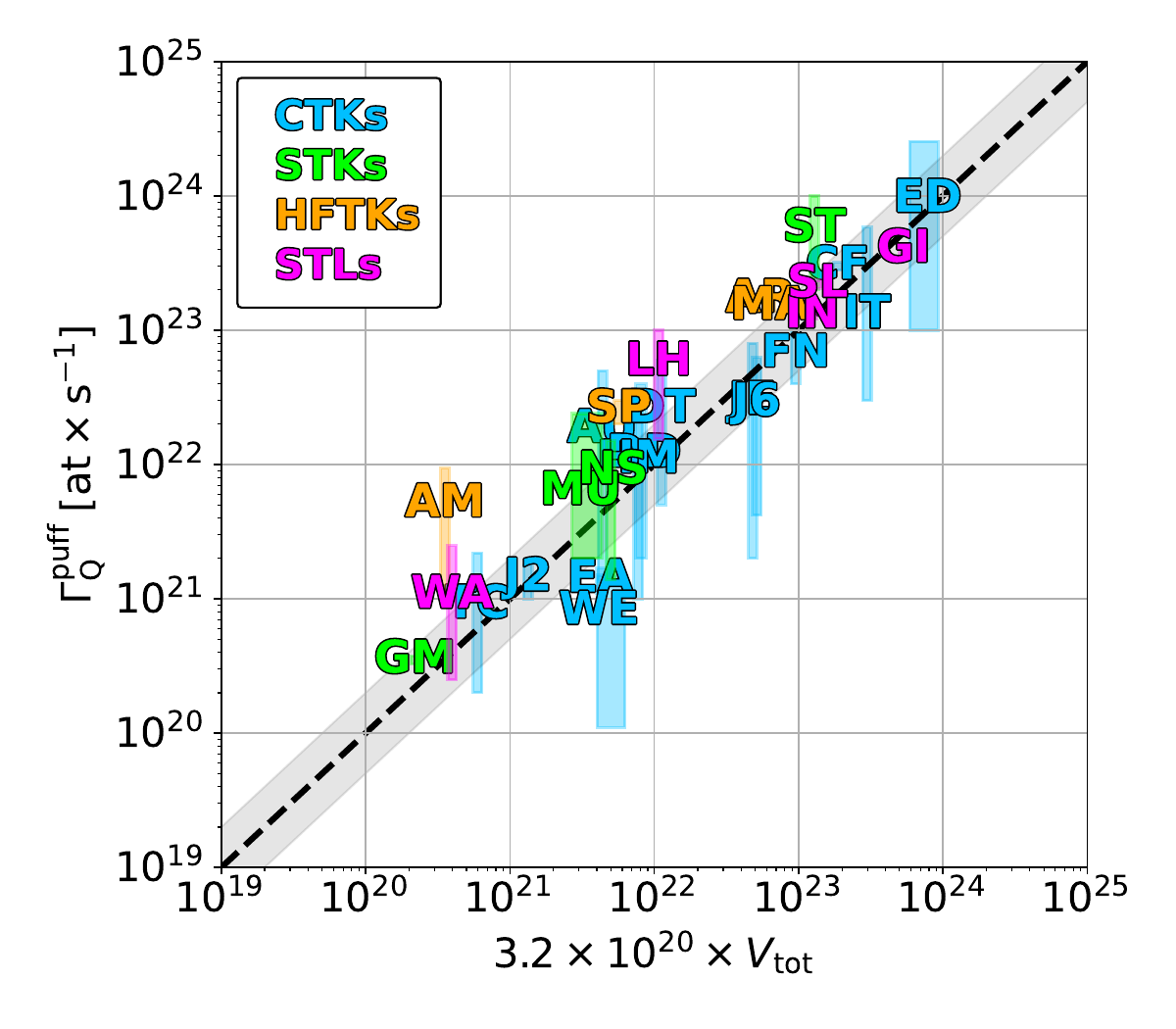}\label{fig:scaling_GammaQpuff_Vtot}}
            \caption{(a) Regression of core fuelling rate $\GammaQcore$ as a function of total volume $\Vtot$. Data are taken from the newly-compiled database \cite{zenodo_repo_v2}. (b) Regression of fuel puffing rate $\GammaQ$ as a function of $\Vtot$, with data taken from the original database \cite{zenodo_repo_v1} of \cite{Moscheni_2026}. This scaling is computed: discarding linear plasma devices (where divertor volume is undefined); and omitting newly-added devices ARC (AR) \cite{sorbom_arc_2015, Hillesheim_2026}, GIGA (GI) \cite{gaussfusion_cdr_exec_2026}, Infinity Two (IN) \cite{Guttenfelder_2025}, MANTA (MA) \cite{Rutherford_2024} and Stellaris (SL) \cite{LION2025114868}. Puffing figures of this set are taken from table 3 of \cite{Moscheni_2026} (equations (10)--(12)) and pictured for reference. The $\tpm 2$ band around the scaling is depicted in grey.}
            \label{fig:scaling_GammaQcore_and_GammaQpuff_Vtot}
        \end{figure*}

        The goodness of fit is comparable, though slightly lower in the volumetric representation---which is represented in figure \ref{fig:scaling_GammaQcore_Vtot}.
        
        As previously observed for puffing rates in \cite{Moscheni_2026}, the scaling holds across different device concepts, evidencing a robust and device-independent ordering of the $\GammaQcore$ injection at this level of detail. A more in-depth assessment is provided throughout appendix \ref{apx:estimation_GammaQcore} and, specifically, in appendix \ref{apx:robustness_scaling_GammaQcore}.
        
        In contrast, gas puffing scales approximately linearly with divertor volume and, to first order, with plasma volume (equation (4) and section 4.4.1 of \cite{Moscheni_2026}). A direct scaling with $\Vtot$ gives ($R^2 = 0.76$):
        
        \begin{equation}\label{eq:scaling_GammaQpuff_Vtot}
            \GammaQ = 3.2 \times 10^{20} \times \Vtot \,,
        \end{equation}

        which is pictured in figure \ref{fig:scaling_GammaQpuff_Vtot}---with additional details on the machine set employed given in the caption. The appropriateness of the regression is verified in appendix \ref{apx:comparison_vs_Moscheni_OPQ}.
        
        Combining the core fuelling and fuel puffing scalings in equations (\ref{eq:scaling_GammaQcore_Vtot}) and (\ref{eq:scaling_GammaQpuff_Vtot}), respectively, yields:
        
        \begin{equation}\label{eq:scaling_ratio_GammaQpuff_GammaQcore}
            \ratioPuffToCore = \frac{\GammaQ}{\GammaQcore} \sim 4.8 \times \Vtot^{0.11} \,.
        \end{equation}

        That the ratio exceeds unity, on average, is clarified by the side-by-side comparison of figures \ref{fig:scaling_GammaQcore_Vtot} and \ref{fig:scaling_GammaQpuff_Vtot}. Additionally, the positive exponent in $\Vtot^{0.11}$ indicates a tendency for the puff-to-core fuelling disparity to increase with device size. Within the present uncertainties, the ratio $\ratioPuffToCore = \GammaQ/\GammaQcore$ can therefore be considered not to decrease with increasing device volume. Consequently, \textit{puffing remains generally larger than core fuelling across the entire range of devices}, hence substantiating the TFC implications.
        
        Notably, all future reactor-scale devices cluster in the region $\GammaQcore \gtrsim 10^{22}\,\mathrm{at \, s^{-1}}$ with $\Vtot \gtrsim 150\,\mathrm{m^3}$. Equation (\ref{eq:scaling_ratio_GammaQpuff_GammaQcore}) therefore indicates $\GammaQ/\GammaQcore \gtrsim 8$ in that regime, placing future TFC requirements firmly within a puffing-dominated region of the design space.

        Further comments in this regard are given in section \ref{sec:discussion__core_fuelling__size}.

 \subsection{Implications of 50:50 D:T puffing on tritium self-sufficiency}\label{subsec:tss_puffing_DT}
 
\begin{figure}
        \centering
        \includegraphics[width=0.9\linewidth]{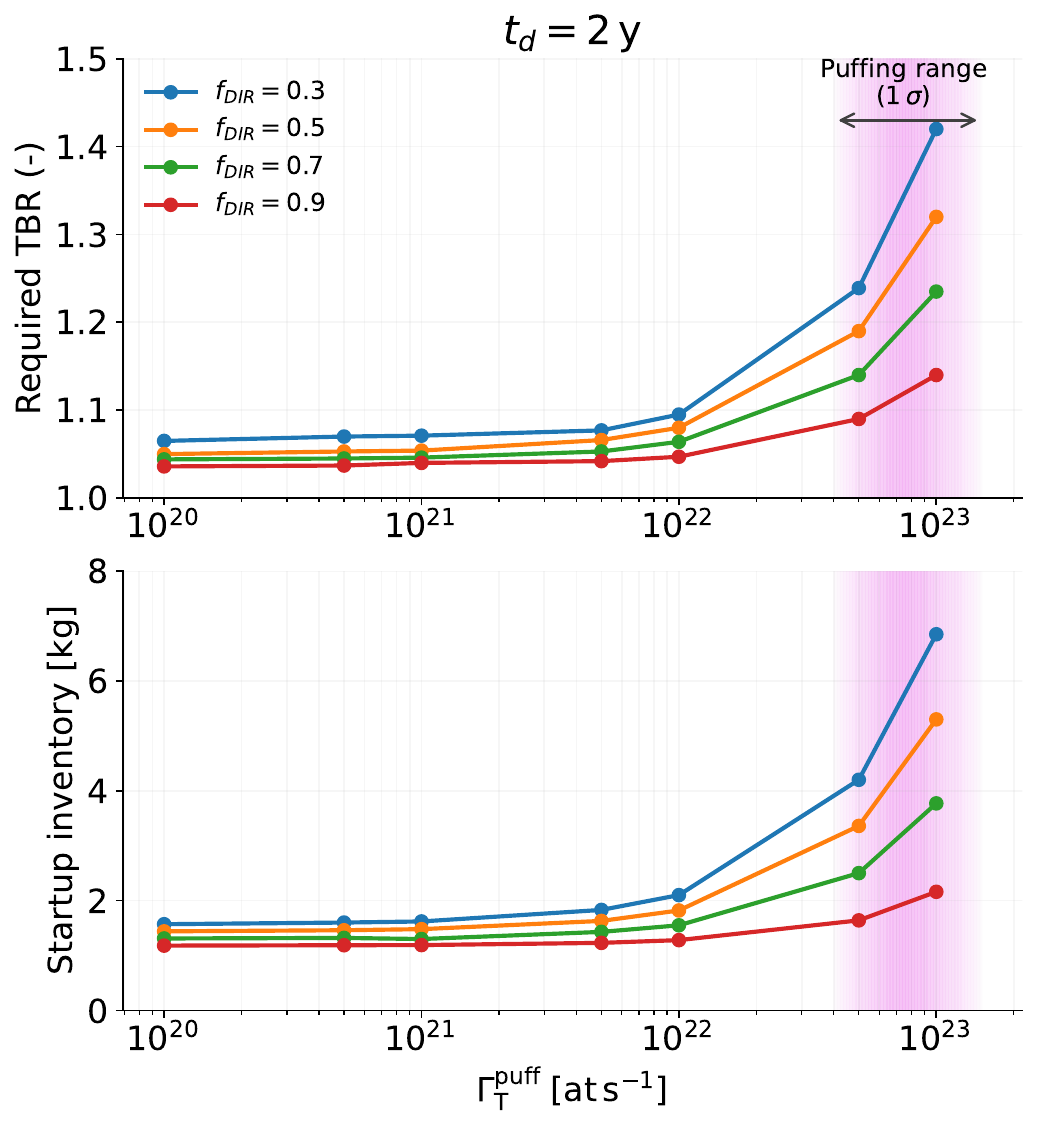}
        \caption{Required TBR and start-up inventory to achieve a doubling time of 2 years as a function of the tritium puff rate. $\pfus = 525 \,\rm MW$, corresponding to $\GammaTcore = 9.3 \times 10^{21}  \, \mathrm{at \, s^{-1}}$. Puffing is performed with 50:50 D:T, that is, $\GammaTpuff = \GammaQ/2$. The puffing range for FPP designs is also depicted, according to the values in table \ref{tab:fueling_stats}.}
        \label{fig:tss_puffing_DT}
    \end{figure}

    The implications of the gas puff rates reported in table \ref{tab:fueling_stats} on tritium self-sufficiency are summarised in figure \ref{fig:tss_puffing_DT}, where $\tbrr$ and $\Ist$ are quantified at increasing values of T puffed into the vacuum chamber, for $t_d = 2 \, \rm y$, AF = 0.9, and a 50:50 D:T core fuelling and puffing ratio. In this simulation $\GammaTcore = 9.3 \times 10^{21}  \, \mathrm{at \, s^{-1}}$ (corresponding to the baseline parameters TBE = 2\% and $\pfus = 525 \, \rm MW$).
    
    The critical value of $\GammaTpuff$ above which both $\tbrr$ and $\Ist$ increase significantly is $\GammaTpuff \simeq 5 \times 10^{21}  \, \mathrm{at \, s^{-1}}$, that is, when its contribution becomes comparable to $\GammaTcore$. Yet, $\GammaTpuff \simeq 5 \times 10^{21}  \, \mathrm{at \, s^{-1}}$ is significantly lower than the required gas puffing rate reported in table \ref{tab:fueling_stats} for comparable $\GammaQcore$ (e.g. JT-60SA, ARC, MANTA, Infinity Two, STEP). For those machines, the lower bound of $\GammaTpuff$ considering 50:50 D:T puffing and the 1/2 uncertainty factor is about $\GammaTpuff \simeq 4 \times 10^{22}  \, \mathrm{at \, s^{-1}}$, while the upper bound, considering the 2$\times$ uncertainty factor, is $\GammaTpuff \simeq 1.5 \times 10^{23}  \, \mathrm{at \, s^{-1}}$. For low DIR fractions ($\fdir = 0.3$), in the most optimistic cases where $\GammaTpuff \simeq 4 \times 10^{22}  \, \mathrm{at \, s^{-1}}$, $\Ist$ increases from $\sim 1 \,\rm kg$ up to $\sim 4 \,\rm kg$, and $\tbrr$ increases from 1.065 (without tritium in the puff stream) up to 1.2. At high DIR fractions ($\fdir = 0.9$) a tritium puffing stream of $\GammaTpuff \simeq 4 \times 10^{22}  \, \mathrm{at \, s^{-1}}$ can be easily sustained with minimal increases in $\tbrr$ and $\Ist$, while the upper bound range, $\GammaTpuff \simeq 1.5 \times 10^{23}  \, \mathrm{at \, s^{-1}}$, requires a significant increases of $\tbrr$ up to 1.14, and a start-up inventory twice the value without tritium in the puffing stream.
    
    Therefore, adopting a 50:50 D:T puffing mixture from the start of operations, at the gas puffing rates expected for future fusion power plants, could impose extremely demanding requirements on the fuel cycle architectures commonly considered in the literature. Section \ref{subsec:optimal_T_fraction_gaspuffing} addresses what would be the steady-state operation for a given gas puffing scenario, without adding further tritium to the start-up inventory to match the imbalance in the D:T core fuelling ratio.

    \subsection{Optimal T fraction for gas puffing without $\GammaTcore$ rebalancing}\label{subsec:optimal_T_fraction_gaspuffing}
        
        In this section we do not seek the optimal value of $\fTpuff$ across the entire design space. The goal is instead to present a coherent picture of the trade-offs that arise when the gas puff rate and its composition are included into fuel cycle analysis, and illustrate the strategy to find the optimal value, which is ultimately dependent on machine-specific parameters and constraints. A qualitative chart linking operating parameters, requirements and constraints is shown in the Appendix, in figure \ref{fig:plasma_tfc_flowchart}.
        
        The optimal T fraction in the gas puffing depends on many fuel cycle parameters (most notably, TBE and $\fdir$). The dependency on TBE is encoded in $\ratioPuffToCore \equiv \GammaQ/\GammaQcore$ because at fixed fusion power and 50:50 D:T fuelling, $\GammaQcore = 2 \pfus /(E_\mathrm{fus}\mathrm{TBE})$. Once exhausted, a fraction $\fdir$ is directly routed to the fuelling system through the DIR. Its D:T ratio, $\fTdir$, depends on both $\ratioPuffToCore$ and the D:T ratio of the puffed gas, $\fTpuff$. 
        
        \subsubsection{Implications on achievable fusion power}\label{subsec:optimal_T_fraction_gaspuffing__implications_Pfus}
        
\begin{figure}
                \centering
                \includegraphics[width=0.95\linewidth]{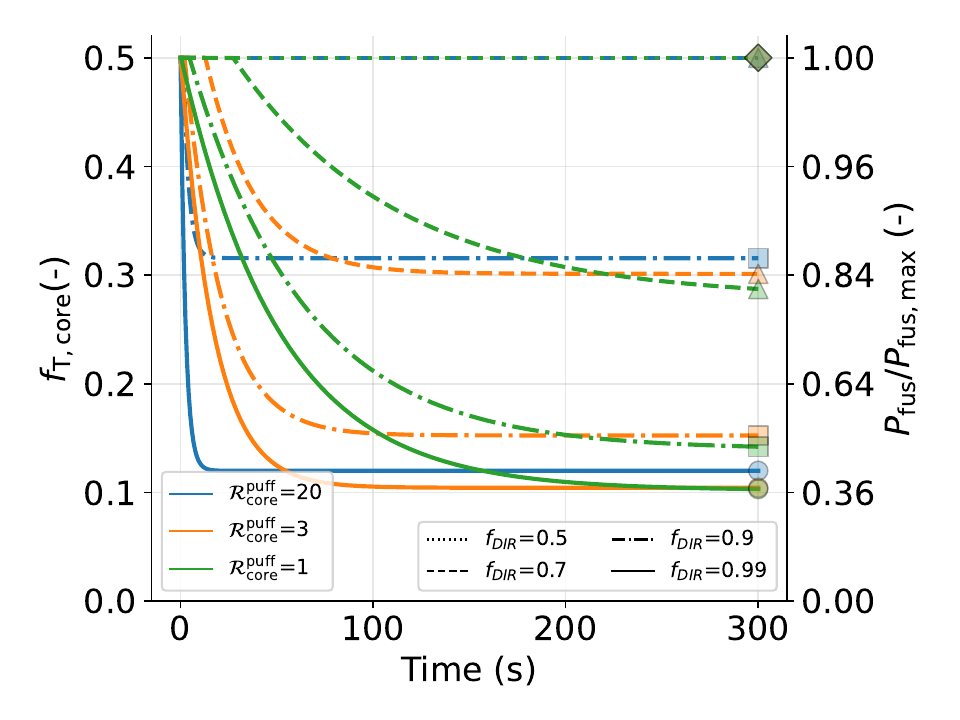}
                \caption{Evolution of the tritium core fraction as a function of $\ratioPuffToCore$ and DIR fraction. Core fuelling is initially performed with 50:50 D:T, while gas puffing with fixed 90:10 D:T. The linestyle encodes the DIR fraction, while the color the ratio between puff rate and fuelling rate. The markers identify the operating conditions used in figure \ref{fig:f_T_rho_TBRr_Ist_90_10}.}
                \label{fig:f_T_Pfus_90D10T}
            \end{figure}

            Figure \ref{fig:f_T_Pfus_90D10T} shows the evolution of $\ftcore$ when gas puffing is injected at 90:10 D:T, together with the fusion power ratio, according to equations \eqref{eq:gamma_vs_ftdir}, \eqref{eq:ftdir}, and \eqref{eq:gamma_vs_ftdir_ftpuff}, and the assumptions defined in Section \ref{sec:methods__TFC_model}. 
            
            At high DIR fractions, the fusion power is reduced by $\sim 60\%$. For $\fdir < 0.7$ a steady state is instead reached for any $\ratioPuffToCore$ where the fusion power drops by a moderate 15\%. As $\ratioPuffToCore$ increases (higher $\GammaQ$) the steady state value of $\ftcore$ approaches 0.5. This stems from the fact that at those high gas puff rates the core fuelling mixture can be easily provided by the (1-$\fdir$) fraction of exhaust from which pure T is extracted, plus a contribution from the DIR stream. 
            
            Nonetheless, increasing $\ratioPuffToCore$ comes at the cost of a higher start-up inventory, which depends on the combined tritium streams $\GammaTpuff + \GammaTcore$, making both $\fTpuff$ and $\fdir$ central to the result. Consequently, the choice of increasing $\ratioPuffToCore$ depends strongly on the DIR fraction, as discussed in the next section.

        \subsubsection{Implications on start-up inventory and $\tbrr$}\label{subsec:optimal_T_fraction_gaspuffing__implications_Ist_TBRr}
        
            Figure \ref{fig:f_T_Pfus_90D10T} suggests to operate at a lower nominal core tritium fraction when $\ratioPuffToCore \gg 1$, since $\ftcore$ would decrease due to tritium dilution in the DIR loop (we recall that, in this section, we are not considering an additional tritium contribution from the start-up inventory to rebalance the core tritium fraction). By doing so, the start-up inventory can be reduced, potentially reaching an optimal operating point defined by $\GammaQ$.
            
            Figure \ref{fig:f_T_rho_TBRr_Ist_90_10} maps the steady-states shown in figure \ref{fig:f_T_Pfus_90D10T} into the tritium self-sufficiency metrics. Each marker corresponds to a scenario with fixed $\ratioPuffToCore$, $\fdir$, $\ftcore$, and $\pfus$. For intermediate DIR fractions ($\fdir = 0.5 - 0.7$), increasing $\ratioPuffToCore$ from 1 to 20 does not increase $\Ist$ by a factor of 20, but only by roughly a factor of 2, because $\GammaTpuff$ is smaller than $\GammaTcore$ at $\ratioPuffToCore=1$ and becomes only of the order of $2\GammaTcore$ at $\ratioPuffToCore=20$ (90:10 D:T ratio). In this regime, the loss in fusion power associated with sub-optimal D:T operation can therefore be reduced substantially at the cost of a twofold increase in start-up inventory.
            
            The situation is different at very high $\fdir$. At $\fdir = 0.9$, the normalized fusion power is 86\% for $\ratioPuffToCore = 20$, while 52\% for $\ratioPuffToCore = 3$ and 48\% for $\ratioPuffToCore = 1$. The higher $\tbrr$ and $\Ist$ for $\ratioPuffToCore = 20$ could be acceptable in light of the smaller fusion power reduction. For $\fdir = 0.99$, increasing $\ratioPuffToCore$ from 1 to 20 leads to a small improvement in achievable fusion power, at the cost of increasing $\Ist$ by about 50\%. At the opposite end, for low $\fdir$, operating at larger gas puff rates provides essentially no benefit to TFC, because $\ftcore$ is already close to 0.5 and $\pfus$ is therefore saturated, while $\Ist$ continues to increase. The existence of an optimal operating region is thus not determined by $\fTpuff$ alone, but by the combined action of $\ratioPuffToCore$, $\fdir$, and the relative weight of $\GammaTpuff$ and $\GammaTcore$ in setting the start-up inventory.
            
            

            A second case with 75:25 D:T puffing is shown in figure \ref{fig:f_T_rho_TBRr_Ist_75_25}. Because the tritium fraction in the puffing stream is increased relative to the 90:10 D:T case, the operating points move closer to maximum fusion power, but also towards higher $\tbrr$ and $\Ist$, consistently with the trends shown in figure \ref{fig:tss_puffing_DT}. In this case, low puffing rates may become preferable around $\fdir \simeq 0.9$: the fusion-power loss remains limited, of order $\sim 15\%$, while the load imposed on the pumping system and IFC is substantially reduced. Contrary to the 90:10 D:T case, even at $\fdir = 0.99$ and $\ratioPuffToCore = 20$, the normalized fusion power remains high (83\%), thus avoiding the increase in $\tbrr$ that would otherwise result from a low neutron and tritium production.

        \subsubsection{Trade-off between start-up inventory and fusion power}\label{subsubsec:optimal_T_fraction_gaspuffing__trade_off}
            
\begin{figure}
                \centering
                \includegraphics[width=0.95\linewidth]{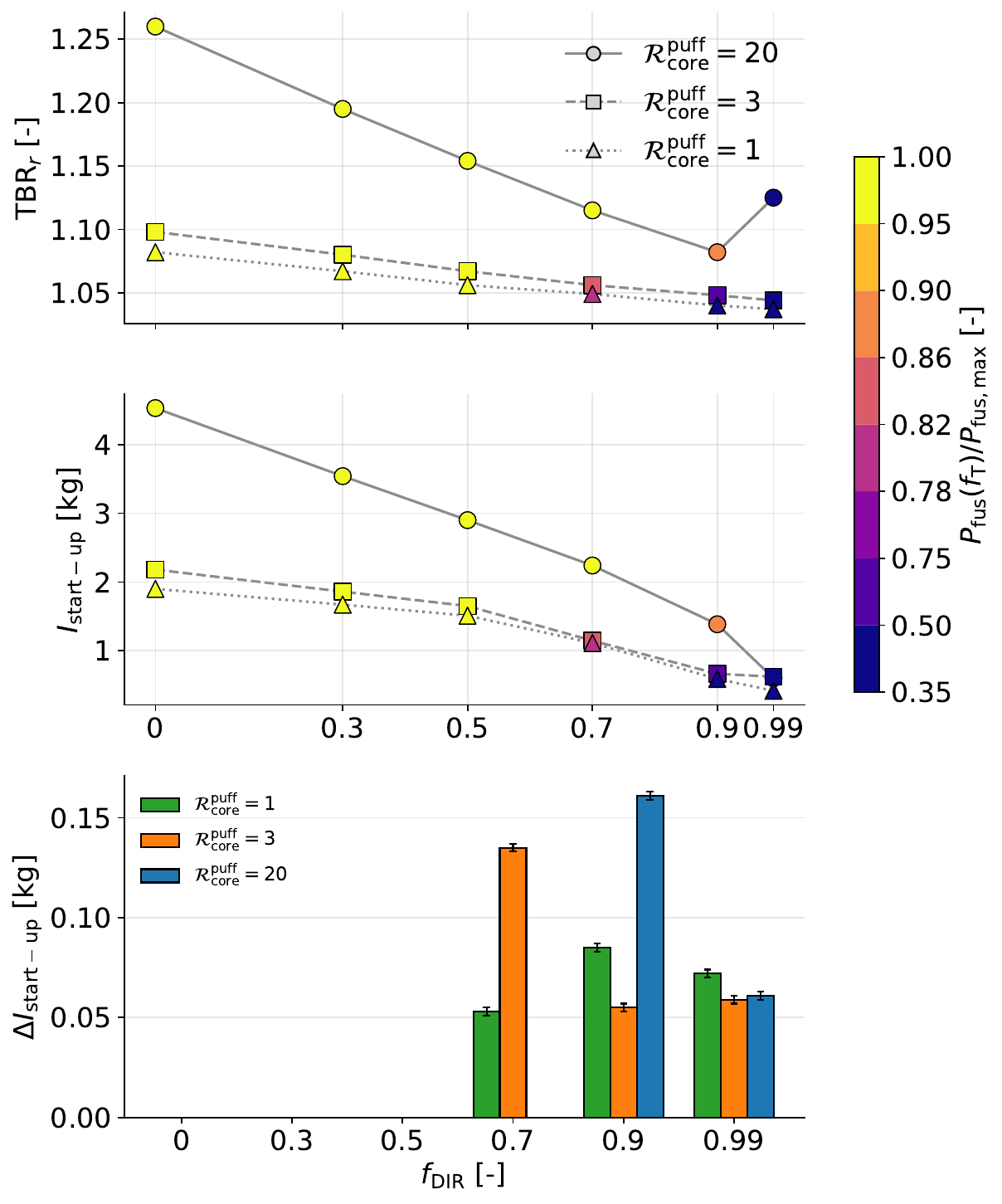}
                \caption{Top: $\tbrr$ and $\Ist$ for different $\ratioPuffToCore$ and $\fdir$. Both $\ftcore$ and the normalised fusion power depends on $\ratioPuffToCore$ per equations \eqref{eq:ftdir} and \eqref{eq:pfus_vs_ftcore}. Gas puffing is performed with 90:10 D:T. Bottom: increase in $\Ist$ to rebalance the core fuelling mixture.}
                \label{fig:f_T_rho_TBRr_Ist_90_10}
            \end{figure}

\begin{figure}
                \centering
                \includegraphics[width=0.95\linewidth]{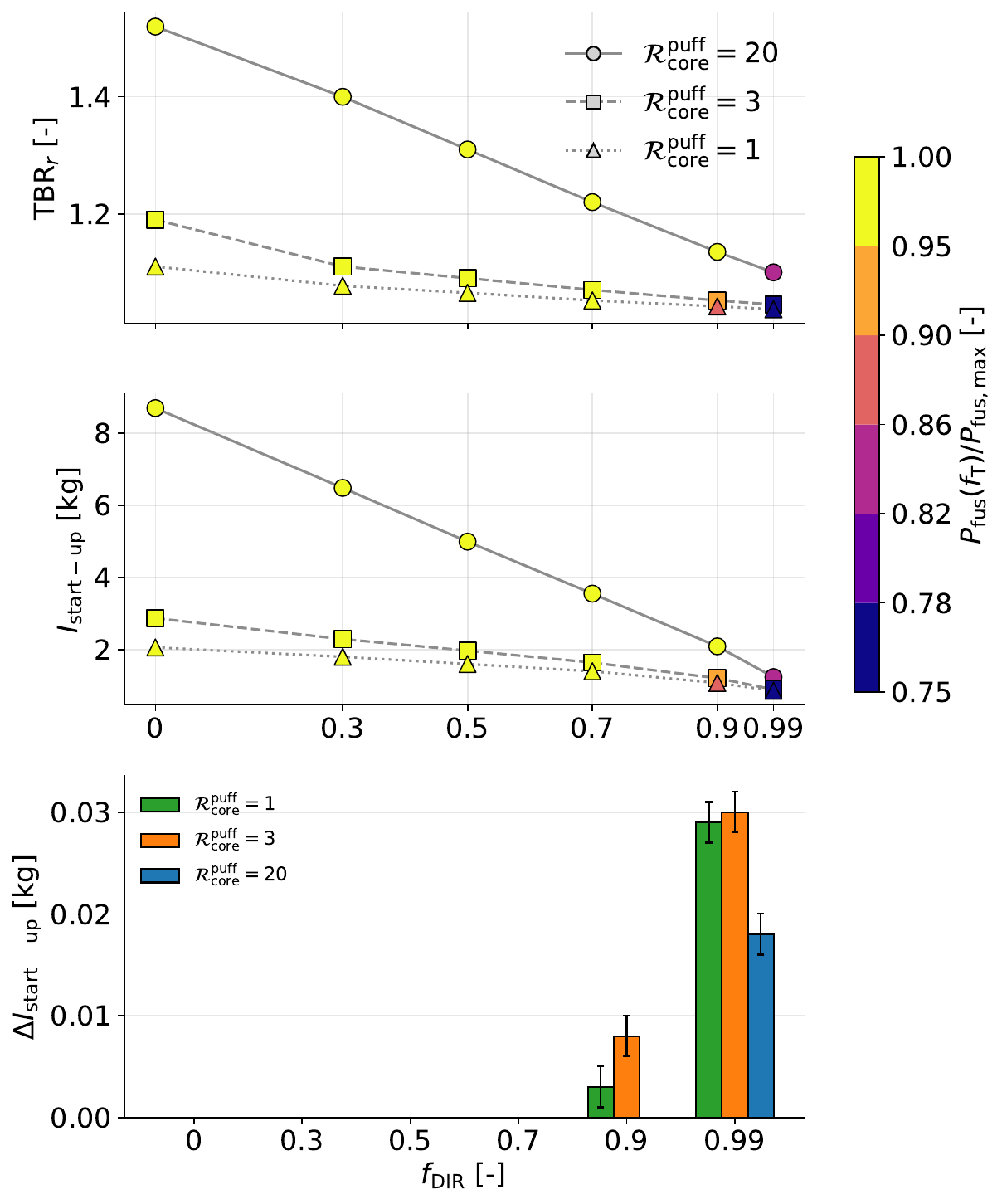}
                \caption{Top: $\tbrr$ and $\Ist$ for different $\ratioPuffToCore$ and $\fdir$. Both $\ftcore$ and the normalised fusion power depends on $\ratioPuffToCore$ per equations \eqref{eq:ftdir} and \eqref{eq:pfus_vs_ftcore}. Gas puffing is performed with 75:25 D:T. Bottom: increase in $\Ist$ to rebalance the core fuelling mixture.}
                \label{fig:f_T_rho_TBRr_Ist_75_25}
            \end{figure}

            Figure \ref{fig:f_T_rho_TBRr_Ist_90_10} highlights that an important relation arises between $\pfus$, $\tbrr$ and $\Ist$ at high gas puffing rates and with TFC implementing a DIR loop. In previous TFC analyses the optimal point for $\fdir$ was found to be the one minimising $\Ist$, thus $\tbrr$  \cite{abdou_physics_2020, meschini_modeling_2023}. For the same $\fdir$, $\Ist$ is minimised by the lowest $\ratioPuffToCore$. However, tritium production is proportional to neutron production (thus fusion power), and the non-linear dependency of the fusion power on $\ftcore$ makes more effective to operate at high gas puff rate since the fusion power reduces significantly less at the cost of tiny increases in $\tbrr$. In other words, the gain in neutron production partially balance the higher $\Ist$ (at fixed doubling time), keeping the required TBR within achievable values for breeding blankets ($\tbrr < 1.2$). Consequently, depending on constraints related to tritium availability for reactor start-up, two opposing gas puffing strategies may be justified: one favouring minimal start-up inventory, and another favouring increased fusion power and tritium production---still ensuring fulfilment of the pumping constraints described in Section \ref{subsec:implication_pumping_IFC}.
    
            This competing behaviour was already identified more than 30 years ago by Reiter \cite{reiter_burn_1990}, through the coupling of the Lawson criterion with the exhaust criterion. Despite its relevance, only a limited number of studies have explicitly addressed this trade-off, e.g. \cite{meschini_modeling_2023}, which compares the requirements for high TBE with those for maximising fusion power. This conflict is inherent and will systematically arise whenever particle management and power production are treated simultaneously, as both are intrinsically coupled through fuel dilution, plasma impurity concentration, and exhaust constraints. Notably, while the TBE was identified as one of the leading parameters to minimise tritium inventories, especially at low values (TBE = 0.5 - 1\%) where small improvements lead to strong gains in tritium self-sufficiency \cite{abdou_physics_2020, meschini_modeling_2023}, it has only a marginal impact in $\ratioPuffToCore \gg 1$ scenarios, where the dominant contribution comes from gas puffing rather than core fuelling, and improvements in TBE do not lower $\GammaQ$.
    
    \subsubsection{Increasing the start-up inventory to rebalance $\GammaTcore$}\label{sec:results_incresing_ist}
    Section \ref{subsec:optimal_T_fraction_gaspuffing} considered cases where no additional tritium is provided by the start-up inventory to rebalance the core fuelling mixture. This section quantifies the additional tritium requirements to maintain the core fuelling mixture at 50:50 D:T. The reactor is thus operating with $\ftcore = 0.5$ and the maximum achievable fusion power. The analysis does not account for the additional components that would be needed to rebalance the D:T stream from the DIR, and the resulting increase in the residence time through the DIR loop. Figures \ref{fig:f_T_rho_TBRr_Ist_90_10} and \ref{fig:f_T_rho_TBRr_Ist_75_25} (bottom panel) show the increase in start-up inventory ($\Delta \Ist$), which is higher for those cases where $\ftcore$ deviates more from $\ftcore = 0.5$, depending also upon the magnitude of the stream coming from the DIR. For a 90:10 D:T puffing mixture, $\Delta\Ist = 50$--$150 \,\rm g$, or 2.5 - 5\% the value of $\Ist$. For a 75:25 D:T puffing mixture, the increase is negligible compared to $\Ist$. Therefore, (a) adding tritium to $\Ist$ to ensure core fuelling at 50:50 D:T looks always justified in light of the magnitude of $\Ist$ when puffing is considered; and (b) although a $\Delta\Ist$ is needed at high $\fdir$, the increase inf $\fdir$ provides a larger reduction in $\Ist$ so that it is generally convenient to target the highest achievable DIR fraction.

    
    \subsection{How low is low impurity stream concentration: semi-empirical reconstruction}\label{sec:results__how_low_impurities}

        After reporting on the tritium related results, below we approximately quantify the order of magnitude behaviour of impurity concentration in the outgoing stream from the divertor.

        \subsubsection{Seeded impurity concentration}\label{sec:results__how_low_impurities__seeding}

            As anticipated in section \ref{sec:methods__database__puffing_and_concentration}, a direct power-law-type regression of seeded impurity concentration $\cZ = \cZ(\Vtot)$ is unsatisfactory---in fact featuring significant scatter and, crucially, varying over a restricted domain. Here this is also caused by the functional form of equation (\ref{eq:impurity_stream_concentrations})---a ratio of quantities in turn depending on $\Vtot$---which can be only asymptotically approximated by a power law. A direct empirical fit is therefore not sufficiently robust to be relied upon.
            
            A semi-empirical approach is instead adopted. Using the \textit{estimated} impurity seeding rate at detachment onset, $\GammaZstar$ (equation (\ref{eq:seeding_rate_GES_Sudo})), a well-defined dependence on total plasma volume is recovered ($R^2 = 0.94$):
            
            \begin{equation}\label{eq:scaling_GammaZpuff_Vtot}
                \GammaZstar = 2.7 \times 10^{19} \times \Vtot^{0.53} \,.
            \end{equation}
            
\begin{figure}
                \centering
                \subfloat{\includegraphics[width=0.45\textwidth]{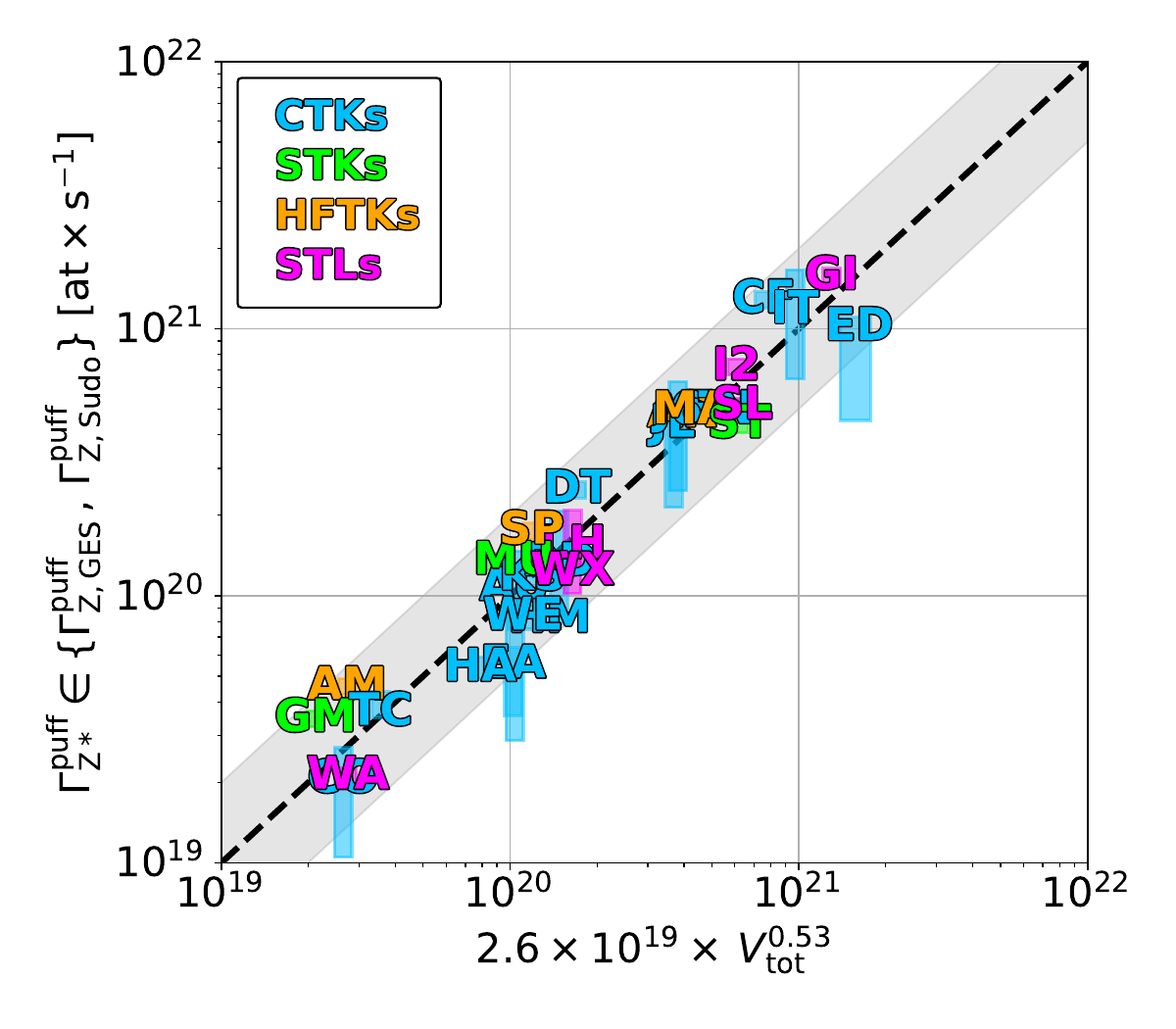}\label{fig:scaling_GammaZpuff_Vtot}}
                \\
                \subfloat{\includegraphics[width=0.45\textwidth]{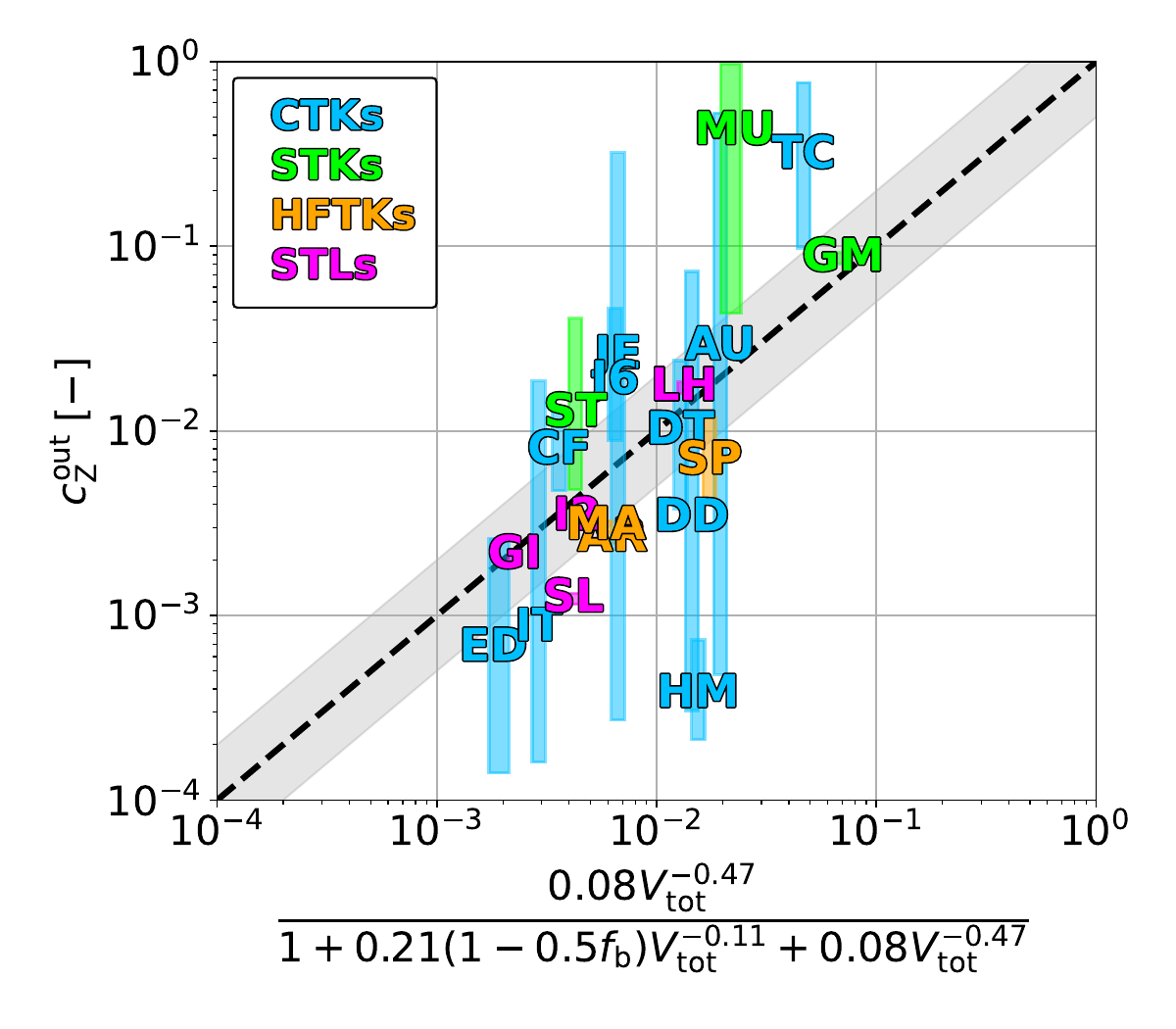}\label{fig:scaling_cZ_Vtot}}
                \caption{(a) Regression of impurity seeding rate $\GammaZstar = \{ \GammaZges , \, \GammaZsudo\}$ as a function of total volume $\Vtot$ ($R^2 = 0.94$). The rate $\GammaZstar$ is computed from equations (11) and (13) of \cite{Moscheni_2026}. (b) Comparison between the actual impurity concentration in the outstream $\cZ$ and its semi-empirical reconstruction as a function of $\Vtot$ and tritium burning efficiency $\TBE$.}
                \label{fig:scaling_GammaZpuff_and_cZ_Vtot}
            \end{figure}

            The corresponding fit is shown in figure \ref{fig:scaling_GammaZpuff_Vtot}.
            
            Substituting, into equation (\ref{eq:impurity_stream_concentrations}), the scaling above, and equations (\ref{eq:scaling_GammaQcore_Vtot}) and (\ref{eq:scaling_GammaQpuff_Vtot}) for $\GammaQcore$ and $\GammaQ$, respectively, yields the impurity concentration:
            
            \begin{equation}\label{eq:scaling_cZ_Vtot}
                \cZ \sim 
                \frac{0.08 \times \Vtot^{-0.47}}
                {1 + 0.21 (1 - 0.5\fBurn) \Vtot^{-0.11} + 0.08 \Vtot^{-0.47}} \,.
            \end{equation}
            
            The comparison of the \textit{actual} concentrations against $\cZ$ as semi-empirically reconstructed is shown in figure \ref{fig:scaling_cZ_Vtot}. The observed vertical scatter---primarily reflecting different degrees of detachment of the actual $\GammaZ$ \cite{Moscheni_2026}---confirms that a purely empirical regression would be unreliable. Nevertheless, the agreement in both slope and magnitude indicates that the derivation of equation (\ref{eq:scaling_cZ_Vtot}) is internally consistent, at least in the regime $\cZ < 0.1$---the region of relevance for reactor-scale devices.
            
            The denominator introduces a correction smaller than $35\%$, the largest deviation occurring for Globus-M2 ($\Vtot \sim 0.70 \, \mathrm{m^3}$), i.e. the smallest-volume device in table \ref{tab:fueling_stats}. Hence, to leading order:
            
            \begin{equation}\label{eq:scaling_cZ_Vtot_asymptotic}
                \cZ \sim 0.08 \times \Vtot^{-0.47},
            \end{equation}
            
            indicating a systematic decrease of impurity concentration with increasing device size. This was expected from the absence of a strong size dependence in the actual $\GammaZ$, on average across different degrees of detachment (figure 4a of \cite{Moscheni_2026}). The size dependence is instead recovered specifically at detachment onset in equation (\ref{eq:scaling_GammaZpuff_Vtot}).
            
        \subsubsection{Helium ash concentration}\label{sec:results__how_low_impurities__helium}

            In analogy with seeded impurities, a direct regression of the helium ash concentration $\cHe = \cHe(\Vtot)$ does not yield a robust scaling. This is primarily due to the limited dynamic range of the available data, with $\langle \cHe \rangle$ varying by less than one order of magnitude across the database.
            
            Therefore, leveraging the proportionality between helium production rate and core fuelling rate $\GammaQcore$ via $\fBurn$ (equation (\ref{eq:gammaQcore_from_helium_production})), the helium concentration can be semi-empirically estimated as:
            
            \begin{equation}\label{eq:scaling_cHe_Vtot}
                \cHe \sim
                \frac{0.11 \times \fBurn \, \Vtot^{-0.11}}
                {1 + 0.21 (1 - 0.5 \fBurn) \Vtot^{-0.11} + 0.08 \Vtot^{-0.47}} \,.
            \end{equation}
            
            To leading order, this reduces to:
            
            \begin{equation}\label{eq:scaling_cHe_Vtot_asymptotic}
                \cHe \sim 0.11 \times \fBurn \, \Vtot^{-0.11},
            \end{equation}
            
            indicating that the helium ash dilutes only mildly with increasing device volume. The dependence on $\Vtot$ is significantly weaker than that found for seeded impurities (equation (\ref{eq:scaling_cZ_Vtot_asymptotic})). This reflects the different scaling of helium production relative to impurity seeding, despite both being of similar magnitude on average (table \ref{tab:fueling_stats}).

\section{Discussion}\label{sec:discussion}

    \subsection{Intrinsic level of uncertainty and TFC design margins}
    \label{sec:discussion__operational_uncertainty}

        Both the puffing data collected in the database and the underlying plasma simulations bear their own intrinsic uncertainty, as discussed in section~2.3.3 of \cite{Moscheni_2026}. Since the present study shows that puffing dominates the global particle throughput across devices, the associated uncertainties also dominate the uncertainty of particle exhaust.
        
        Importantly, even assuming complete validation of present edge modelling tools against existing devices, extrapolation to a reactor-scale burning plasma remains subject to irreducible first-of-a-kind uncertainty. Since puffing constitutes one of the principal particle controls, sufficient operational flexibility in this channel is essential.
        
        In this respect, next-step burning plasma devices such as SPARC \cite{creely_overview_2020}---which aim to access regimes with significant self-heating---would offer a suitable testbed. Focused investigation of injection system, detachment robustness \cite{Cowley_2022}, and achievable puffing margins in such regimes would be beneficial to reduce extrapolation uncertainty for reactor-scale TFC design.
        In this regard, the results presented in sections \ref{subsec:tss_puffing_DT} and \ref{subsec:optimal_T_fraction_gaspuffing} extend classical TFC analyses by explicitly including gas puffing and its composition. Doing so introduces at least two additional degrees of freedom, the puffing rate and the tritium fraction in the puffed stream, into a design space that is already densely populated by fuel cycle parameters. Nevertheless, these two variables open operating regimes and trade-offs that are conceptually (and quantitatively) different from those explored in earlier fuel cycle studies. The reason is that \textit{gas puffing is not simply an additional throughput term to be carried in the balance, but a proxy for the full complexity of edge plasma and divertor physics and the associated operational constraints}. The leading design principle of minimising throughputs (particularly tritium throughputs) can become less robust when divertor constraints impose high puffing rates. Under such conditions, introducing a finite tritium fraction in the puff mixture can even become beneficial in order to maintain a favourable exhaust isotopic composition and sustain efficient DIR operation. Similarly, maximising $\fdir$ is not always optimal, since the core tritium fraction $\ftcore$ may be driven to unacceptably low values, directly penalizing fusion power through fuel dilution. Pumping constraints provide another example of this shift in perspective. When only core fuelling is considered, pumping requirements may appear non-limiting. Once gas puffing is included, the increased throughput can make the installed pumping capacity and associated footprint a primary design constraint.
        
        In conclusion, the present results indicate a particle throughput on the TFC largen than anticipated under current modelling capabilities, and also highlight the necessity of incorporating adequate design margins and control flexibility to accommodate both intrinsic plasma physics and modelling uncertainties. A natural way to address this issue is to integrate simplified, yet self-consistent, representations of core and edge plasma behaviour into existing fuel cycle models. This can be achieved either by detailing recent models that capture the dominant physics \cite{hattab_analysis_2025, Morandi_2025_SOFE}, or by leveraging established tritium transport tools already embedded within multiphysics frameworks \cite{simon2025moose}.

    \subsection{Core fuelling behaviour}\label{sec:discussion__core_fuelling}

        \subsubsection{Emergent size ordering of fuelling requirements}\label{sec:discussion__core_fuelling__size}

            Although TFC constraints are relevant to next-generation reactors, meaningful scaling trends could not be extracted by considering future devices alone. The presently proposed reactor concepts span only slightly more than an order of magnitude in $\Vtot$, insufficient to establish robust ordering. Inclusion of current-day devices therefore constitutes an enabling asset, providing the dynamic range necessary to identify systematic trends.
            
            Notably, the lowest $\ratioPuffToCore$ among the future devices---while still significantly exceeding unity---are obtained when $\GammaQ$ is predicted from equations (10) and (12) of \cite{Moscheni_2026} (grey entries in table \ref{tab:fueling_stats}). This corresponds to the maximum puffing level required to, however, only access detachment. By contrast, \textit{the remaining reactors are modelled via SOLPS-ITER \cite{XavierBONNIN2016}, achieve deeper detachment and tend to exhibit higher puff-to-core fuelling ratios} [$(1.6 \times 10^1) \tpm 3.9$ vs.\footnote{Notice that, even without the exceptional EU-DEMO value $\ratioPuffToCore = (7.1 \times 10^1) \tpm 2.4$, the average of the SOLPS-ITER-modelled reactors would read $(9.8 \times 10^0) \tpm 2.6$.} $(7.4 \times 10^0) \tpm 1.9$]. This is consistent with the high-fuel-puffing operational benefits discussed in sections \ref{sec:intro_matter_injection_plasma__puffing} and \ref{sec:intro_matter_injection_plasma__synthesis}.
            
            From the perspective of core fuelling, the existence of a well-defined ordering of $\GammaQcore$ values across future FPP is not surprising. Reactor concepts must satisfy broadly similar constraints, including (i) sufficient power density to ensure economic viability, and (ii) sustainable helium ash removal without excessive fuel dilution in the core plasma \cite{reiter_burn_1990, Putterich_2019, Angioni_2026}. These constraints apply irrespective of device type. A comparable trend is also observed in present-day machines. After all, future reactors are designed around plasma scenarios that are presently regarded as successful, and experimental campaigns tend to converge toward such regimes (section 4.1 of \cite{Moscheni_2026}).
            
            More noteworthy, instead, is that geometry alone appears capable of strongly ordering the $\GammaQcore$ datapoints across devices. This is reminiscent of empirical threshold scalings such as the Martin L--H power threshold \cite{Martin_2008}, which behaves like a global throughput condition:
            \begin{equation}\label{eq:scaling_Martin_Plh}
                P_{\mathrm{LH}} \propto \neavg^{0.72} \, B_T^{0.80} \, \Splasma^{0.94} \,.
            \end{equation}
            In that case geometry enters nearly linearly through the plasma surface $\Splasma$, not dissimilarly to the surface representation of $\GammaQcore$ in equation (\ref{eq:scaling_GammaQcore_Stot}).

        \subsubsection{Physical core fuelling rate}\label{sec:discussion__core_fuelling__physics}

            According to its definition in section \ref{sec:methods__database__core__fuelling}, $\GammaQcore$ represents the \textit{engineering} core fuelling rate. For the main purposes of the present study this definition is sufficient. Nevertheless, in the physics of particle balance, the \textit{physical} core fuelling rate primarily matters, and is therefore analysed next.

            \paragraph{0D particle inventory description.}\label{sec:discussion__core_fuelling__physics__confinement_time}

                A natural starting point is the commonly adopted 0D particle-inventory description
                \begin{equation}\label{eq:definition__GammaQcorePhys_nV_tau}
                    \GammaQcorePhys \sim \frac{nV}{\tauP}
                    = \GammaQcore + \etaPuffToCore \,\GammaQ \,,
                \end{equation}
                where $nV$ denotes the total plasma particle inventory, $\tauP$ the particle confinement time for Q, and $\etaPuffToCore$ the effective efficiency with which puffed neutrals contribute to core fuelling. The second equality is the decomposition proposed by \cite{leoni2024scrape}. This definition isolates the contribution of puffed fuel and neglects any direct contribution from recycled Q at the divertor (figure~11a of \cite{Zito_2025}), an approximation expected to improve with machine size.
                
\begin{figure}
                    \centering
                    \includegraphics[width=0.45\textwidth]{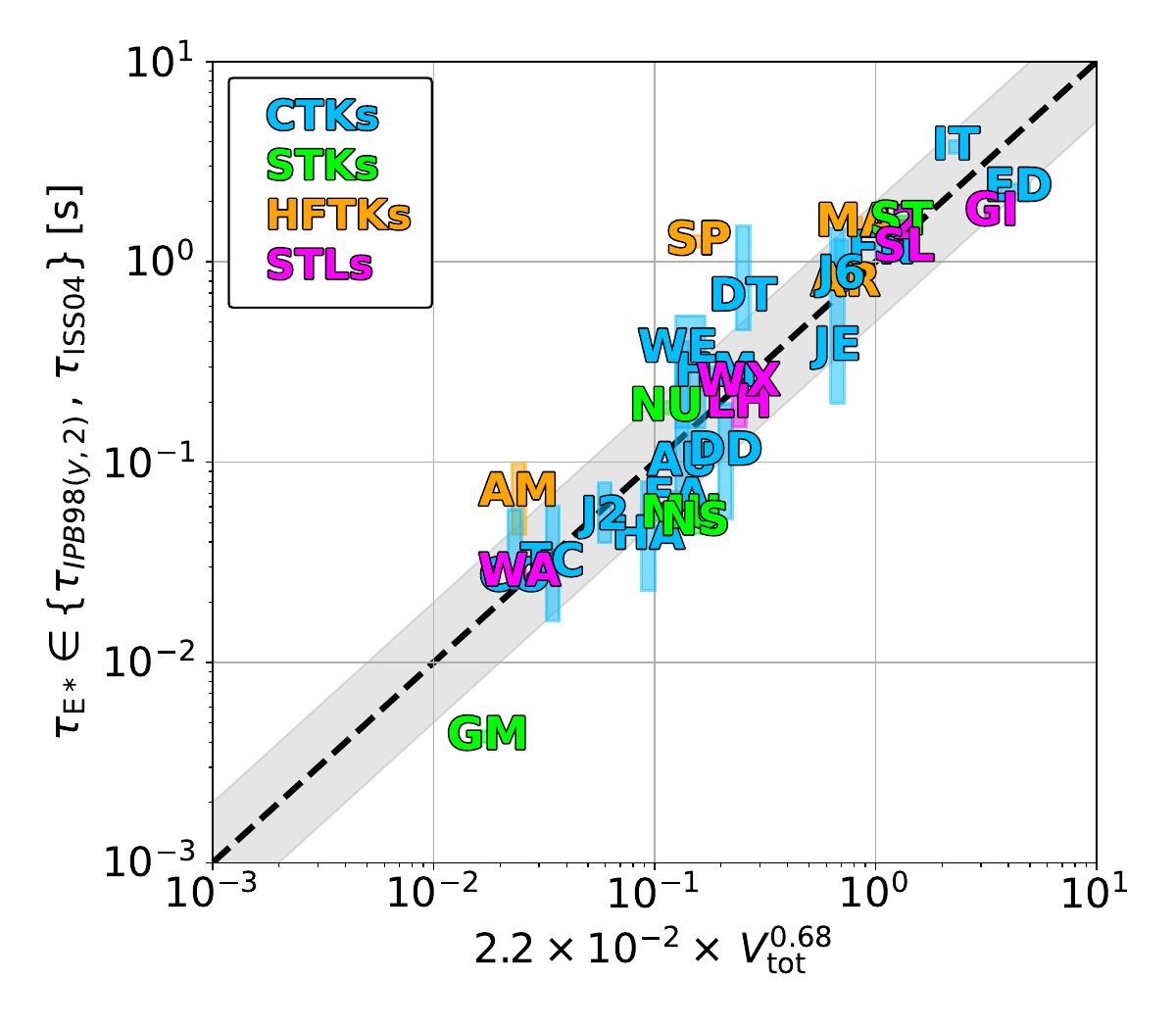}
                    \caption{Regression of effective energy confinement time $\tauStar$ estimated from $\tauIPB$ in tokamaks \cite{ITER_confinement_1999} and from $\tauISS$ for stellarators \cite{Yamada_2005}, as a function of $\Vtot$.}
                    \label{fig:scaling_tau_effective_Vtot}
                \end{figure}

                Because empirical expectations for $\tauP$ are considerably less established than for the energy confinement time, we estimate it from an effective energy-confinement proxy $\tauStar$, assuming $\tauP \propto \tauStar$. Here $\tauStar$ denotes $\tauIPB$ for tokamaks \cite{ITER_confinement_1999} and $\tauISS$ for stellarators \cite{Yamada_2005}, with details of the corresponding confinement scalings, assumptions, and database treatment reported in appendix \ref{apx:tau_effective}. A direct fit gives ($R^2=0.82$):
                \begin{equation}\label{eq:scaling_tau_effective_Vtot}
                    \tauStar = 0.022 \times \Vtot^{0.68} \,.
                \end{equation}
                The graphical representation is shown in figure \ref{fig:scaling_tau_effective_Vtot}, which confirms the robustness of the fit\footnote{The high-field tokamak outliers are expected, as explained in section 5 of \cite{Angioni_2026} for Alcator C-Mod. Similarly, figure 11 therein \cite{Angioni_2026} shows how a small-sized spherical tokamak like START may also violate general trends, as found in figure \ref{fig:scaling_tau_effective_Vtot} for Globus-M2.}. Once more, no significant difference is observed between tokamaks and stellarators within the approximations of the present study---most of the variation in $\tauStar$ is already captured by plasma volume alone. Additionally, the exponent remains consistent with figures 11a--11c of \cite{Angioni_2026}. 

                Using the fact that the effective density limit $\neStar$ does not increase systematically with machine size across the database\footnote{Similarly, figure 12 of \cite{Angioni_2026} shows lack of a clear size dependence in line-average densities across the entire ITPA global confinement database \cite{ITER_confinement_1999, Verdoolaege_2021}.} ($\sim 1.3 \times 10^{20}\,\mathrm{m^{-3}}$ on average), substitution into equation (\ref{eq:definition__GammaQcorePhys_nV_tau}) gives
                \begin{equation}\label{eq:scaling_GammaQcorePhys_nV_tau_Vtot}
                    \GammaQcorePhys \sim \frac{\neStar \Vtot}{\tauStar} \propto \Vtot^{0.32} \,.
                \end{equation}
                
                This trend is substantially weaker than the observed engineering scaling $\GammaQcore \propto \Vtot^{0.89}$ in equation (\ref{eq:scaling_GammaQcore_Vtot}), and would be even weaker if alternative confinement scalings were used. The discrepancy is therefore much more naturally attributed to the puff-penetration term $\etaPuffToCore\,\GammaQ$ than to a radically different and currently unsupported size scaling of $\tauP$.
                
            \paragraph{Effective puff penetration contribution.}\label{sec:discussion__core_fuelling__physics__penetration}
                
                Equations (\ref{eq:definition__GammaQcorePhys_nV_tau}) and (\ref{eq:scaling_GammaQcorePhys_nV_tau_Vtot}) imply that, in the regime where puff penetration contributes materially to fuelling the confined region, the effective penetrative source term must scale as
                \begin{equation}\label{eq:scaling_etaPuffToCore_GammaQpuff_Vtot}
                    \etaPuffToCore \,\GammaQ \sim \GammaQcorePhys - \GammaQcore \,,
                \end{equation}
                so that, to leading order in the small-device regime where the correction is most relevant,
                \begin{equation}
                    \etaPuffToCore \,\GammaQ \propto \Vtot^{0.32} \sim \Stot^{-1}\,.
                \end{equation}
                Using equation (\ref{eq:scaling_GammaQpuff_Vtot}) also implies an effective penetration efficiency decreasing with size, but the important quantity for the present discussion is the penetrative puff source term itself.
                
                This inferred trend is not inconsistent with dedicated multi-machine edge-transport modelling by Pacher \textit{et al.} \cite{PACHER2003657}. In B2-EIRENE (SOLPS4.2) \cite{SCHNEIDER1992810, Reiter_1991} simulations of AUG, JET, ITER and DEMO, the inward D:T neutral flux across the separatrix at incipient detachment was found to scale approximately linearly with machine size ($\sim R_0^{1.03}$ in table I of \cite{PACHER2003657}). Under approximate geometric similarity, $a \sim R_0$ and $\Vtot \propto a^2 R_0 \sim R_0^3$, so that
                \begin{equation}
                    \etaPuffToCore \,\GammaQ \propto \Vtot^{0.32} \sim R_0^{0.96} \,,
                \end{equation}
                i.e.\ a leading-order dependence very similar to that reported by \cite{PACHER2003657}. This comparison should however be interpreted cautiously, since: (i) the separatrix-crossing neutral influx does not necessarily result in core fuelling; (ii) the present scaling mainly applies to the regime where puff penetration remains appreciable, i.e.\ primarily smaller machines.
                
                Notably, the dependence of $\etaPuffToCore \,\GammaQ$ on size does not contradict the fact that puff penetration becomes less important in larger devices. Indeed, even if the \emph{absolute} penetrative puff source increases with machine size, its relative importance compared with $\GammaQcore$ decreases in equation (\ref{eq:definition__GammaQcorePhys_nV_tau}) because $\GammaQcore$ itself rises more steeply. As a result,
                \begin{equation}
                    \GammaQcorePhys \to \GammaQcore
                \end{equation}
                as machine size increases.
                
                More detailed quantification of neutral penetration into the confined plasma region would however be required, and this remains an active area of research \cite{Wilkie_2026} (section \ref{sec:intro_matter_injection_plasma__puffing__fuel__plasma}). The present section merely stresses the distinction between the engineering and physical values of the core fuelling rate.

        \subsection{High-density operating scenarios: feasible, but with reservations}\label{sec:discussion__high_density_operation}
    
            Exceptions to section \ref{sec:results__how_high_fuel_puffing}---and to the general finding that most operating points are localised in the fuel-puffing-dominated region of parameter space---do exist.
            
            The high-density pellet-fuelled scenarios reported by Lang \textit{et al.} \cite{Lang_2018} for ASDEX Upgrade (AUG) feature a core pellet fuelling rate $\GammaQcore \sim 2.5\times10^{22}\,\mathrm{at \, s^{-1}}$, comparable to the concomitant gas puffing rate $\GammaQ \sim 2\times10^{22}\,\mathrm{at \, s^{-1}}$ ($\ratioPuffToCore \sim 1$). In \cite{Lang_2020}, similar core fuelling levels are achieved at lower gas puffing, such that $\GammaQcore$ exceeds $\GammaQ$ by up to a factor of 3 (figures 2e and 4c therein). These cases contribute to the significant vertical extent of the AUG bar in figure \ref{fig:scaling_GammaQcore_Vtot} and demonstrate that high-average-density operation can indeed move the system away from the usual puff-dominated ordering.
            
            This possibility is appealing also from the reactor viewpoint. Angioni \textit{et al.} \cite{Angioni_2026} discuss compact, high-average-density operation as a promising route for future power plants, although explicit fuelling-actuator levels are not reported therein. Complementary data from Tholerus \textit{et al.} \cite{Tholerus_2024} (especially tables 6 and 7) suggest that, in STEP, moving from a low-density (EC-LD) to a high-density (EC-HD) scenario with only $+40\%$ in line-average density may require an increase of order $\sim 200$ in $\GammaQcore$. The reduced size envisaged in \cite{Angioni_2026} partly compensates through the volume dependences in equations (\ref{eq:scaling_GammaQcore_Vtot}) and (\ref{eq:scaling_GammaQpuff_Vtot}), but whether high core density can be accompanied by sufficiently high separatrix density for detached operation at modest fuel puffing remains to be assessed.
            
            The broader conclusion of the present work nevertheless still holds: \textit{high-density regimes require significant overall particle throughput}. In the AUG cases of \cite{Lang_2018,Lang_2020}, the gas puffing levels remain well within the detachment-level range of table \ref{tab:fueling_stats} ($\sim 1.3\times10^{22}\,\mathrm{at \, s^{-1}}$, with spread of order 2.5), whereas the corresponding $\GammaQcore$ lies roughly a factor $\sim 30$ above AUG's average core fuelling rate via NBI (appendix \ref{apx:estimation_GammaQcore__procedure__NBI}). Lang’s scenarios can therefore be categorised as ``average $\GammaQ$ and high $\GammaQcore$''. In other words, $\GammaQcore$ is not higher than $\GammaQ$ because fuel puffing is unusually low, but because the core fuelling level itself becomes comparable to that of ITER-/STEP-class devices.
            
            Additionally, in the same machine---AUG---stably detached scenarios at $\GammaQ \sim 5\times10^{22}\,\mathrm{at \, s^{-1}}$ have been obtained \cite{Kallenbach_2015}, i.e.\ at puffing levels exceeding even the high core fuelling rates of \cite{Lang_2018,Lang_2020}. This highlights that, from a throughput standpoint, \textit{gas puffing alone can---seemingly universally---reach or surpass the core fuelling rate}. Therefore, gas puffing must always be included in any particle balance. In this respect, \cite{Moscheni_2026} further notes that fuel puffing values in MARFE-limited cases may be used to place absolute upper bounds on reactor-relevant puffing requirements. The MARFE-onset database assembled by Giacomin \textit{et al.} \cite{Giacomin_2022} would be uniquely suited to constrain such bounds across machines.
            
            Overall, this section should not discourage high-density operating points, but rather emphasise the engineering requirements they may impose.

            A similar caution applies to lithium-based plasma-facing components and their tritium retention properties \cite{deCastro_2021}. Large fuel puffing rates are reported in \cite{Emdee_2023}, whereas Emdee \textit{et al.} \cite{Emdee_2026} show that these values depend strongly on the degree of deuterium absorption---and that for saturated-lithium-like values the corresponding throughputs may become manageable relative to machine size. The implications for tritium retention and TFC operation would therefore require dedicated assessment.

        \subsection{Revisiting edge plasma scenarios: a staged rather than exclusive approach}\label{sec:discussion__different_plasma_scenarios}

            Lore \textit{et al.} \cite{Lore_2022} in ITER, and Henderson \textit{et al.} in STEP \cite{Henderson_2025}, already recognised that different branches of edge-plasma operating space should be  explored beyond a single preferred trajectory. In particular, the advantages and limitations of Lore's triangle series---where $\GammaZ$ is increased at constant $\GammaQ$---are discussed therein and summarised in section \ref{sec:intro_matter_injection_plasma__puffing__impurities}. Although such impurity-dominated scenarios are not ideal from the viewpoint of fusion-power production, they remain burning plasma operating points which satisfy divertor requirements. As such, they suggest a broader system-level interpretation once TFC constraints are included.
            
            The central point is that the plasma scenarios discussed in the literature need not be interpreted as mutually exclusive stationary alternatives. Rather, they may be viewed as different operating points of the same broader fuelling space, to be used at different stages of reactor operation. In particular, the puffing composition itself---i.e.\ the relative balance of deuterium, tritium, and impurities in the injected mixture---may evolve in time. This could arise from control strategies, or naturally if the puff stream is supplied by the DIR stream, as discussed in section \ref{sec:methods__TFC_model}. A fuel-dominated puffing regime may still remain the ultimate target, owing to its more attractive power production performance, provided that the tritium inventory increase and the trade-offs with the achievable DIR fraction remain appealing (section \ref{fig:tss_puffing_DT} and \ref{subsec:optimal_T_fraction_gaspuffing}).However, this does not imply that reactor operation must begin directly in that state.
            
            More generally, at least two distinct start-up directions can be envisaged.
            
            One is leveraging a deuterium-dominated initial phase, in which a fuel-dominated puffing is maintained but with a deliberately low tritium fraction. Since the start-up inventory depends on both $\GammaTcore$ and $\GammaTpuff$, a substantial reduction of $\Ist$ may be achieved\footnote{Although this strategy is beneficial for decreasing $\Ist$, the steady-state tritium inventory depends on the long term fuelling strategy. If $\fTpuff$ is later increased, additional tritium must be supplied, likely drawing from the bred tritium.}.
            
\begin{figure}
                \centering
                \includegraphics[width=0.98\linewidth]{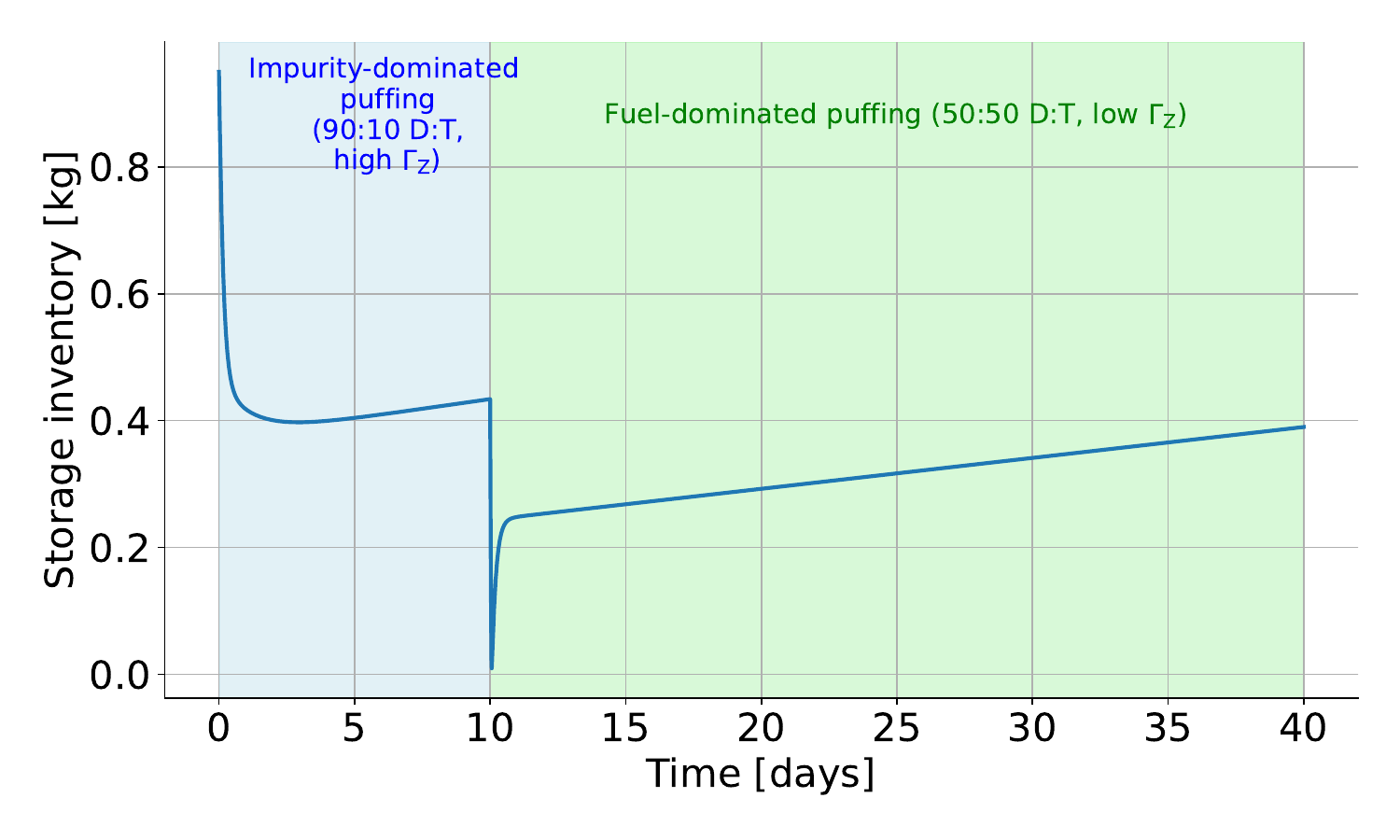}
                \caption{Illustrative staged-operation strategy. An initial impurity-dominated detached phase with reduced tritium puffing demand allows partial saturation of fuel cycle inventories. After this transient, the reactor can switch to a more fuel-dominated D:T puffing with a substantially milder tritium inventory drop than would occur under direct start-up in that regime. The radiation fraction shown in the plot is computed as $f_\mathrm{rad}=P_\mathrm{rad}/P_\alpha$, and stems from imperfect neon separation in the DIR loop. 
                }
                \label{fig:storage_inventory_puff_switch}
            \end{figure}

            Another option is an impurity-dominated start-up, illustratively shown in figure \ref{fig:storage_inventory_puff_switch}. Here we drop the assumption of a single operating point as described in section \ref{sec:methods__TFC_model}, and we consider a transition from gas puffing (at $\ratioPuffToCore=20$) with high $\GammaZ$ (impurity-dominated) and 90:10 D:T, to gas puffing with 50:50 D:T (fuel-dominated) and lower impurity fraction. Hence, at start-up $\GammaZ$ is comparatively high while deuterium puffing and the DIR fraction ($\fdir = 0.3$) are kept low, according to the discussion in section \ref{subsec:optimal_T_fraction_gaspuffing}, that is, to avoid excessive deviation from $\ftcore = 0.5$ due to the low fraction of tritium in the puffed stream. The resulting start-up inventory is relatively low as there is little tritium in $\GammaQ$, with the only contribution coming from $\GammaTcore$. During the initial transient, the fuel cycle components progressively build up a small tritium inventory \cite{abdou_deuterium-tritium_1986}. As the operations switch to a fuel-dominated puffing, the strong tritium demand from the gas puffing system leads to a drop in the storage inventory, although much lower than what would be experienced if the operations were started with fuel-dominated puffing. The DIR fraction is increased from 0.3 to 0.9, drastically reducing the tritium flow through the tritium processing plant. The rebound of the storage inventory following the drop associated with the switch in puffing strategy is indeed related to the recovery of some of the tritium in the exhaust processing and isotope separation system once $\fdir$ is increased to 0.9. For comparison, continued operations at $\fdir=0.3$ with impurity-dominated puffing would require $\Ist = 450 \,\rm g$, at the cost of higher radiative losses; with the switch from impurity-dominated to fuel-dominated $\Ist = 950 \,\rm g$ (in which $\fdir$ is increased from $\fdir = 0.3$ to $\fdir = 0.9$ as the puffing transition to 50:50 D:T) - although $\Ist$ is higher than the impurity-dominated puffing case, most of the tritium in the IFC is recovered once $\fdir$ increases up to 0.9, resulting in a lower steady-state tritium inventory; and by starting up with fuel-dominated, $\Ist = 1.6 \,\rm kg$. This approach is conceptually similar to a power ramp-up strategy \cite{konishi_myth_2017} and a tritium-lean start-up \cite{morandi2026multispecies}, in which the start-up transient evolves slower than a start-up at full power, thus decreasing the start-up inventory\footnote{The steady-state tritium inventory in fuel cycle components depends on the tritium throughput through them, which, with all other parameters held constant, scales with the fusion power. In a full power start-up, the transient is too short for bred tritium to give a significant contribution in satisfying the fuelling demand. If the transient is slower, however, the contribution from bred tritium increases progressively and begins to support the fuelling demand, thus reducing the required start-up inventory}.

            Because the start-up transient lasts only a few days in the absence of strong tritium trapping \cite{meschini_impact_2024}, any impurity-dominated strategy need not compromise tritium self-sufficiency, although some tritium retention in the gas puffing system is unavoidable. Such phases would also likely remain less favourable in terms of instantaneous fusion power, and may require additional auxiliary heating during the start-up transient.
            
            The key advantage, however is that it avoids the otherwise large tritium start-up in a 50:50 D:T puffing regime whenever $\fdir$ does not approach unity. 

            This staged interpretation can be further supplemented with the high-density operation discussed in section \ref{sec:discussion__high_density_operation} and/or the low pumping avenues described in section \ref{sec:discussion__pressure_with_pumping_speed}.
                        
            Such trade-offs involving impurities are not methodologically unprecedented. For instance, EU-DEMO studies envisage the use of krypton and xenon as core radiators \cite{SICCINIO2020111603}, primarily to alleviate the divertor power handling problem \cite{Kallenbach_2022}. Although this is usually motivated from the viewpoint of power exhaust, in conjunction with the above it also illustrates a broader principle: how the relevant design problem is no longer the identification of a single best stationary puffing composition---which may coincide with the maximally pure or most intuitively desirable plasma composition. Rather, it is \textit{the identification of a time-dependent trajectory in D:T:Z space leading to the preferred long-term operating point through a start-up that minimizes $\Ist$.}
            
            This also reinforces the need for fuel cycle models capable of accounting for multiple species other than tritium, both within the fuel cycle itself and in its coupling to the plasma, as recently addressed by \cite{hattab_analysis_2025, morandi2026multispecies}.
            
            Quantifying the corresponding trade-offs in start-up inventory, transient duration, achievable fusion power, and fuel cycle dynamics would be a natural next step, but lies beyond the scope of the present work.

            \subsubsection{Impurity dynamics in stellarators.}

                A related consideration concerns whether stellarator edge impurity dynamics could modify the balance between fuel puffing and impurity seeding. Feng \textit{et al.}~\cite{Feng_2006} report that, in high-recycling conditions, the relation between thermal and friction forces acting on impurities can differ from the tokamak case (section \ref{sec:intro_matter_injection_plasma__puffing}). In particular, impurity transport in the stellarator edge may become friction-dominated, impeding upstream propagation and thereby reducing core contamination.
                
                Within the present system-level picture, such behaviour would be favourable: for a given acceptable core impurity concentration, a more effective divertor impurity retention could allow a lower fuel throughput. This would directly alleviate the TFC burden identified in the present work.
                
                However, the limited stellarator entries available in the database~\cite{Moscheni_2026, zenodo_repo_v2}, and \cite{MASUZAKI2013S133} in particular, do not show a systematic departure from tokamak trends when fuel and impurity puffing are simultaneously present. This may reflect the scarcity and heterogeneity of available stellarator data rather than the absence of a physical difference.
                
                Dedicated stellarator-specific studies including puffing and its location are therefore required before this possible advantage can be included in reactor-level TFC assessments.
        
    \subsection{Implications on the the inner fuel cycle and the primary pumping system}\label{subsec:implication_pumping_IFC}
    
        The results from the TFC model underscore the need for a review of the DIR loop approach envisioned in \cite{day_smart_2019} and related works. The leading idea of a DIR is to minimize tritium throughput and inventories. However, the results in sections \ref{subsec:tss_puffing_DT} and \ref{subsec:optimal_T_fraction_gaspuffing} showed the following. The DIR introduces a strong coupling between puffed gas and core fuelling via exhaust recycling. Ideally, the gas puffing should not include tritium, so that the tritium throughput is minimized (all the other parameters held constant). However, a 100\%D puffing mixture would quickly imbalance the D:T core fuelling mixture. As $\ftpuff$ increases, this effect vanishes, at the cost of a higher $\Ist$ and larger tritium throughput across the plant. This is counterintuitive, as $\ftpuff$ is being increased to make the DIR compatible with high gas puffing rates, while the DIR goal is to minimize tritium usage. Limiting $\ftpuff$ either lowers the fusion power due to $\ftcore < 0.5$, or requires the introduction of a core mixture rebalancing unit. The first option is unattractive from an economic perspective, the second is sub-optimal because it partially erodes the advantages of the DIR loop. Alternatively, the DIR could be employed to sustain the puffing stream only, decoupling it from the core fuelling. This would result in tritium self-sufficiency scenarios that corresponds to the $\fdir = 0$ case in most of the public available analyses \cite{abdou_physics_2020, coleman_demo_2019, meschini_modeling_2023, jin2026cfedr, chen_tritium_2016}, which showed that the start-up inventory grows to extremely high values without a DIR providing recycled tritium for core fuelling. More nuanced architectures could be envisioned, such as the one coupling a direct by-pass with a downstream DIR loop presented in \cite{hattab_analysis_2025} for GIGA and DEMO. Overall, a revision of the DIR loop concept appears needed in light of the high gas puff rates of future fusion power plants.

        Regardless of the strategy used to handle the exhaust downstream of the torus pumps, the high exhaust throughput resulting from high gas puff rates challenges the primary vacuum pump system, being the technology options for high throughput, tritium compatible pumps limited to cryopumps, with diffusion pumps envisioned as possible alternative \cite{teichmann_particle_2021, teichmann_simulation_2022}. This implication is addressed only qualitatively in this work. Taking as example ITER primary vacuum pump system, it features 8 large cryopumps ($15 \,\rm m^3$ each) in staggered mode to provide a continuous pumping speed of about $100 \, \rm m^3 \, s^{-1}$ during plasma operations \cite{pearce_design_2025}. This translates to a throughput $Q = 100$--$1000 \, \rm Pa \, m^3 \,s^{-1}$ for divertor pressures in the range $\pdiv = 1$--$10 \,\rm Pa$. The required throughput (from table \ref{tab:fueling_stats}) is $Q_\mathrm{req} \simeq 264$--$1062 \, \rm Pa \, m^3 \,s^{-1}$. Therefore, particle exhaust can be satisfied by this design operating at $\pdiv \geq 2.7 \,\rm Pa$, increasing the divertor pressure as the gas puffing rate increases.
        
        We highlight here an important concept. Each point-design scenario is associated with its own gas puffing rate. Therefore, when we refer to ``increasing the gas puff rate'', we do not necessarily mean moving from a scenario A to an otherwise identical scenario B in which only the gas puff rate (and thus the divertor pressure) is changed. In practice, increasing the puffing rate corresponds
        to a different operating scenario (e.g., with a different separatrix density and impurity concentration), for which the divertor pressure is not necessarily higher. For example, in \cite{Lore_2022} (figure 3 therein), $\pdiv$ increases sub-linearly with the puffing rate.

    This directly affects the required number of pumps, which must increase to accommodate higher gas puff rates while remaining compatible with integration constraints. In principle, the pumps themselves could be located farther from the machine and made arbitrarily large. In practice, however, this is not a straightforward solution, because increasing the distance from the vacuum vessel requires longer ducts, which lowers the conductance and therefore reduces the effective pumping speed seen by the divertor. Additional constraints also arise from the maximum tritium inventory that can be allowed in a single component, which may limit the admissible size of each pump unit and favour the installation of multiple units in parallel. Lastly, each pump duct introduces streaming pathways for neutrons and complicates radiation shielding \cite{Pitcher_1997}.
    The same considerations apply to DIR technologies. Whether MSCs or MFPs are used, additional installed capacity must be provided to accommodate high gas puff rates.
    These geometrical constraints are design dependent: some stellarator layouts may offer greater flexibility in pump placement, whereas tokamaks, especially compact tokamaks, are generally more constrained.
        
        For instance, some smaller machines (e.g. ARC \cite{sorbom_arc_2015, Hillesheim_2026}, STEP \cite{Chapman_2024}, MANTA \cite{Rutherford_2024}) require comparable puffing rates to ITER but have lower room for pumps and ducting. Others are comparable in size with ITER, but require even higher gas puff rates (e.g. EU-DEMO, GIGA \cite{gaussfusion_cdr_exec_2026}). The increase of gas throughput due to puffing impacts all the downstream subsystems with a magnitude proportional to $\ratioPuffToCore$. In future machines, $\ratioPuffToCore \simeq 10$ for ARC, MANTA, STEP, CFETR, and EU-DEMO, $\ratioPuffToCore \simeq 5$ for Infinity Two and GIGA, and $\ratioPuffToCore \simeq 3$ for Stellaris (table \ref{tab:fueling_stats} and section \ref{sec:discussion__core_fuelling__size}).
        
        Publicly available fuel cycle analysis of some of these machines often considered the core fuelling term only, or they strongly underestimated gas puff rates \cite{day_smart_2019, schwenzer_operational_2022, coleman_demo_2019, abdou_physics_2020, chen_tritium_2016}, sometimes with direct implications to component sizing. Examples include early sizing studies on cryogenic distillation columns by \cite{kinoshita_computer_1981}, whose results are later used in \cite{teichmann_fusion_2025}; the cryopump designs in \cite{pearce_gas_2012,PEARCE2013809}, which assume a $200 \, \rm Pa \, m^3 \,s^{-1}$ 50:50 D:T exhaust stream and are subsequently adopted in \cite{pearce_design_2025}; the assumption in \cite{ren_overview_2025} that the ITER TEP subsystem processes an off-gas stream of roughly 50:50 D:T; the target of at least 80\% DIR fraction in \cite{clark_breeder_2025}, which may not be achievable at the corresponding puffing rates; and analyses such as \cite{Bader} that do not include gas puffing explicitly in the primary pumping requirements or only include it for ramp-up and -down \cite{Ploeckl19052021}, although for with a magnitude that is much lower than the one estimated in table \ref{tab:fueling_stats} ($2.2 \times 10^{22} \, \rm at \,s^{-1}$ in \cite{PLOECKL2017186} vs. $10^{23}$--$10^{24} \, \rm at \, s^{-1}$ from the raw data populating the database \cite{zenodo_repo_v2}). 

        The first point to address, in light of the particle flows reported in table \ref{tab:fueling_stats}, is whether those high puff rates are really needed (section \ref{sec:discussion__pressure_with_pumping_speed}). If so, what strategies could be implemented to reduce the burden on the torus pump (sections \ref{sec:discussion__pressure_with_pumping_speed__lover} and \ref{sec:discussion__pressure_with_pumping_speed__bypass}) while satisfying detached divertor requirements.

    \subsection{Lowering effective pumping speed: is high fuel puffing intrinsically required?}\label{sec:discussion__pressure_with_pumping_speed}

        This study is motivated by the \textit{empirical} observation that---compared to core fuelling---high fuel puffing is both traditionally employed and projected for future devices. In other words, this is the direction presently taken by operational and modelling practice in the community, rather than an assumption introduced by the present regressions or extrapolations (section \ref{sec:results__how_high_fuel_puffing}).
        
        In the present section, however, the appropriateness of this projection is deliberately questioned. This is done setting aside the possible adverse impacts on detachment and density control, which would nevertheless require a dedicated assessment \cite{Henderson_2025}. In the same spirit, \cite{Henderson_2025} also identifies changes in divertor geometry as an additional knob for future investigation. Here, however, the focus is restricted to pumping speed.

        Within this framework, we examine the possibility---and the associated implications---of \textit{reducing the effective pumping speed while maintaining constant divertor pressure}. A reduced $\Seff$ in equation (\ref{eq:pdiv_divertor_pressure}) implies, to leading order, a reduced fuel puffing rate to sustain a fixed $\pdiv$---thereby alleviating the TFC burden, provided the exhaust criterion remains satisfied. Depending on how this reduction is implemented, it may correspond either to a genuinely lower overall particle throughput or merely to a redistribution of that throughput between external fuelling and internal recirculation.
        
        In turn, a decrease in $\Seff$ can be achieved---not equivalently---by decreasing the conductance $C$ and/or the nominal pumping speed $\Spump$ in equation (\ref{eq:effective_pumping_speed}). Reducing $C$ (e.g.\ via throttling) can lead to a decrease in fuel puffing demand. Instead, reducing $\Spump$ can additionally relax the installed pumping capacity, thus offering a potential two-fold advantage.
        
        However, caution is required in interpreting these equivalences. In the conductance-limited regime ($C < \Spump$) the effective pumping speed $\Seff$ depends primarily on $C$ via equation (\ref{eq:effective_pumping_speed}) \cite{PEARCE2013809}. As a result, variations in $\Spump$ do not translate linearly into variations of $\Seff$---notwithstanding pressure dependencies in conductances \cite{Bosch_1997, Pitcher_2000, Van_Oost_2023, Litovoli_2026} (section \ref{sec:intro_matter_injection_plasma__puffing__fuel__exhaust}).
        
        In the following, we first discuss the physical implications and plasma constraints associated with reduced $\Seff$. Then we examine two representative engineering implementations: one which modifies $C$ with decreasing fuel puffing, and another aimed at reducing $\Spump$ at constant throughput.

        \subsubsection{Plasma physics viewpoint: similar pressure does not imply similar plasma}\label{sec:discussion__pressure_with_pumping_speed__physics}
            
            Examples exist suggesting that comparable plasma conditions may be recovered by reducing fuel puffing (and total throughput) at constant divertor pressure via a lower effective pumping speed \cite{PITTS2019100696, park_full_2024}. This possibility is crucial, but its applicability and generality should not be overstated.
            
            A first fundamental limitation of the ``same pressure, same plasma'' interpretation arises with helium. Higher deuterium throughput is recognised to favour helium exhaust (section \ref{sec:intro_matter_injection_plasma__puffing__fuel__exhaust}). Therefore, two operating points with different fuel throughput cannot, in general, sustain the same helium balance. In reactor-relevant conditions, this already \textit{excludes strict plasma equivalence}: a different helium exhaust balance implies a different plasma state.
            
            A second limitation arises with seeded impurities. Different studies agree that $\pdiv$ sets the gross detachment regime \cite{Bosch_1996, PITTS2019100696, KAVEEVA2023101424}. However, this should not be over-interpreted as implying that $\pdiv$ uniquely determines the plasma state. In reactor-relevant conditions, impurity seeding can significantly modify separatrix conditions while contributing only weakly to the divertor pressure \cite{Lore_2022}, since typically $\GammaQ \gg \GammaZ$ (section \ref{sec:intro_matter_injection_plasma__puffing__impurities}). Thus, two plasmas with nearly identical $\pdiv$ but different impurity content need not be physically equivalent. In the limiting case, a pure deuterium plasma and a deuterium--impurity plasma with similar divertor pressure are evidently not the same plasma, even if the impurity particle source is negligible in relative magnitude.
            
            A third limitation concerns the spatial distribution of the neutral source. Fuel puff injection location is known to affect divertor behaviour and strongly influence impurity transport \cite{Schaffer_1995, McCracken_1997, Wade_1998, WEST199944, Pitcher_2000, PETRIE2007416, Senichenkov_2019, Emdee_2023, ZHOU2022113222, KAVEEVA2023101424, Park_2024_ITER, Osawa_2024, Tao_2024, Lee_2026}. If recycling were to substitute for lower fuel puffing, the resulting neutral source would more closely resemble a downstream or private flux region contribution than a conventional upstream scrape-off-layer injection\footnote{Additionally featuring broader angular distribution and higher neutral temperature than the $\sim 0.03$ eV characteristic of room-temperature puffed gas.}, at least in vertical target configurations \cite{GROTH2015471}. Even at the same divertor pressure, this change in source topology need not preserve the same SOL flow structure, impurity screening, or edge-plasma conditions.
            
            Accordingly, identical divertor pressure should be regarded as a necessary indicator of comparable neutral conditions, but not as a sufficient condition for plasma equivalence\footnote{An instructive analogy can be drawn with human energy balance \cite{Westerterp2018}: the same steady body weight can arise from very different combinations of intake and expenditure, corresponding to different physiological states. For instance, high intake combined with high expenditure leads to a metabolically active and generally healthier body composition. Likewise, in edge plasmas, identical macroscopic conditions (e.g.\ pressure) can be realised by different fuelling--pumping combinations that are not necessarily physically equivalent.}. Bosch \textit{et al.} \cite{Bosch_1996} indeed found near-equivalent detached AUG plasmas when varying pumping speed and adjusting the deuterium puff to maintain the same divertor neutral density. This is consistent with a regime dominated by internal divertor recycling, in which externally imposed deuterium throughput remains subdominant (fuel puffing in \cite{Bosch_1996} being an order of magnitude lower than the average in table \ref{tab:fueling_stats}). A related but even more permissive result was reported by Pitcher \textit{et al.} \cite{Pitcher_2000} in Ohmic deuterium plasmas in Alcator C-Mod, where divertor target conditions remained comparatively similar despite a substantial change in neutral pressure. By contrast, Kaveeva \textit{et al.} \cite{KAVEEVA2023101424} show that, even at fixed divertor pressure, different neutral source distributions can correspond to distinct plasma solutions in impurity-seeded ITER conditions.
            
            These differences are plausibly connected to the relative magnitude of the relevant fluxes. This is precisely the point highlighted by Kaveeva \textit{et al.} \cite{KAVEEVA2023101424}: once the puff-driven source in the SOL becomes comparable to the upstream flows, it can materially alter impurity retention and separatrix conditions even at fixed divertor pressure. In the same spirit, impurity seeding can significantly modify edge-plasma conditions while remaining only a small correction to the total divertor pressure balance.
            
            A final note of caution is that the studies compared above do not adopt identical impurity-seeding strategies. In Bosch \textit{et al.} \cite{Bosch_1996}, the seeding is feedback-controlled and thus time varying, whereas in Kaveeva \textit{et al.} \cite{KAVEEVA2023101424} the seeding is held fixed when comparing different fuelling configurations. Since separatrix conditions are known to depend sensitively on impurity seeding, while divertor pressure does not when $\GammaQ \gg \GammaZ$ (section \ref{sec:intro_matter_injection_plasma__puffing__impurities}), this methodological difference adds further nuance to the comparison. A more systematic characterisation of these regimes, especially under varying impurity seeding and varying relative source strengths, remains needed and is an active area of research \cite{Moscheni_2026_PSI}.

        \subsubsection{Considerations on louver implementation: lowering conductance and fuel puffing}\label{sec:discussion__pressure_with_pumping_speed__lover}

            Irrespective of the arguments above, a possible implementation of the ``lower-pumping lower-throughput'' state employs a louver. This has been recently explored by Park \textit{et al.} \cite{park_full_2024}. In this configuration, the effective pumping speed $\Seff$ is reduced by introducing a movable slit that physically throttles the conductance between the sub-divertor volume and the pump hardware.
            
            The authors report only minor changes in the main plasma parameters, with the primary difference being a less uniform pressure distribution between the strike point and the louver \cite{park_full_2024}. Although such a configuration introduces additional engineering complexity---movable components being less favorable in reactor environments---it offers a clear potential benefit in alleviating TFC burden.
            
            However, the results of \cite{park_full_2024} are obtained in a simplified plasma scenario that does not include impurities. In light of the discussion in section \ref{sec:discussion__pressure_with_pumping_speed__physics}, the apparent invariance of plasma parameters should therefore be interpreted with caution. 
            
            Further work is required to evaluate the robustness of the louver concept in more realistic, impurity-seeded plasmas. The approach nonetheless remains an possible candidate for reducing effective pumping speed through conductance control.
    
        \subsubsection{Considerations on pump-bypass TFC architecture: lowering nominal pumping speed while maintaining high throughput}\label{sec:discussion__pressure_with_pumping_speed__bypass}

            The well-established benefits of high fuel puffing are described in section \ref{sec:intro_matter_injection_plasma__puffing__fuel}. Yet, some of such actually require high fuel throughput, not high fuel puffing specifically. A promising concept able to separate the two is described below.
            \paragraph{Fundamentals.}\label{sec:discussion__pressure_with_pumping_speed__bypass__fundamentals}

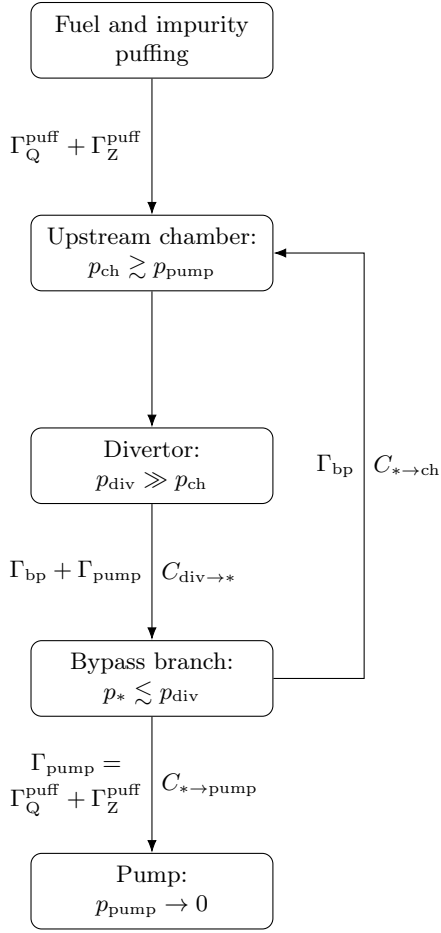
\begin{figure}[h]
\centering
\begin{tikzpicture}[
    font=\small,
    node distance=1.8cm and 2.8cm,
    box/.style={
        draw,
        rectangle,
        minimum width=3.2cm,
        minimum height=1.0cm,
        align=center,
        rounded corners
    },
    group/.style={
        draw,
        dashed,
        rounded corners=3pt,
        inner sep=8pt
    },
    >={Latex}
]

\node[box] (puff) {Fuel and impurity\\puffing};
\node[box, below=of puff] (up) {Upstream chamber:\\$\pChamber \gtrsim \pPump$};
\node[box, below=of up] (div) {Divertor:\\$\pdiv \gg \pChamber$};
\node[box, below=of div] (branch) {Bypass branch:\\$p_\ast \lesssim \pdiv$};
\node[box, below=of branch] (pump) {Pump:\\$\pPump \to 0$};

\draw[->] (puff) --
    node[left] {$\GammaQ + \GammaZ$}
    (up);

\draw[->] (up) -- (div);

\draw[->] (div) --
    node[left] {$\GammaBypass+\GammaPump$}
    node[right] {$C_{\rm div \to \ast}$}
    (branch);
\draw[->] (branch.east) -- ++(1.2,0) coordinate (aux1)
          -- (aux1 |- up.east) coordinate (aux2)
          node[midway, left] {$\GammaBypass$}
      node[midway, right] {$C_{\ast \to \rm ch}$}
          -- (up.east);
\draw[->] (branch) --
    node[left] {\shortstack{$\GammaPump=$\\$\GammaQ + \GammaZ$}}
    node[right] {$C_{\ast \to \rm pump}$}
    (pump);

\end{tikzpicture}
\caption{Simplified hydraulic network for a pump-bypass loop alike \cite{KAYE1984115, Aratari_1990, Pitcher_2000, Pitcher_2001}, as devised by Igitkhanov \textit{et al.} \cite{igitkhanov_new_2018} to favour the tritium fuel cycle. Only fuel puffing and impurity seeding are included for the sake of simplicity.}
\label{fig:bypass_logic}
\end{figure}

                Pump bypasses have been long employed in tokamak divertors \cite{KAYE1984115, Aratari_1990, Pitcher_2000, Pitcher_2001}. Their specific application aimed at reducing the burden on both tritium throughput and installed pumping capacity was proposed\footnote{More elaborate bypass configurations could also be envisaged, such as the one described in \cite{hattab_analysis_2025}. However, \textit{a bypass located downstream of the pumps would not reduce the required pumping capacity}. From a conceptual standpoint, such a scheme is closer to the DIR loop model described in Section \ref{sec:methods__TFC_model}, and therefore carries essentially the same implications. The pump-bypass concept proposed in \cite{igitkhanov_new_2018}, by contrast, constitutes a different and compelling alternative.} by Igitkhanov \textit{et al.} \cite{igitkhanov_new_2018}. In this configuration, a pump bypass from the divertor plenum or vacuum duct feeds a fraction of the exhaust back into the plasma chamber (or SOL-side of the divertor volume) rather than towards the pumps. This fraction is denoted as $\fbp$ (as $R$ in \cite{igitkhanov_new_2018}). A simplified schematic is pictured in figure \ref{fig:bypass_logic} which, as in \cite{igitkhanov_new_2018}, implicitly implements a fixed bypass configuration.

                To leading order, if $p_\ast$ denotes the pressure at the branch point, the fluxes into the bypass branch (``bp'', returning towards the chamber, ``ch'') and into the pump branch obey:
                \begin{equation}
                \begin{split}
                    \Gamma_{\rm bp}   &\sim C_{\rm \ast \rightarrow ch}   \, (p_\ast-p_{\rm ch}) \,, \\
                    \Gamma_{\rm pump} &\sim C_{\rm \ast \rightarrow pump} \, (p_\ast-p_{\rm pump}) \,,
                \end{split}
                \end{equation}
                so that:
                \begin{equation}
                    \fbp \equiv \frac{\Gamma_{\rm bp}}{\Gamma_{\rm bp}+\Gamma_{\rm pump}} \,.
                \end{equation}
                Here, $C_{\rm \ast \rightarrow ch}$ and $C_{\rm \ast \rightarrow pump}$ are the conductances of the return branch and of the path towards the primary pump, respectively, while $p_{\rm ch}$ and $p_{\rm pump}$ are the downstream pressures seen by each branch. The gas temperature in both such non-plasma-facing conduits is assumed equal (room temperature), so that throughput $[\rm Pa \, m^3 \, s^{-1}]$ and particle fluxes $[s^{-1}]$ remain proportional.

                The appeal of the concept is immediate. In the limiting case $\fbp \rightarrow 1 \Leftrightarrow \GammaBypass \gg \GammaPump$, \textit{the externally required fuel puffing and the net pumped flux may both be reduced, while the total fuel throughput circulating locally through the divertor remains high}. In that sense, the pump bypass does not eliminate the beneficial high-throughput divertor state (e.g. for helium exhaust, section \ref{sec:intro_matter_injection_plasma__puffing__fuel__exhaust}), but rather replaces part of the externally imposed puffing by passive internal recirculation. In puff-dominated regimes ($\ratioPuffToCore \gg 1$), to leading order $\GammaQ \propto (1-\fbp)$.
                
                Despite this apparent simplicity, several discussion points arise.
            
            \paragraph{Exhaust criterion, hydraulics, and recycling requirements.}

                A pump bypass is only meaningful if it remains compatible with the exhaust criterion. The limit $\fbp=1$ with $\GammaPump=0$ is clearly unworkable: the relevant question is whether an optimal $\fbp<1$ exists such that $\GammaHeCore$ is exhausted together with the minimum collateral fuel and impurity exhaust. This is non-trivial because a passive bypass provides no species separation. To first approximation, each branch carries the same mixture composition, contrary to the DIR concept in section \ref{sec:intro_matter_injection_TFC__DIR}. The concept is therefore counterproductive if helium, or seeded impurities, simply remain entrained in the deuterium--tritium stream and are passively recirculated rather than exhausted.
                
                Even in the realistic case $\fbp<1$, this fraction is not an independently prescribed design parameter. Rather, it emerges self-consistently from the neutral hydraulic network through the combined effect of conductances, downstream pressures, and the operating plasma state. Achieving a large $\fbp$ therefore requires the bypass leg to be hydraulically favourable compared with the path to the pumps, i.e.\ $C_{\rm \ast \rightarrow ch}\gg C_{\rm \ast \rightarrow pump}$, while $C_{\rm \ast \rightarrow pump}$ must nonetheless remain sufficient to guarantee adequate exhaust. This point is not quantified in \cite{igitkhanov_new_2018}, where $\fbp$ is effectively scanned as a scenario parameter rather than derived from a conductance-based hardware design. Yet the concept becomes attractive only for rather high recirculation, $\fbp \gtrsim 0.5$, so if only modest values were practically achievable the gain would be correspondingly weaker.
                
                A large bypassed flux $\GammaBypass$ is likewise not generated by the bypass itself. It must be drawn from the pre-existing divertor exhaust stream sustained by recycling at the divertor surface. Accordingly, a necessary condition for achieving $\GammaBypass \gg \GammaPump$ is a large \emph{absolute} recycling flux, because $\GammaPump$ must remain high enough.
                
                Pitcher \textit{et al.} \cite{Pitcher_2000} provide useful evidence of plausibility, although not validation. In Alcator C-Mod they quantify $C_{\rm \ast \rightarrow ch}\sim 23\,\rm m^3\,s^{-1}$, increasing with pressure, i.e.\ of the same order as the $C_{\rm \ast \rightarrow pump}$ reported for AUG by Kallenbach \textit{et al.} \cite{Kallenbach_2018}. They also report bypass fluxes large enough to become comparable to divertor recycling fluxes after detachment onset (figure~3 therein \cite{Pitcher_2000}). This suggests that the required ordering of conductances and fluxes is not, in itself, unattainable. However, those results correspond to ohmic deuterium plasmas only, and the level of active fuelling control required to access and sustain such a state is not quantified. They should therefore be interpreted as evidence of plausibility rather than as a direct validation of the concept in reactor-relevant conditions.
                
            \paragraph{Impurities and reinjection location.}
                
                Because the pressure difference must favour reinjection into a lower-pressure region, the return flow must be reintroduced on the chamber/SOL side (at least in vertical target configurations \cite{Loarte_2001}) rather than deep into the divertor private-flux region, where the pressure would remain too high, $\pdiv \gtrsim p_\ast \gg \pChamber$. This is in tension with standard fuelling and seeding strategies (section \ref{sec:discussion__pressure_with_pumping_speed__physics}), where impurities are typically seeded in the divertor while fuel is at least partially injected upstream \cite{McCracken_1997, Osawa_2024, KAVEEVA2023101424, Park_2024_ITER, Tao_2024}. As a result, impurities are recirculated together with fuel, potentially increasing upstream contamination. This concern is supported by Pitcher \textit{et al.} \cite{Pitcher_2000}, where krypton is seen to recirculate upstream via the bypass and the corresponding core krypton density increases by roughly a factor of two (figures~7--8 therein \cite{Pitcher_2000}).
                
            \paragraph{Geometrical complexity.}
                
                A further difficulty concerns the required volume and integration of the bypass ducts. In present-day tokamaks and stellarators, relatively large ``empty'' volumes behind and around divertor components are often available (e.g.\ \cite{Pitcher_2001, DEKEYSER2017899, Park_2018, Gao_2021, WU2023114023, Varoutis_2024}), and parasitic bypass leakages may even arise naturally \cite{HAUER2009903, varoutis_effect_2019, Tantos_2024}. In reactor-scale devices, however, such inert, non-shielding volume within the vacuum vessel is costly and tightly constrained. Achieving sufficiently high $C_{\ast \to \rm ch}$ would therefore require large cross-sectional areas and/or short paths, both of which compete with blanket, shielding, and structural requirements \cite{YOU2022113010}. In addition, the increased surface area associated with a more complex duct network may enhance parasitic tritium retention \cite{meschini_impact_2024}, partially offsetting the intended benefit of reducing overall tritium throughput.
                
            \paragraph{Synthesis.}
                
                A pump bypass appears as a promising strategy to reduce the load on gas puffing and pumping systems. However, this comes at the cost of increased complexity, effectively embedding a passive feedback loop into divertor operation---the robustness of which remains to be quantified.
                
                Simplified vacuum network models \cite{Litovoli_2026, HAUER2009903, Yagasaki01112024}---supplemented by high-fidelity modelling whether self-consistently coupled \cite{Lore_2023} or not \cite{varoutis_effect_2019, Tantos_2022} to the plasma, with \cite{park_full_2024, Cowley_2022} or without time-dependence \cite{Zito_2025}---would be natural, important next steps. This, however, lies beyond the scope of the present study.

\section{Conclusion}\label{sec:conclusion}

    This work shows that fuel gas puffing for detached divertor operation is one of principal drivers of the tritium fuel cycle (TFC).

    Building on the expanded puffing database released in \cite{zenodo_repo_v2} and on the analyses of \cite{Moscheni_2026}, we compare fuel puffing against a newly systematised estimate of the core fuelling rates. In doing so, we fill a logical gap that has often remained implicit in simplified fuel cycle treatments: \textit{detached operation requires large, externally imposed puffing rates---and, in reactor-relevant operation with a TFC employing a direct internal recycling loop, that puffing must contain tritium}.
    
    We find that fuel puffing systematically exceeds core fuelling by about one order of magnitude across tokamaks and stellarators of any size, with detached operating points clustering in a puffing-dominated region of parameter space. \emph{The dominant, externally imposed particle injection in a fusion reactor is thus not the stream directly sustaining fusion, but the one required to sustain edge-plasma and divertor conditions}. This directly challenges the historical assumption---common in literature, simplified fuel cycle treatments---that core fuelling dominates the particle balance.
            
    Because fuel puffing contains a non-negligible tritium fraction, the consequence for the TFC is substantial: it is no longer a downstream engineering subsystem that adapts to a prescribed plasma scenario; instead, \emph{the fuel cycle becomes a system that could force trade-offs in the plasma operation itself}.
    
    Albeit promising overall, fuel cycle architectures relying on direct internal recycling are shown to be highly sensitive to both the magnitude and isotopic composition of the puffed stream. Near 50:50 D:T puffing strongly increases tritium inventory, while strongly D-rich puffing undermines the isotopic balance required for efficient DIR.
    
    Although a DIR fraction approaching unity would allow 50:50 D:T with drawbacks limited to the pumping system, affecting only negligibly tritium self-sufficiency, the feasibility of such a high DIR fraction in a fusion reactor has never been demonstrated.
    
    Intermediate choices can define non-trivial compromise points: in the present TFC model---which does not aim to find a specific optimal operating point---a 75:25 D:T puffing ratio can still admit broadly acceptable TFC design parameters while either retaining about 85\% of the nominal fusion power, or by increasing the start-up inventory accordingly to the magnitude of the tritium puffed and the DIR fraction. Likewise, reducing fuel puffing by compensating with stronger impurity seeding may relieve some TFC constraints, but only at the price of increased dilution and radiative losses in the core plasma. This trade-off is not alien to reactor design: accepting some loss in fusion performance to make the overall plant workable is already familiar from the use of core radiators to ease power exhaust. Here, the same systems logic extends to particle exhaust and tritium handling. Divertor detachment strategy, torus pumping, and fuel cycle design must therefore be treated as a single integration problem, with explicit trade-offs between plasma performance, particle throughput, and tritium inventories.
            
    A second outcome of the present work is methodological. We quantitatively distinguish the \emph{engineering} core fuelling rate---here intended as the throughput delivered at the vacuum chamber boundary---from the corresponding \emph{physical} core fuelling rate. This distinction is essential for TFC analysis, since the plant is constrained by the throughput handled by real injection systems, not by the fraction ultimately deposited in the core plasma. Within this engineering viewpoint, \emph{the scaling laws obtained here for core fuelling and exhaust stream concentrations provide first-order estimates for reactor studies and for the preliminary scoping of more detailed tools}. At the same time, the present formulation does not yet include ``pipeline'' inefficiencies between the fuelling system and plasma-facing injection. As such, it should be regarded as, likely, a lower bound on the throughput that the TFC must ultimately provide. Its purpose is therefore not detailed design, but rather to establish the magnitude, direction, and system-level importance of the problem, while calling attention to the need for dedicated assessment of injection inefficiencies in future work. 
        
    The present results also highlight that ``small'' impurity fractions in the exhaust stream are not necessarily benign from a control perspective. In reactor-relevant conditions, values of order 0.1\% are low in absolute terms, but precisely for that reason they leave little room for uncontrolled departure. \emph{Maintaining acceptable operation requires a high degree of precision in both impurity seeding and separation performance}.
        
    More generally, this work reinforces the idea that, \emph{given the growing number of fusion pilot plant design, there must be a continuous bridge between physics targets and engineering demands}. For instance, interest in high density core plasma scenarios and lithium-based plasma-facing components is entirely justified from a plasma performance/power exhaust perspective. However, the corresponding demand they impose on TFC operation must be assessed with equal explicitness if those scenarios are to be accessed and maintained in practice.
        
    The broader implication is that the long-recognised need for core--edge integration must now be extended to \emph{core--edge--TFC integration}. In that sense, the present study provides a concrete example of the under-explored plasma--TFC interface identified by the IAEA as critical to achieve magnetic-confinement fusion. More generally, it suggests that \textit{particle balance, particle confinement and exhaust deserve a level of attention closer to that traditionally reserved for their energy counterparts and power exhaust.}
    
        
    Among the possible mitigation strategies, the pump-bypass concept emerges as particularly promising \cite{igitkhanov_new_2018, KAYE1984115, Aratari_1990, Pitcher_2000, Pitcher_2001}. In principle, it could preserve a high local divertor throughput while reducing the externally imposed fuelling burden and the pump throughput. However, the present analysis also shows that this concept remains at an early stage of maturity: its compatibility with impurity control and exhaust, geometry and hydraulic constraints requirements remains to be demonstrated.
        
    For these reasons, we conclude that future work should not focus solely on refining numerical values, but on improving the physical fidelity of the coupled picture itself. What we believe is needed next is targeted cross-comparison between assumptions, methods, databases, and models of plasma, neutrals, and fuel cycle---both tokamak- and stellarator-specific. This would allow possible hidden inconsistencies to be uncovered, and pave the way for investigation of alternative operating paradigms that are not locally but \emph{globally} optimised. A key point of the present work is not only that current estimates are uncertain, but that the uncertainty sits in a region of parameter space where silent assumptions can qualitatively change the conclusion. \emph{The problem is therefore not only one of accuracy, but of consistency}. A viable fusion reactor will require the core plasma, the edge, and the fuel cycle to be conceived and integrated together from the outset.

\section{Data Availability}

    The data supporting the findings of this study are openly available in the updated Zenodo repository \cite{zenodo_repo_v2}. This repository includes the expanded database used in the present work. The original database compiled in \cite{Moscheni_2026} remains separately available at \cite{zenodo_repo_v1}.

\section{Author information}

    The views and opinions expressed in this work are those of the authors. Neither S. Meschini nor M. Moscheni are part of the Gauss Fusion GmbH's Tritium Fuel Cycle Division. The work presented here does not refer to, nor is informed by, Gauss Fusion GmbH's tritium fuel cycle developed by the Tritium Fuel Cycle Division.
    
    S. Meschini developed the tritium fuel cycle aspects of the work and extended the corresponding modelling framework. M. Moscheni developed the plasma-oriented aspects of the study, including the scaling-law framework and database-driven analysis. S. Meschini and M. Moscheni jointly designed the overall study, interpreted the results, and wrote the paper---and as such are recognised as equal-first authors.

\section{Acknowledgements}

    The authors gratefully acknowledge the diligent work of the many colleagues whose reporting of experimental, modelling, and hardware data made possible the expansion of the original database \cite{zenodo_repo_v1} into its updated compilation \cite{zenodo_repo_v2}.
    
    The authors also thank M.~Cox, R.~Neu and O. Meneghini for early useful discussions that first highlighted the inconsistency between plasma and tritium fuel cycle assumptions, and helped motivate the present study; P.~Staniec,  S.~Lazerson, and A.~Morandi for feedback; F.~Sauerh\"ofer Rodrigo, J.~Sindermann for valuable discussions; and C.~Soika and F.~Elattab for the ever-present support.
    
    The authors further acknowledge the use of ChatGPT for language refinement and data visualisation support.

\appendix

\numberwithin{equation}{section}
\counterwithin{figure}{section}
\counterwithin{table}{section}

\renewcommand{\theequation}{\Alph{section}.\arabic{equation}}
\renewcommand{\thefigure}{\Alph{section}.\arabic{figure}}
\renewcommand{\thetable}{\Alph{section}.\arabic{table}}

\section*{Appendix}

\section{List of symbols and units}\label{apx:list_of_symbols}

    \begin{description}

        \item[\ensuremath{\Stot \, [\mathrm{m^2}]}]: total surface
        \item[\ensuremath{\Vtot \, [\mathrm{m^3}]}]: total volume
        \item[\ensuremath{\Splasma \, [\mathrm{m^2}]}]: plasma surface
        \item[\ensuremath{\Vplasma \, [\mathrm{m^3}]}]: plasma volume
        \item[\ensuremath{\Sdiv \, [\mathrm{m^2}]}]: divertor surface
        \item[\ensuremath{\Vdiv \, [\mathrm{m^3}]}]: divertor volume
        \item[\ensuremath{\Ldiv \, [\mathrm{m}]}]: characteristic divertor length
        \item[\ensuremath{\Rdiv \, [\mathrm{m}]}]: characteristic divertor radius
        \item[\ensuremath{\ftor \, [-]}]: divertor toroidal coverage
        \item[\ensuremath{\Seff \, [\mathrm{m^3 \, s^{-1}}]}]: effective pumping speed
        \item[\ensuremath{\pdiv \, [\mathrm{Pa}]}]: divertor neutral pressure

        \item[\ensuremath{\GammaTot \, [\mathrm{s^{-1}}]}]: total particle source rate
        \item[\ensuremath{\GammaPump \, [\mathrm{s^{-1}}]}]: pumping particle rate
        \item[\ensuremath{\GammaCore \, [\mathrm{s^{-1}}]}]: total core particle source

        \item[\ensuremath{\GammaQ \, [\mathrm{s^{-1}}]}]: puffing rate of species Q
        \item[\ensuremath{\GammaZ \, [\mathrm{s^{-1}}]}]: puffing rate of impurity species Z
        \item[\ensuremath{\GammaNepuff \, [\mathrm{s^{-1}}]}]: neon puffing rate
        \item[\ensuremath{\GammaTPuff \, [\mathrm{s^{-1}}]}]: tritium puffing rate
        \item[\ensuremath{\GammaDPuff \, [\mathrm{s^{-1}}]}]: deuterium puffing rate

        \item[\ensuremath{\GammaQcore \, [\mathrm{s^{-1}}]}]: core fuelling rate of species Q
        \item[\ensuremath{\GammaTcore \, [\mathrm{s^{-1}}]}]: core fuelling rate of tritium
        \item[\ensuremath{\GammaQcoreNBI \, [\mathrm{s^{-1}}]}]: core fuelling rate from NBI
        \item[\ensuremath{\GammaQcoreNBImax \, [\mathrm{s^{-1}}]}]: maximum NBI fuelling rate
        \item[\ensuremath{\GammaQcorePellet \, [\mathrm{s^{-1}}]}]: pellet injection rate
        \item[\ensuremath{\GammaQcorePelletNominal \, [\mathrm{s^{-1}}]}]: reference pellet injection rate
        \item[\ensuremath{\GammaQcorePelletMax \, [\mathrm{s^{-1}}]}]: maximum pellet injection rate
        \item[\ensuremath{\GammaQcoreSim \, [\mathrm{s^{-1}}]}]: true core source rate

        \item[\ensuremath{\GammaHeCore \, [\mathrm{s^{-1}}]}]: helium core source rate
        \item[\ensuremath{\GammaNeCore \, [\mathrm{s^{-1}}]}]: neon core source rate

        \item[\ensuremath{\GammaQex \, [\mathrm{s^{-1}}]}]: exhaust rate of species Q
        \item[\ensuremath{\GammaTex \, [\mathrm{s^{-1}}]}]: exhaust rate of tritium
        \item[\ensuremath{\GammaDex \, [\mathrm{s^{-1}}]}]: exhaust rate of deuterium

        \item[\ensuremath{\GammaZges \, [\mathrm{s^{-1}}]}]: GES impurity puffing rate
        \item[\ensuremath{\GammaZsudo \, [\mathrm{s^{-1}}]}]: Sudo impurity puffing rate
        \item[\ensuremath{\GammaZstar \, [\mathrm{s^{-1}}]}]: effective impurity puffing rate

        \item[\ensuremath{\cZ \, [-]}]: impurity concentration in exhaust
        \item[\ensuremath{\cHe \, [-]}]: helium concentration in exhaust
        \item[\ensuremath{\fBurn \, [-]}]: tritium burn-up fraction
        \item[\ensuremath{\TBE \, [-]}]: tritium burn efficiency
        \item[\ensuremath{\tbrr \, [-]}]: tritium breeding ratio (required)
        \item[\ensuremath{\Ist \, [\mathrm{kg}]}]: start-up tritium inventory
        \item[\ensuremath{\fdir \, [-]}]: direct internal recycling fraction
        \item[\ensuremath{\fTdir \, [-]}]: tritium fraction in DIR stream
        \item[\ensuremath{\fTpuff \, [-]}]: tritium puffing fraction
        \item[\ensuremath{\ftcore \, [-]}]: tritium fraction in the core
        \item[\ensuremath{\taudir \, [\mathrm{s}]}]: direct internal recycling time
        \item[\ensuremath{\sne \, [-]}]: neon separation sharpness
        \item[\ensuremath{\necore \, [-]}]: number of neon particles in core
        \item[\ensuremath{\nediv \, [-]}]: number of neon particles in divertor

        \item[\ensuremath{\nesep \, [\mathrm{m^{-3}}]}]: separatrix electron density
        \item[\ensuremath{\neavg \, [\mathrm{m^{-3}}]}]: line-averaged electron density
        \item[\ensuremath{\neGW \, [\mathrm{m^{-3}}]}]: Greenwald density limit
        \item[\ensuremath{\neSu \, [\mathrm{m^{-3}}]}]: Sudo density limit
        \item[\ensuremath{\neStar \, [\mathrm{m^{-3}}]}]: effective density limit

        \item[\ensuremath{\tauIPB \, [\mathrm{s}]}]: IPB98(y,2) confinement time
        \item[\ensuremath{\tauISS \, [\mathrm{s}]}]: ISS04 confinement time
        \item[\ensuremath{\tauStar \, [\mathrm{s}]}]: effective confinement time
        \item[\ensuremath{\tauP \, [\mathrm{s}]}]: particle confinement time
        \item[\ensuremath{\tauE \, [\mathrm{s}]}]: energy confinement time

        \item[\ensuremath{\lqEich \, [\mathrm{m}]}]: heat flux decay length
        \item[\ensuremath{\fScara \, [-]}]: Scarabosio correction factor
        \item[\ensuremath{\BT \, [\mathrm{T}]}]: toroidal magnetic field
        \item[\ensuremath{\Pin \, [\mathrm{W}]}]: input heating power
        \item[\ensuremath{\Pnbi \, [\mathrm{W}]}]: neutral beam injection power
        \item[\ensuremath{\Enbi \, [\mathrm{J}]}]: neutral beam injection energy
        \item[\ensuremath{\pfus \, [\mathrm{W}]}]: fusion power
        
    \end{description}

\section{Estimation of Core Particle Fluxes}\label{apx:estimation_GammaQcore}

    \subsection{Contribution of puffing to core fuelling}\label{apx:estimation_GammaQcore__puffing_to_core_unimportant}

        The intrinsic difficulty in experimentally quantifying neutral gas sources inside the separatrix \cite{Tholerus_2024, leoni2024scrape, Wilkie_2026}, and more generally in closing the particle balance \cite{Na_2019, Moulton2015Pumping}, implies that the \textit{physical} value of $\GammaQcore$ is often not directly accessible.
        
        As discussed in section \ref{sec:intro_matter_injection_plasma__puffing__fuel__plasma}, the contribution of fuel puffing to core fuelling is non-zero, but tends to become increasingly negligible as reactor-relevant conditions are approached. Conversely, only a fraction of pellet or NBI fuelling is deposited directly in the core, while the remainder is left in the edge plasma during penetration. These pathways are depicted in figure \ref{fig:tfc_fuelling_pathways}, where the greyed boxes identify the quantities retained in the present study.
        
        Specifically, the present framework does not account for inefficiencies occurring upstream of the plasma-facing opening of the injection systems, such as losses in gas-puff pipelines, or pellet extrusion in PI and ion beam neutralisation in NBI. These neglected inefficiencies are nevertheless important for a detailed TFC mass balance. In particular, parasitic losses along injection lines or within pellet handling systems can affect the true inventory requirements (section \ref{sec:intro_matter_injection_plasma__core_fuelling}). However, they remain difficult to quantify consistently at the present stage and are therefore left outside the scope of this work.
        
        Instead, the quantities considered here are the particle flowrates delivered at the plasma-facing openings themselves, which may therefore be termed \textit{engineering} injection rates. This is the relevant level of description for the present purpose, namely to compare the order of magnitude of the throughputs that the TFC must deliver to the plasma chamber through the different injections. The procedure adopted to estimate this quantity is described in the following.

        \usetikzlibrary{arrows.meta, positioning, calc}
\definecolor{lightgrey}{RGB}{200,200,200}

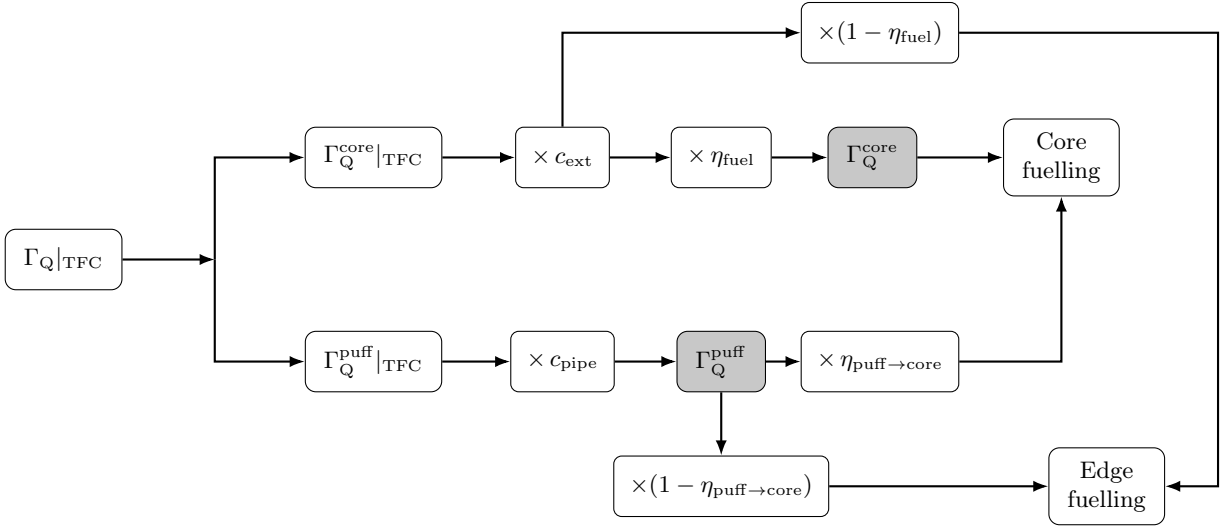
\begin{figure*}[t]
\centering
\begin{tikzpicture}[
    font=\small,
    >={Latex[length=2.0mm]},
    line/.style={->, thick},
    block/.style={
        draw,
        rounded corners=4pt,
        inner xsep=7pt,
        inner ysep=5pt,
        minimum height=8mm,
        align=center
    },
    op/.style={
        draw,
        rounded corners=3pt,
        inner xsep=6pt,
        inner ysep=4pt,
        minimum height=8mm,
        align=center
    }
]

\node[block] (src) at (0,0)
    {$\Gamma_\mathrm{Q}|_{\mathrm{TFC}}$};

\coordinate (split) at (2.0,0);

\node[block] (gcoretfc) at (4.1, 1.35)
    {$\GammaQcore|_{\mathrm{TFC}}$};

\node[block] (gpufftfc) at (4.1,-1.35)
    {$\GammaQ|_{\mathrm{TFC}}$};

\node[op] (cextcore) at (6.6,1.35)
    {$\times\, c_{\mathrm{ext}}$};

\node[op] (etacore) at (8.7,1.35)
    {$\times\, \eta_{\mathrm{fuel}}$};

\node[block, fill=lightgrey] (gcore) at (10.7,1.35)
    {$\GammaQcore$};

\node[block, align=center] (corefuel) at (13.2,1.35)
    {Core\\fuelling};

\node[op] (cpipe) at (6.6,-1.35)
    {$\times\, c_\mathrm{pipe}$};

\node[block, fill=lightgrey] (gpuff) at (8.7,-1.35)
    {$\GammaQ$};

\node[op] (etapuff) at (10.8,-1.35)
    {$\times\, \etaPuffToCore$};

\node[op] (losspuff) at (8.7,-3.0)
    {$\times(1-\etaPuffToCore)$};

\node[block, align=center] (edgefuel) at (13.8,-3.0)
    {Edge\\fuelling};

\node[op] (losscore) at (10.8,3.0)
    {$\times(1-\eta_{\mathrm{fuel}})$};

\draw[line] (src.east) -- (split);

\draw[line] (split) |- (gcoretfc.west);
\draw[line] (split) |- (gpufftfc.west);

\draw[line] (gcoretfc.east) -- (cextcore.west);
\draw[line] (cextcore.east) -- (etacore.west);
\draw[line] (etacore.east) -- (gcore.west);
\draw[line] (gcore.east) -- (corefuel.west);

\draw[line] (cextcore.north) |- (losscore.west);
\coordinate (edgefuelEast) at ($(edgefuel.east)+(0.7,0)$);

\draw[line]
    (losscore.east)
    -| (edgefuelEast)
    -- (edgefuel.east);

\draw[line] (gpufftfc.east) -- (cpipe.west);
\draw[line] (cpipe.east) -- (gpuff.west);

\draw[line] (gpuff.east) -- (etapuff.west);
\draw[line] (etapuff.east) -- (corefuel.south |- etapuff.east) -- (corefuel.south);

\draw[line] (gpuff.south) -- (losspuff.north);
\draw[line] (losspuff.east) -- (edgefuel.west);

\end{tikzpicture}
\caption{Schematic decomposition of the TFC throughput into core and puff fuelling pathways, neglecting possible wall outgassing. Greyed boxes identify the ``engineering'' quantities referred to in the present study. The efficiency of the extrusion/neutralisation processes for PI/NBI ($0 \leq c_\mathrm{ext} \leq 1$) and of the pipeline for puffing ($0 \leq c_\mathrm{pipe} \leq 1$) are important specifications, but of non-trivial quantification and hence not explicitly included herein.}
\label{fig:tfc_fuelling_pathways}
\end{figure*}

    \subsection{Estimation procedure}\label{apx:estimation_GammaQcore__procedure}

        Here the approximate procedure introduced in section \ref{sec:methods__database__core__fuelling} is detailed, in order to define in a coherent and transparent manner the computation of the engineering core fuelling rate. The procedure followed is schematically depicted in figure \ref{fig:estimation_GammaQcore__procedure__logic}.
    
        \begin{figure*}[t]
\centering

\begin{tikzpicture}[
  font=\small,
  node distance=10mm and 18mm,
  >=Latex,
  box/.style={draw, rounded corners, align=center, inner sep=6pt},
  decision/.style={draw, diamond, aspect=2.2, align=center, inner sep=3pt},
  repeatbox/.style={draw, ellipse, align=center, inner sep=3pt},
  arrow/.style={->, line width=0.6pt},
  dashedbox/.style={draw, dashed, rounded corners, inner sep=10pt},
  dashedmini/.style={draw, dashed, rounded corners, align=center, inner sep=4pt},
  labelplain/.style={font=\small, inner sep=0pt}
]

\node[box] (entry) {Database entry};

\node[decision, below=of entry] (hasSim)
{$\GammaQcoreSim$ quoted?};

\node[box, below left=of hasSim, xshift=-6mm] (useSim)
{$\GammaQcore = \GammaQcoreSim$};

\node[box, below right=of hasSim, xshift=6mm] (useProxy)
{$\GammaQcore = \GammaQcoreNBI + \GammaQcorePelletNominal$};

\node[dashedmini, above=of useProxy] (pelDef)
{Define once\\[1mm]
$\GammaQcorePelletNominal$ and $\GammaQcorePelletMax$\\[1mm]
(appendix \ref{apx:estimation_GammaQcore__procedure__PI})};

\draw[->, dashed] (pelDef.south) -- (useProxy.north);

\node[box, below=18mm of hasSim] (merge)
{Populate $\left\{ \GammaQcore\right\}$};

\node[repeatbox, below=of merge] (repeat)
{Repeat};

\node[box, at={(useSim |- repeat)}] (storeSim)
{Populate $\left\{\GammaQcoreSim\right\}$};

\node[box, at={(useProxy |- repeat)}] (storeNbi)
{Populate $\left\{\GammaQcoreNBI\right\}$};

\draw[arrow] (entry) -- (hasSim);

\draw[arrow] (hasSim) -- node[midway, above, sloped]{yes} (useSim.north east);
\draw[arrow] (hasSim) -- node[midway, above, sloped]{no}  (useProxy.north west);

\draw[arrow] (useSim.south) -- (storeSim.north);
\draw[arrow] (useProxy.south) -- (storeNbi.north);

\draw[arrow] (storeSim) |- (merge);
\draw[arrow] (storeNbi) |- (merge);
\draw[arrow] (merge) -- (repeat);

\node[dashedbox,
  label={[align=center]north:
    For a given machine:
  },
  fit=(entry) (hasSim) (useSim) (useProxy) (storeSim) (storeNbi) (merge) (repeat)
] (loopbox) {};

\node[box, below=20mm of loopbox.south] (meanBox)
{Machine-specific average:\\[1mm]
$\langle \GammaQcore \rangle$};

\draw[->, dashed] 
(loopbox.south) -- 
node[midway, right, align=left, text width=8cm]
{$\left\{\GammaQcore\right\}$, $\left\{\GammaQcoreSim\right\}$ and $\left\{\GammaQcoreNBI\right\}$ populated,\\[1mm]
$\GammaQcorePelletNominal$ and $\GammaQcorePelletMax$ defined for given machine}
(meanBox.north);

\node[box, below=14mm of meanBox] (spanBox)
{Variation range figure \ref{fig:scaling_GammaQcore_Vtot}:\\[1mm]
$\min\!\,[\{\GammaQcoreSim \}\,\cup\, \{\{\GammaQcoreNBI\} + \GammaQcorePelletNominal \}]$\\[1mm]
$\max\!\,[\, \{\GammaQcoreSim \}\,\cup\, \{\{\GammaQcoreNBI\} + \GammaQcorePelletMax \}]$
};

\draw[arrow] (meanBox) -- (spanBox);

\node[box, right=20mm of meanBox] (tableBox)
{Variation range table \ref{tab:fueling_stats}:\\[1mm]
$\tpm 1\sigma$};

\draw[arrow] (meanBox) -- (tableBox);

\end{tikzpicture}

\caption{Decision logic for computing $\GammaQcore$ for each database entry (dashed loop), followed by aggregation into a mean value and variation range for a given machine, with corresponding $\pm 1\sigma$ reported in table \ref{tab:fueling_stats}. $\GammaQcoreNBI$ is computed via equation (\ref{eq:GammaQcoreNBI}) with the power $\Pnbi$ utilised in the specific database entry (if available),  $\cup$ represents the union of two sets $\{\cdot\}$, and min/max errorbar computation involve the set where either $\GammaQcorePelletNominal$ or $\GammaQcorePelletMax$ is added to every element of $\left\{\GammaQcoreNBI\right\}$.}
\label{fig:estimation_GammaQcore__procedure__logic}
\end{figure*}
        
        For a given machine and a given database entry within that machine, the left branch identifies \textit{simulated} cases where $\GammaQcoreSim$ is explicitly reported in the literature (e.g. boundary condition in edge plasma models). In these cases, the reported value directly populates the collection $\{\GammaQcore\}$. Importantly, this value still represents the engineering core fuelling: puffed/outgassed neutral particles penetrating till the core-facing boundary of edge simulations are added at run-time on top of the specified boundary condition and are therefore not accounted for in the specified boundary condition. \textit{Experimental} entries can rarely quantify this quantity directly due to the uncertainties associated with closing the particle balance \cite{Tholerus_2024, leoni2024scrape}. It is nevertheless recognised that $\GammaQcoreSim$ does not necessarily represent the ``true'' core fuelling rate; it is labelled as such only to indicate that it is the value most consistent with the other features of the analysed entry (e.g. the corresponding puffing rate or boundary conditions). As anticipated in section \ref{sec:methods__database__core__fuelling}, and further discussed below, even in this case the value may depart from the engineering reality.
        
        The right branch of figure \ref{fig:estimation_GammaQcore__procedure__logic} instead describes the procedure used to estimate the core fuelling rate when $\GammaQcoreSim$ is not available. In this context, priority is given to data reported in commissioning results or post-campaign summaries, e.g. \cite{Streibl01112003}. However, a fraction of the information is retrieved from hardware design studies, including conceptual design phases \cite{lang_considerations_2015}. Refined specifications following detailed engineering design, as well as values eventually achieved during experimental operation, may differ from these conceptual estimates. Fully characterising this sequence of values is beyond the scope of the present work, as the resulting uncertainty likely remains within the error bars in this study.

    \subsubsection{Neutral beam injection (NBI).}\label{apx:estimation_GammaQcore__procedure__NBI}

        In machines featuring NBI heating, its contribution to core fuelling is estimated from the nominal beam energy and the beam heating power corresponding to the operating point considered. Namely:
        \begin{equation}\label{eq:GammaQcoreNBI}
            \GammaQcoreNBI = \frac{\Pnbi}{\Enbi} \,,
        \end{equation}
        with all quantities expressed in SI units. Hardware specifications are reported in table \ref{tab:nbi_sources} and in the database \cite{zenodo_repo_v2}. When multiple NBI sources are present (e.g. in \cite{Streibl01112003}), an effective mean beam energy is used so as to preserve the same total particle source rate.
        
        Given a total input power $P_{\mathrm{in}}$ for a given database entry, care is taken to discriminate between the relative contributions of NBI and radio-frequency heating---the latter not contributing directly to particle fuelling. Where this separation is not explicitly stated in the literature, the entry is discarded as a precaution\footnote{Assuming $\Pnbi = P_{\mathrm{in}}$ in presence of unquantified radio-frequency heating gives a difference of $\lesssim 5\%$ in the pre-factor and exponent of equation (\ref{eq:scaling_GammaQcore_Vtot}).}. Ohmic heating is instead neglected where not explicitly reported, and the full injected power is appointed to NBI \cite{FEVRIER2021100977, Yang_2023}---a conservative choice in the context of the present study, whose objective is to assess whether puffing dominates over core fuelling sources. Accordingly, $100\%$ beam absorption is assumed (no shinethrough losses, $\eta_\mathrm{fuel}|_\mathrm{NBI} = 1$) \cite{Joffrin_2017, Valovic_2009}.
        
        Overall, \textit{the NBI contribution represents the most case-specific component of core fuelling, as it depends directly on the heating power of the particular discharge considered}, although it does not necessarily dominate the total core fuelling budget (see below).
        
        Inspection of the data in table \ref{tab:nbi_sources} suggests that the NBI particle source is correlated with $\Vtot$, albeit with not insignificant scatter. A power-law fit yields an approximate scaling ($R^2 = 70\%$):
        \begin{equation}\label{eq:scaling_GammaQcoreNBI_Vtot}
            \GammaQcoreNBI = 9.9 \times 10^{19} \times \Vtot^{0.41} \,,
        \end{equation}
        for devices up to JET \cite{Valovic_2024, GERAUD20131064} and JT-60SA \cite{LANG2017167}---beyond which, excepting ITER \cite{COMBS2012634}, only pellet fuelling is leveraged.
        
        \begin{table}[t]
\centering
\small
\setlength{\tabcolsep}{6pt}
\renewcommand{\arraystretch}{1.15}
\begin{tabular}{l c c c}
\toprule
\multirow{2}{*}{Machine} & $\Enbi$ & $\Pnbi^{\mathrm{max}}$ & $\GammaQcoreNBImax$ \\
 & [keV] & [MW] & [$\mathrm{at \times s^{-1}}$] \\
\midrule
AUG & $76$ \cite{Streibl01112003} & $20$ \cite{Streibl01112003} & $1.6 \times 10^{21}$ \\
DIII-D & $75$ \cite{SCOVILLE20196} & $20$ \cite{SCOVILLE20196} & $1.7 \times 10^{21}$ \\
DTT & $300$ \cite{AGOSTINETTI2019441} & $16$ \cite{AGOSTINETTI2019441} & $3.3 \times 10^{20}$ \\
EAST & $80$ \cite{Hu_2012} & $8.0$ \cite{Hu_2012} & $6.2 \times 10^{20}$ \\
Globus-M2 & $45$ \cite{Shchegolev2023} & $1.0$ \cite{Shchegolev2023} & $1.4 \times 10^{20}$ \\
HL-2M & $80$ \cite{Zhong_2022} & $15$ \cite{Zhong_2022} & $1.2 \times 10^{21}$ \\
ITER & $1000$ \cite{Hemsworth_2017} & $33$ \cite{Hemsworth_2017} & $2.1 \times 10^{20}$ \\
JET & $125$ \cite{CIRIC2011509} & $34$ \cite{CIRIC2011509} & $1.7 \times 10^{21}$ \\
JFT-2M & $32$ \cite{Ida_1992} & $0.60$ \cite{Ida_1992} & $1.2 \times 10^{20}$ \\
JT-60SA & $117$ \cite{HANADA2011835} & $30$ \cite{HANADA2011835} & $1.6 \times 10^{21}$ \\
KSTAR & $80$ \cite{NA2022113320} & $5.5$ \cite{NA2022113320} & $4.3 \times 10^{20}$ \\
LHD & $107$ \cite{Takeiri01082010} & $23$ \cite{Takeiri01082010} & $1.3 \times 10^{21}$ \\
MAST-U & $75$ \cite{MCADAMS2026115408} & $5.0$ \cite{MCADAMS2026115408} & $4.2 \times 10^{20}$ \\
NSTX & $80$ \cite{Ono_2001} & $5.0$ \cite{Ono_2001} & $3.9 \times 10^{20}$ \\
NSTX-U & $100$ \cite{menard2012overview} & $15$ \cite{menard2012overview} & $9.4 \times 10^{20}$ \\
TCV & $36$ \cite{LISTOPAD2025114867} & $2.4$ \cite{LISTOPAD2025114867} & $4.2 \times 10^{20}$ \\
W7-AS & $50$ \cite{Baldzuhn_2003} & $3.0$ \cite{OTT1997641} & $3.8 \times 10^{20}$ \\
\bottomrule
\end{tabular}
\caption{Neutral beam injection (NBI) specifications present in the database. Machines not reported implicitly carry zero.}
\label{tab:nbi_sources}
\end{table}

    \subsubsection{Pellet injection (PI)}\label{apx:estimation_GammaQcore__procedure__PI}

        To the NBI contribution, a reference PI source $\GammaQcorePelletNominal$ is added for machines equipped with pellet fuelling. For each device, a single representative value is selected from the literature, typically corresponding to commonly reported or mid-range operating conditions. Therefore, \textit{PI lacks specificity to the particular database entry}. When several baseline operating points are reported for a given machine (e.g. in \cite{lang_considerations_2015}), each is included as a separate entry in the database so that the averaging procedure described in section \ref{sec:methods__database__statistics} retains the information associated with the different fuelling scenarios. For future burning plasma scenarios, $\GammaQcorePelletNominal$ is estimated according to equation (\ref{eq:gammaQcore_from_helium_production}).
        
        In addition, the maximum deliverable pellet fuelling rate $\GammaQcorePelletMax$ is retrieved wherever available---otherwise assumed to match $\GammaQcorePelletNominal$. This quantity provides an estimate of the upper bound of the engineering fuelling capability and defines the upper extents of the variation ranges in figure \ref{fig:scaling_GammaQcore_Vtot}.
        
        Both the nominal and maximum pellet fuelling rates are reported in table \ref{tab:pi_sources} and in the database \cite{zenodo_repo_v2}. For burning-plasma scenarios, pellets are assumed to consist of a $50{:}50$ D:T mixture, whereas a $100{:}0$ D:T composition is assumed for present-day devices.
        
        \begin{table}[t]
\centering
\small
\setlength{\tabcolsep}{6pt}
\renewcommand{\arraystretch}{1.15}
\begin{tabular}{l c c}
\toprule
\multirow{2}{*}{Machine} & $\GammaQcorePelletNominal$ & $\GammaQcorePelletMax$ \\
 & [$\mathrm{at \times s^{-1}}$] & [$\mathrm{at \times s^{-1}}$] \\
\midrule
ARC & $1.2 \times 10^{22}$ \cite{sorbom_arc_2015} & -- \\
AUG & $0.0$ \cite{Kallenbach_2018} & $3.6 \times 10^{22}$ \cite{Lang_2018} \\
CFETR & $1.5 \times 10^{22}$ \cite{McClenaghan_2023} & $1.5 \times 10^{22}$ \cite{McClenaghan_2023} \\
EAST & $0.0$ \cite{Wang_2019} & $4.0 \times 10^{21}$ \cite{LI201499} \\
EU-DEMO & $4.5 \times 10^{22}$ \cite{lang_considerations_2015} & $1.6 \times 10^{23}$ \cite{lang_considerations_2015} \\
GIGA & $7.1 \times 10^{22}$ \cite{gaussfusion_cdr_exec_2026} & -- \\
Infinity Two & $2.2 \times 10^{22}$ \cite{Guttenfelder_2025} & $2.2 \times 10^{22}$ \cite{Guttenfelder_2025} \\
ITER & $1.2 \times 10^{23}$ \cite{COMBS2012634} & $3.5 \times 10^{23}$ \cite{COMBS2012634} \\
JET & $1.1 \times 10^{22}$ \cite{Valovic_2024} & $6.3 \times 10^{22}$ \cite{GERAUD20131064} \\
JT-60SA & $8.4 \times 10^{21}$ \cite{LANG2017167} & $1.3 \times 10^{22}$ \cite{PLOECKL2025114829} \\
KSTAR & $3.0 \times 10^{21}$ \cite{PARK2017163} & $7.4 \times 10^{21}$ \cite{PARK2017163} \\
LHD & $6.0 \times 10^{21}$ \cite{YAMADA200311} & $6.0 \times 10^{21}$ \cite{YAMADA200311} \\
MANTA & $1.1 \times 10^{22}$ \cite{Rutherford_2024} & -- \\
Stellaris & $6.4 \times 10^{22}$ \cite{LION2025114868} & -- \\
STEP & $1.0 \times 10^{22}$ \cite{Tholerus_2024} & $1.0 \times 10^{22}$ \cite{Tholerus_2024} \\
W7-X & $2.0 \times 10^{21}$ \cite{Baldzuhn_2020} & $7.5 \times 10^{21}$ \cite{Baldzuhn_2020} \\
\bottomrule
\end{tabular}
\caption{Pellet injection (PI) core-fueling specifications present in the database. Machines not reported implicitly carry zero. $\GammaQcorePelletNominal$ in future reactors not reporting $\GammaQcorePelletMax$ is estimated from equation (\ref{eq:gammaQcore_from_helium_production}), with data retrieved from the cited reference.}
\label{tab:pi_sources}
\end{table}

        The data shows that pellet fuelling scales with the total machine volume ($R^2=0.77$):
        \begin{equation}\label{eq:scaling_GammaQcorePelletNominal_Vtot}
            \GammaQcorePelletNominal = 3.1 \times 10^{20} \times \Vtot^{0.71} \,,
        \end{equation}
        which implicitly contains an error tied to the choice of the reference values themselves (lack of specificity). A weaker correlation is instead observed for $\GammaQcorePelletMax \sim 1.7 \times 10^{21} \times \Vtot^{0.51}$ ($R^2=0.42$).

    \subsection{Comparison: neutral-beam vs. pellet injection}\label{apx:comparison_GammaQcorePelletNominal_GammaQcoreNBI}

        Overall, the results above confirm that pellet injection typically provides a significantly larger fuelling source than NBI, consistent with previous observations in the literature (e.g. \cite{geulin_pellet_2022}).

        In particular:
        \begin{equation}\label{eq:scaling_ratio_GammaQcorePelletNominal_GammaQcoreNBI}
            \frac{\GammaQcorePelletNominal}{\GammaQcoreNBI} \sim 3.1 \times \Vtot^{0.30} \,,
        \end{equation}
        approximately, with a relatively restricted domain where both sources coexist in the same machine (i.e. common devices in tables \ref{tab:nbi_sources} and \ref{tab:pi_sources}).

    \subsection{Overall robustness and uncertainty of the core-fuelling scaling}\label{apx:robustness_scaling_GammaQcore}

        Importantly, NBI is comparatively well characterised and primarily constrains the cluster of small devices in figure \ref{fig:scaling_GammaQcore_Vtot}. Instead, pellet fuelling---though subject to uncertainty---exhibits comparatively limited variability for large devices, partly reflecting the more limited exploration of operating space in this regime. This effectively anchors both the lower and upper ends of the $\GammaQcore(\Vtot)$ scaling in equation (\ref{eq:scaling_GammaQcore_Vtot}), thereby enhancing the robustness of the fitted exponent.
        
        \begin{table}[t]
\centering
\small
\begin{tabular}{lcc}
\toprule
$\GammaQcore$ & n. entries & $\langle\ratioPuffToCore\rangle \, [-]$ \\
\midrule
$\GammaQcoreSim$ & 89 & $(1.8 \times 10^{1}) \tpm 2.5$ \\
$\GammaQcoreNBI$ & 157 & $(1.5 \times 10^{1}) \tpm 2.7$ \\
$\GammaQcorePelletNominal \neq 0$ & 108(72) & $(2.3 \times 10^{0}) \tpm 2.2$ \\
\bottomrule
\end{tabular}
\caption{Break-down of the resulting $\GammaQcore$. In the first row entries feature a directly available core-fuelling rate $\GammaQcoreSim$; in the second, $\GammaQcore$ is estimated from NBI alone; in the third the estimate involves a finite pellet fuelling (and may hence rely on an \textit{assumed} $\GammaQcorePelletNominal$). However, 72 entries are computed from equation (\ref{eq:gammaQcore_from_helium_production}). The ratio column reports the geometric mean and multiplicative $1\sigma$ spread of $\ratioPuffToCore = \GammaQ / \GammaQcore$.}
\label{tab:fueling_breakdown}
\end{table}

        This interpretation is further supported by the breakdown reported in table \ref{tab:fueling_breakdown}: 89 out of 354 entries have an explicitly available value of $\GammaQcore$, i.e. $\GammaQcoreSim$. A further 157 out of 354 do not report such a value but feature NBI-only fuelling, so that 246 out of 354 entries are fully case-specific and based on the most accurate information directly tied to the corresponding operating point. Of the remaining 108 entries involving pellet injection ($\GammaQcorePelletNominal \neq 0$), 72 correspond to D--T reactor scenarios in which $\GammaQcore$ is either computed in a case-specific manner from equation (\ref{eq:gammaQcore_from_helium_production}) or directly available in the literature. Therefore, only the final 36 entries---in mid-sized machines---rely on an \textit{assumed} mid-range literature value $\GammaQcorePelletNominal$ (table \ref{tab:pi_sources}) and therefore carry the largest uncertainty. Notably, these latter cases also tend to exhibit smaller $\ratioPuffToCore$ than the more case-specific entries, consistently with section \ref{sec:discussion__core_fuelling__size}. This again makes the procedure conservative from the perspective of demonstrating puffing dominance.
        
\begin{figure}
            \centering
            \includegraphics[width=0.45\textwidth]{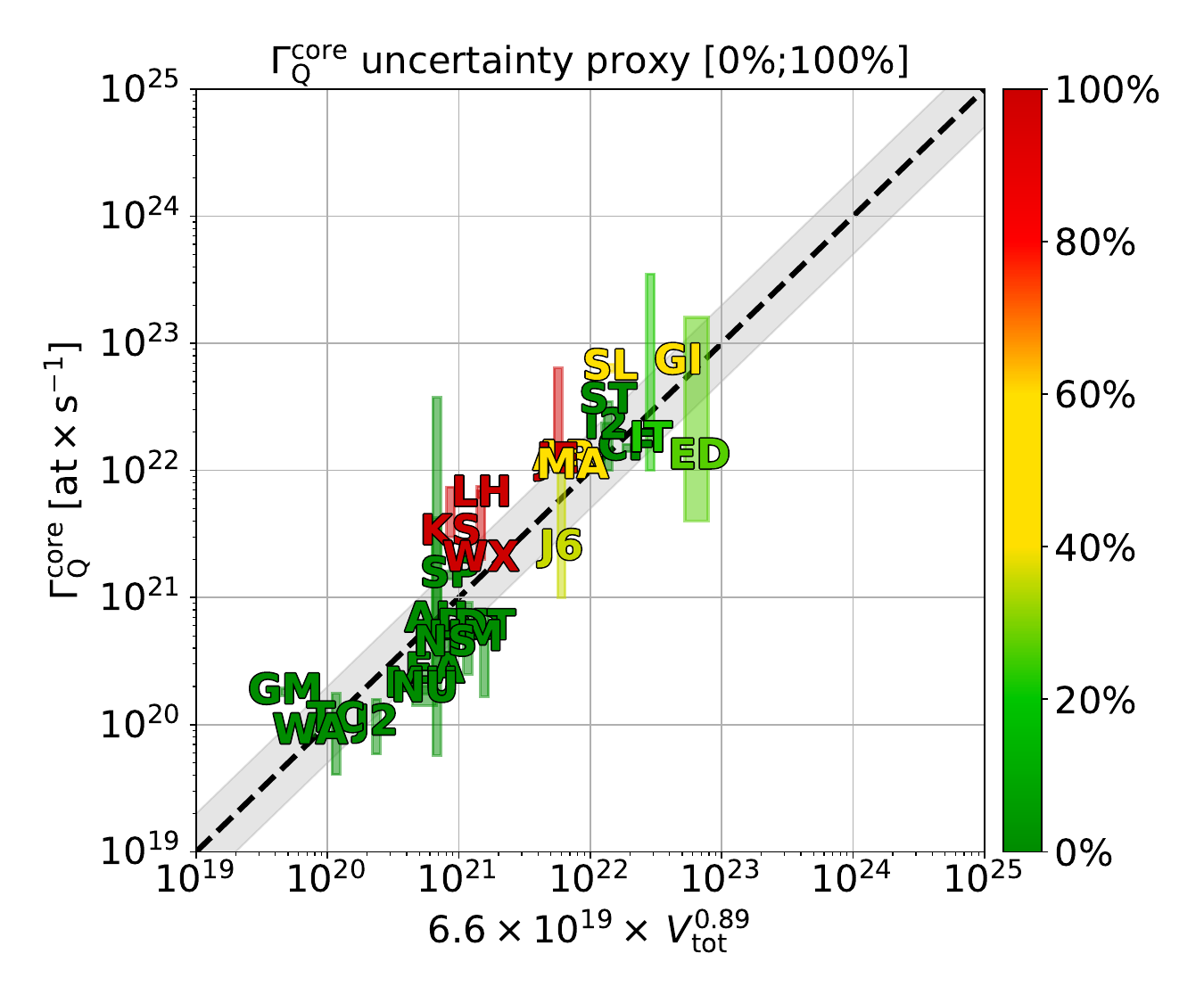}
            \caption{Same as figure \ref{fig:scaling_GammaQcore_Vtot}, but with colouring approximately reflecting the uncertainty in $\GammaQcore$ according to equation (\ref{eq:GammaQcore_machine_uncertainty}). Devices relying on equation (\ref{eq:gammaQcore_from_helium_production}) are assigned a nominal $50\%$ uncertainty, reflecting the uncertainty associated with $\fBurn$. Accordingly, only the red-coloured devices systematically feature a pellet injection rate estimated from literature values rather than tied to case-specific operating points.}
            \label{fig:GammaQcore_uncertainty}
        \end{figure}

        An indicative uncertainty proxy for the $i^{\mathrm{th}}$ machine can therefore be defined as
        \begin{equation}\label{eq:GammaQcore_machine_uncertainty}
            \mathrm{uncertainty}_i = \frac{N_i(\GammaQcorePelletNominal \neq 0)}{N_{i,\mathrm{tot}}} \,,
        \end{equation}
        where $N_i(\GammaQcorePelletNominal \neq 0)$ denotes the number of entries of the $i^{\mathrm{th}}$ machine for which a non-zero nominal pellet contribution is involved, and $N_{i,\mathrm{tot}}$ is the total number of entries for that machine. Further details are provided in the caption of figure \ref{fig:GammaQcore_uncertainty}, which supports the interpretation above.
        
        Overall, \textit{while approximate, the procedure is constrained by comparatively high-accuracy information over most of the dataset---especially at the small- and large-device ends that anchor the scaling.}
        
    \subsection{Noteworthy cases}\label{apx:estimation_GammaQcore__notable_cases}

        Some entries in the database illustrate the intrinsic variability associated with core fuelling estimates. A notable example is AUG, which most commonly operates without pellet fuelling, yet a number of high-density pellet-fuelled discharges have been analysed in \cite{Lang_2018, Lang_2020} (section \ref{sec:discussion__high_density_operation}). Including these cases defines the top-end value of the AUG bar represented in figure \ref{fig:scaling_GammaQcore_Vtot} (``AU'' marker), thereby clarifying the full variability range associated with the device. At the same time, the average AUG core fuelling value (from NBI) lies significantly below the upper bound (30-fold difference).
        
        As anticipated in section \ref{sec:methods__database__core__fuelling} for STEP, future reactor designs are also subject to variability across different design iterations \cite{SICCINIO2022113047}. Another representative example is EU-DEMO. Pellet fuelling rates reported in \cite{lang_considerations_2015} consistently exceed later simulations. The lowest value therein, $4\times10^{22}$, is comparable to the $3.5\times10^{22}$ enforced in \cite{Xiang_2021}. Subsequent figures decrease further, reaching $1.4\times10^{22}$ in \cite{geulin_pellet_2022}, $7\times10^{21}$ in \cite{subba_solps-iter_2021}, and $4\times10^{21}$ in \cite{Korzueva_2023}. This can be partially appointed to differences implicitly assumed in the pipeline and penetration efficiencies.
        
        Overall, these examples illustrate an unavoidable level of variability in the estimation of core fuelling rates. This variability is nevertheless commensurate with the approximations inherent in the procedure adopted in the present study.

\section{Effective energy confinement time}\label{apx:tau_effective}

    The present study estimates $\tauP$ from an effective energy-confinement proxy $\tauStar$, assuming $\tauP \propto \tauE$ (section \ref{sec:discussion__core_fuelling__physics}). This appendix summarises the rationale and defines the corresponding confinement scalings used for tokamaks and stellarators.
    
    Empirical expectations for $\tauP$ are considerably less established than for the energy confinement time $\tauE$. The estimate $\tauP \lesssim \tauE$ is operated in \cite{igitkhanov_new_2018}. Instead, values of $\tauP$ roughly 3--4 times larger than $\tauE$ have been reported in JET \cite{Stork_2005, Zastrow_2004}, and similar assumptions are adopted in reactor studies such as \cite{Tholerus_2024} across the STEP design space. Still, irrespective of the particular pre-factor, proportionality between the two timescales persists---also invoked for helium transport \cite{reiter_burn_1990, Angioni_2026}. In the absence of systematic multi-machine particle confinement databases, we similarly adopt the simplifying assumption $\tauP \propto \tauE$.
    
    For tokamaks, the energy confinement time is estimated using the ITER98(y,2) scaling \cite{ITER_confinement_1999}:
    \begin{equation}\label{eq:scaling_tau_ipb98}
        \begin{split}
            \tauIPB = & \, 0.562 \, I_p^{0.93} B_T^{0.15} \neStar^{0.41} P^{-0.69} \\
            & \times R^{1.97} \varepsilon^{0.58} \kappa^{0.78} M^{0.19} \,.
        \end{split}
    \end{equation}
    For stellarators, the ISS04 scaling is used \cite{Yamada_2005}:
    \begin{equation}\label{eq:scaling_tau_iss04}
        \begin{split}
            \tauISS = & \, 0.134 \, a^{2.28} R_0^{0.64} P^{-0.61} \\
            & \times \neStar^{0.54} \BT^{0.84} \iota_{2/3}^{0.41} \,.
        \end{split}
    \end{equation}
    
    The density limit $\neStar$ from equation (\ref{eq:effective_density_limit}) is used as a proxy\footnote{Using only $\neavg$ would reduce the dataset size while producing negligible changes in the resulting scaling (only $\sim0.05$ change in the $\Stot$ exponent). Supplementing missing $\neavg$ values with $n_{\rm GW}$ similarly produces minimal differences but reduces transparency of the procedure.} for $\neavg$ \cite{Eich_2013}. All confinement times are expressed in seconds. The average mass number $M$ is 2.5 for future tokamak power plants (circa-50:50 D:T), and 2.0 otherwise (100:0 D:T). The reader is then redirected to \cite{ITER_confinement_1999, Yamada_2005} for the meaning of the remaining symbols.
    
    Overall:
    \begin{equation}\label{eq:tau_effective}
        \tauStar \in \{\tauIPB ,\, \tauISS\} \,.
    \end{equation}

\section{Comparison with equation (5) of \cite{Moscheni_2026}}\label{apx:comparison_vs_Moscheni_OPQ}
    
    A key result of section \ref{sec:results__how_high_fuel_puffing} is that fuel puffing rates are systematically larger than the corresponding core fuelling rates, both across existing devices and when extrapolated to reactor-scale machines. This conclusion relies on the exponents appearing in equation (\ref{eq:scaling_GammaQcore_Vtot}) for $\GammaQcore$ and in equation (\ref{eq:scaling_GammaQpuff_Vtot}) for the puffing rate $\GammaQ$. 
    
    While the puff regression obtained here suggests $\GammaQ \propto \Vtot$, the determination of the exponent is affected by the inherent scatter of the puffing database, as discussed in section 2.3.3 of \cite{Moscheni_2026}. A further consistency check is therefore warranted.
    
\begin{figure}
        \centering
        \includegraphics[width=0.45\textwidth]{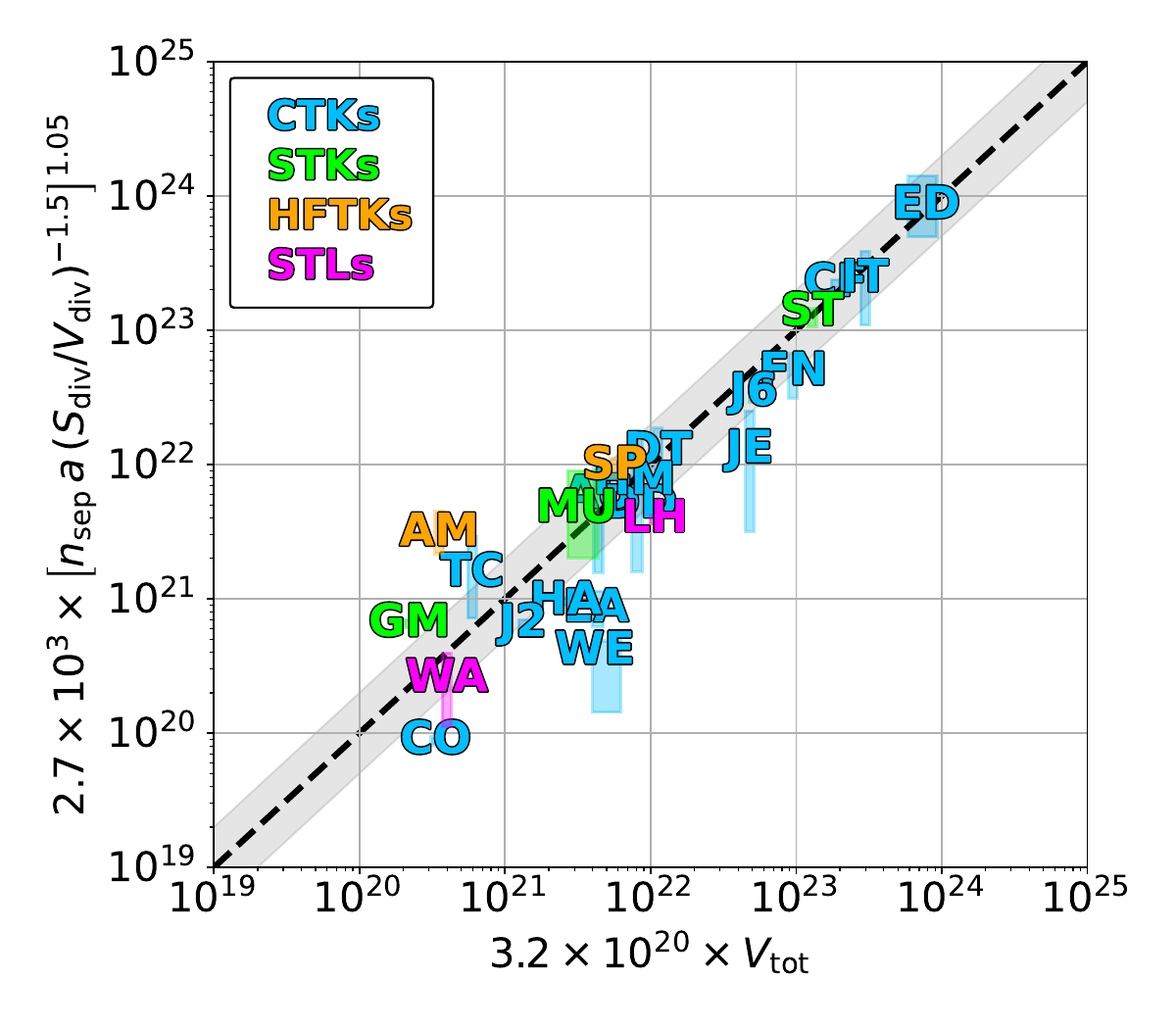}
        \caption{The most accurate regression for $\GammaQ$ in \cite{Moscheni_2026}, i.e. Equation (5) therein, vs. equation (\ref{eq:scaling_GammaQpuff_Vtot}) herein. The original database \cite{zenodo_repo_v1} is used for consistency. The satisfactory agreement sanity-checks the linear dependence between $\GammaQ$ and $\Vtot$, especially in the reactor-relevant region of the parameter space.}
        \label{fig:comparison_vs_Moscheni_OPQ}
    \end{figure}

    For this purpose, figure \ref{fig:comparison_vs_Moscheni_OPQ} directly compares the scaling for $\GammaQ(\Vtot)$ in equation (\ref{eq:scaling_GammaQpuff_Vtot}) with the most accurate regression reported in \cite{Moscheni_2026}, specifically equation (5) therein. The comparison shows a satisfactory agreement ($R^2 = 0.98$), particularly in the parameter range corresponding to next-step devices that are of primary relevance to the present study.
    
    This cross-check confirms that the puffing rate is well approximated by a linear dependence on the total plasma volume, $\GammaQ \propto \Vtot$. Consequently, the conclusions drawn in section \ref{sec:results__how_high_fuel_puffing} regarding the persistent disparity between puffing and core fuelling rates with machine size remain robust.

\section{DIR and fuelling modelling}\label{appendix:DIR_model}

Equations \eqref{eq:gamma_vs_ftdir} and \eqref{eq:ftdir} are derived in this appendix. Two support variables are introduced: the particle flow entering the tritium plant, $\GammaTsep$, and the exhaust particle flow entering the vacuum pumps, $\GammaTex$. We also assume for simplicity $\etaFuel = 100\%$, that is, TBE $\simeq \fBurn$ and $\Gamma_\mathrm{T}^\mathrm{inj} \simeq \GammaTcorePI \simeq \GammaTcore$. Since $\GammaZ \ll \GammaPuff$, it is neglected for simplicity. The derivation for $\etaFuel < 100\%$ and non-negligible $\GammaZ$ is formally equivalent. We recall that this derivation considers no additional tritium provided by the storage system to rebalance the core fuelling mixture. Otherwise, additional process components would be needed downstream of the DIR loop, and additional tritium to start-up the reactor. The scenarios accounting for a rebalancing of the core fuelling mixture do not need the equations presented in this appendix because the core tritium fraction remains unaltered, and the results are presented in section \ref{subsec:tss_puffing_DT}.

Tritium flows downstream of the divertor can be written as:

\begin{equation}\label{eq:GammaTex}
    \GammaTex = \GammaTcore (1-\TBE)
\end{equation}
\begin{equation}\label{eq:GammaTsep}
    \GammaTsep = (1-\fdir)\GammaTex
\end{equation}

The total core fuelling rate is kept fixed and is provided by mixing all the available tritium stream from the tritium plant (100\%T) with a fraction $\alpha$ of the stream coming from the DIR (with variable DT composition) to balance the required fuelling rate:

\begin{equation}\label{eq:GammaQcore_mixing}
    \GammaQcore = \GammaTsep + \alpha \GammaQDIR
\end{equation}

\noindent The tritium core fuelling rate can be written in a similar way as:

\begin{equation}\label{eq:GammaTcore_mixing}
    \GammaTcore = \GammaTsep + \alpha \fTdir \GammaQDIR
\end{equation}

\noindent Plugging equation \eqref{eq:GammaQcore_mixing} into equation \eqref{eq:GammaTcore_mixing}:

\begin{equation}
    \GammaTcore = \GammaTsep(1-\fTdir) + \fTdir \GammaQcore
\end{equation}

\noindent and using equations \eqref{eq:GammaTex} and \eqref{eq:GammaTsep} one finds:

        \begin{equation}
            \GammaTcore(t) = \frac{\fTdir(t) \GammaQcore}{1-k_1(1-\fTdir(t))}
        \end{equation}   

\noindent that is equation \eqref{eq:gamma_vs_ftdir} in the text. The time dependency of $\GammaTcore$ comes from $\fTdir(t)$, which is defined in the following way.

Let $\nqdir (t)$ be the total hydrogenic inventory in the DIR loop, and $\ntdir(t)$ the tritium inventory in the same loop. The tritium fraction in the DIR stream is then defined as:
\begin{equation}
\fTdir(t) \equiv \frac{\ntdir(t)}{\nqdir(t)}.
\end{equation}

A fraction $\fdir$ of the exhaust is routed into the DIR. Therefore, the tritium inflow to the DIR is
\begin{equation}
\dot N_{\mathrm{T,in}}^\mathrm{DIR} = \fdir \GammaTex.
\end{equation}

If the total hydrogenic outflow from the DIR is $\Gamma_Q^{DIR,out}$, the corresponding tritium outflow is
\begin{equation}
\dot N_{\mathrm{T,out}}^\mathrm{DIR} = \fTdir \Gamma_Q^\mathrm{DIR,out}.
\end{equation}

The tritium inventory balance is thus
\begin{equation}
\frac{d\ntdir}{dt} = \fdir \GammaTex - \fTdir \Gamma_Q^\mathrm{DIR,out}.
\end{equation}

The total hydrogenic inflow to the DIR is
\begin{equation}
\dot N_{\mathrm{Q,in}}^\mathrm{DIR} = \fdir \GammaQex.
\end{equation}

Therefore, the total hydrogenic inventory in the DIR evolves according to
\begin{equation}
\frac{d\nqdir}{dt} = \fdir \GammaQex - \Gamma_Q^\mathrm{DIR,out}.
\end{equation}

Because the DIR must feature a short residene time, we can assume that the total inventory in the DIR loop quickly reaches a steady state:
\begin{equation}
\nqdir(t)=N_{DIR}=\mathrm{const.}
\end{equation}
Hence,
\begin{equation}
\Gamma_Q^\mathrm{DIR,out}=\fdir\GammaQex.
\end{equation}

Substituting the previous result into the tritium balance yields
\begin{equation}
\frac{d\ntdir}{dt} = \fdir \GammaTex
- \fTdir\fdir\GammaQex.
\end{equation}

From the definition of $\fTdir$ one gets:
\begin{equation}
\frac{d\fTdir}{dt}
=
\frac{1}{N_{DIR}}\frac{dN_T^{DIR}}{dt}.
\end{equation}

Substituting the tritium inventory balance gives:
        
        \begin{equation}\label{eq:ftdir_apx}
            \frac{d\fTdir(t)}{dt} = \fdir \frac{\left(\GammaTex(t) - \fTdir(t)\GammaQex\right)}{N_\mathrm{DIR}}
        \end{equation}

\begin{landscape}
    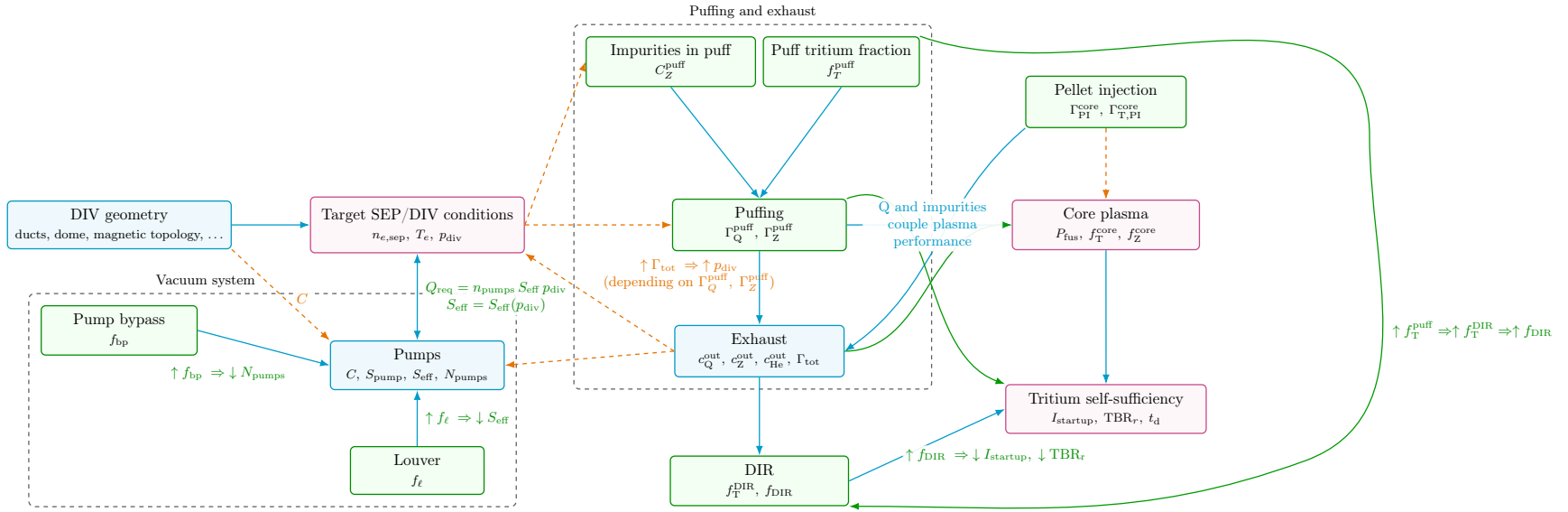
\begin{figure}[p]
\centering
\makebox[\linewidth][c]{%
\begin{adjustbox}{max totalsize={0.99\linewidth}{0.95\textheight},center}
\begin{tikzpicture}[
    scale=1.08,
    transform shape,
    node distance=1.25cm and 1.9cm,
    font=\normalsize,
    box/.style={
        draw,
        rounded corners=3pt,
        align=center,
        minimum width=3.8cm,
        minimum height=1.0cm,
        inner sep=5pt,
        line width=0.8pt,
        fill opacity=0.16,
        text opacity=1
    },
    softbox/.style={
        box,
        draw=cyan!75!black,
        fill=cyan!35
    },
    auxbox/.style={
        box,
        draw=green!55!black,
        fill=green!25
    },
    outcome/.style={
        box,
        draw=magenta!80!black,
        fill=magenta!22
    },
    group/.style={
        draw=gray!75!black,
        dashed,
        rounded corners=4pt,
        inner sep=8pt,
        line width=0.8pt
    },
    line/.style={-Latex, thick, draw=cyan!80!black},
    req/.style={-Latex, thick, dashed, draw=orange!90!black},
    feedback/.style={-Latex, thick, draw=red!80!black},
    reqfeedback/.style={-Latex, thick, dashed, draw=red!80!black},
    softline/.style={-Latex, thick, draw=green!60!black},
    bothline/.style={<->, thick, >=Latex, draw=cyan!80!black},
    note/.style={
        font=\small,
        inner sep=2pt,
        fill=white,
        fill opacity=0.92,
        text opacity=1,
        rounded corners=1pt,
        align=center
    },
]

\node[auxbox, minimum width=3.9cm] (puffsplit)
{Puffing\\{\footnotesize $\GammaQ,\; \GammaZ$}};

\node[auxbox, anchor=south east] (imp)
    at ([xshift=-0.75mm,yshift=2.5cm]puffsplit.north)
    {Impurities in puff\\{\footnotesize $C_Z^{\mathrm{puff}}$}};

\node[auxbox, anchor=south west, minimum width=3.2cm] (ftpuff)
    at ([xshift=0.75mm,yshift=2.5cm]puffsplit.north)
    {Puff tritium fraction\\{\footnotesize $f_T^{\mathrm{puff}}$}};

\node[softbox, below=1.65cm of puffsplit, minimum width=3.8cm] (qz)
{Exhaust\\{\footnotesize $c_{\rm Q}^{\rm out} ,\; \cZ,\; \cHe,\; \Gamma_\mathrm{tot}$}};

\node[auxbox, below=1.75cm of qz, minimum width=4.0cm] (dir)
{DIR\\{\footnotesize $f_{\rm T}^{\mathrm{DIR}},\; f_\mathrm{DIR}$}};

\node[outcome, right=3.7cm of puffsplit, minimum width=4.2cm] (core)
{Core plasma\\{\footnotesize $P_{\mathrm{fus}},\; f_{\rm T}^{\mathrm{core}},\; f_{\rm Z}^{\mathrm{core}}$}};

\node[auxbox, above=1.6cm of core, minimum width=3.6cm] (pi)
{Pellet injection\\{\footnotesize $\Gamma_{\mathrm{PI}}^\mathrm{core}, \; \Gamma_{\mathrm{T,PI}}^\mathrm{core}$}};

\node[outcome, below=3.0cm of core, minimum width=4.5cm] (tss)
{Tritium self-sufficiency\\{\footnotesize $I_{\mathrm{startup}},\; \mathrm{TBR}_r,\; t_\mathrm{d}$}};

\node[outcome, left=3.3cm of puffsplit, minimum width=4.8cm, minimum height=1.25cm] (target)
{Target SEP/DIV conditions\\{\footnotesize $n_{e,\mathrm{sep}},\; T_e,\; p_{\mathrm{div}}$}};

\node[softbox, left=1.75cm of target, minimum width=4.4cm] (divgeom)
{DIV geometry\\{\footnotesize ducts, dome, magnetic topology, $\ldots$}};

\node[auxbox, below=1.25cm of divgeom, minimum width=3.5cm] (bypass)
{Pump bypass\\{\footnotesize $\fbp$}};

\node[softbox, below=1.95cm of target, minimum width=3.9cm] (pumps)
{Pumps\\{\footnotesize $C, \; S_\mathrm{\rm pump}, \; S_{\mathrm{eff}}, \; N_\mathrm{\rm pumps}$}};

\node[auxbox, below=1.25cm of pumps, minimum width=3.0cm] (louver)
{Louver\\{\footnotesize $f_{\ell}$}};

\node[group, fit=(pumps)(louver)(bypass),
      label={[font=\small, xshift=-1.5cm]above:{Vacuum system}}] (gvac) {};

\node[group, fit=(imp)(ftpuff)(puffsplit)(qz),
      label={[font=\small]above:{Puffing and exhaust}}] (gbalance) {};

\coordinate (ftpuff_loop1) at ($(core.east)+(3.8,2.0)$);
\coordinate (ftpuff_loop2) at ($(tss.south east)+(2.2,-0.6)$);

\draw[softline]
    (ftpuff.north east)
    to[out=-20,in=90] (ftpuff_loop1)
    to[out=-90,in=20]
    node[pos=0.5, right=4pt, note, text=green!55!black]
    {$\uparrow f_\mathrm{T}^\mathrm{puff} \Rightarrow \uparrow f_\mathrm{T}^\mathrm{DIR} \Rightarrow \uparrow f_\mathrm{DIR}$}
    (ftpuff_loop2)
    to[out=200,in=0] (dir.south east);

\draw[line] (imp.south) -- (puffsplit.north);
\draw[line] (ftpuff.south) -- (puffsplit.north);
\draw[line] (divgeom.east) -- (target.west);

\draw[req] (divgeom.south east) -- (pumps.north west)
    node[pos=0.72, above=5pt, note, text=orange!90!black]
    {$C$};

\draw[line]
    (bypass.east) -- (pumps.west)
    node[pos=0.72, below left=5pt and 2pt, note, text=green!55!black]
    {$\uparrow \fbp \;\Rightarrow\; \downarrow N_\mathrm{pumps}$};

\draw[bothline] (target.south) -- (pumps.north)
    node[midway, right=3pt, note, text=green!55!black]
    {$Q_\mathrm{req} = n_\mathrm{pumps} \, S_\mathrm{eff} \, p_\mathrm{div}$\\$S_\mathrm{eff} = S_\mathrm{eff}(p_\mathrm{div})$};

\draw[req] (qz.west) -- (pumps.east);

\draw[req]
    (qz.west) --
    node[midway, above right=4pt and 0.01pt, note, text=orange!90!black]
    {$\uparrow \Gamma_{\mathrm{tot}} \;\Rightarrow\; \uparrow p_{\mathrm{div}}$\\(depending on $\Gamma_Q^{\mathrm{puff}},\; \Gamma_Z^{\mathrm{puff}}$)}
    (target.south east);

\draw[line] (puffsplit.south) -- (qz.north);
\draw[line] (qz.south) -- (dir.north);
\draw[line] (puffsplit.east) -- (core.west);
\draw[line] (core.south) -- (tss.north);
\draw[line] (dir.east) -- (tss.west);

\draw[req] (target.east) -- (puffsplit.west);
\draw[req] (target.east) -- (imp.west);

\draw[softline, rounded corners=10pt]
    (qz.east) to[out=0,in=180] (core.west);

\draw[req] (pi.south) -- (core.north);

\draw[line]
    (pi.south west) to[out=-140,in=35] (qz.east);

\coordinate (pi_loop) at ($(tss.east)+(1.6,0.9)$);

    

\draw[softline]
    (puffsplit.north east) to[out=20,in=160] (tss.north west);

\draw[line]
    (louver.north) --
    node[midway, right=3pt, note, text=green!55!black]
    {$\uparrow f_{\ell} \;\Rightarrow\; \downarrow S_{\mathrm{eff}}$}
    (pumps.south);

\node[note, text=green!55!black, above right=10pt and 1.2cm of dir.east]
    {$\uparrow f_\mathrm{DIR} \;\Rightarrow\; \downarrow I_{\mathrm{startup}},\; \downarrow \mathrm{TBR}_\mathrm{r}$};

\node[note, text=cyan!80!black, text width=3cm]
    at ($(puffsplit.east)!0.52!(core.west)$)
    {Q and impurities\\couple plasma performance};

\end{tikzpicture}
\end{adjustbox}%
}
\caption{
Dashed arrows denote \emph{requirements/constraints}; solid arrows denote
\emph{direct physical or system coupling}. Green boxes: operational knobs. Magenta boxes: design targets. Cyan boxes: system/plasma-response quantities.}
\label{fig:plasma_tfc_flowchart}
\end{figure}
\end{landscape}

\bibliographystyle{IEEEtran}
\bibliography{00_references, 00_ref_sm}

\end{document}